\documentclass{article}

\usepackage{amssymb,amsmath,amsbsy,bm}
\usepackage{natbib}
\usepackage{graphicx}
\usepackage{caption}
\usepackage{float}
\usepackage{subcaption}
\usepackage{algorithm}
\usepackage{authblk}
\usepackage{algorithmicx}
\usepackage{algpseudocode}

\usepackage{array}
\usepackage{makecell}
\usepackage[percent]{overpic}
\usepackage{pict2e}
\usepackage{xcolor}
\usepackage[export]{adjustbox}
\usepackage{hyperref}
\hypersetup{colorlinks=true}

\usepackage{tikz}
\usetikzlibrary{shapes,arrows,positioning,fit}

\newcommand{\bd}{\mathbf{d}}

\newcommand{\bu}{\mathbf{u}}
\newcommand{\bv}{\mathbf{v}}

\newcommand{\bk}{\mathbf{k}}
\newcommand{\bx}{\mathbf{x}}
\newcommand{\btau}{\pmb{\tau}}
\newcommand{\bphi}{\pmb{\varphi}}
\newcommand{\btheta}{\pmb{\theta}}

\newcommand{\nr}{n_r}

\newcommand{\nx}{n_{\bx}}
\newcommand{\nt}{n_t}

\newcommand{\ii}{\mathrm{i}}
\newcommand{\FF}{\mathcal{F}}

\newcommand{\CC}{\mathbb{C}}

\newcommand{\RR}{\mathbb{R}}

\renewcommand{\SS}{\mathcal{S}}

\newcommand{\norm}[1]{\left\lVert#1\right\rVert}

\newcommand{\tr}{\mathrm{H}}

\usepackage{xcolor}
\usepackage[normalem]{ulem}

\begin{document}

\title{Towards retrospective motion correction and reconstruction for clinical 3D brain MRI protocols with a reference contrast}
\date{}
\author[1,2]{\small Gabrio Rizzuti}
\author[2]{\small Tim Schakel}
\author[2]{\small Niek R. F. Huttinga}
\author[2]{\small Jan Willem Dankbaar}
\author[1,3]{\small Tristan van Leeuwen}
\author[2]{\small Alessandro Sbrizzi}
\affil[1]{\footnotesize Utrecht University, Utrecht, 3584 CS, The Netherlands}
\affil[2]{\footnotesize Universitair Medisch Centrum Utrecht, Utrecht, 3584 CX, The Netherlands}
\affil[3]{\footnotesize Centrum Wiskunde \& Informatica, Amsterdam, 1098 XG, The Netherlands}
\maketitle

\begin{abstract}
Motion artifacts often spoil the radiological interpretation of MR images, and in the most severe cases the scan needs be repeated, with additional costs for the provider. We discuss the application of a novel 3D retrospective rigid motion correction and reconstruction scheme for MRI, which leverages multiple scans contained in a MR session. Typically, in a multi-contrast MR session, motion does not equally affect all the scans, and some motion-free scans are generally available, so that we can exploit their anatomic similarity. The uncorrupted scan is used as a reference in a generalized rigid-motion registration problem to remove the motion artifacts affecting the corrupted scans. We discuss the potential of the proposed algorithm with a prospective in-vivo study and clinical 3D brain protocols. This framework can be easily incorporated into the existing clinical practice with no disruption to the conventional workflow.
\end{abstract}

\section*{Keywords}

Image Reconstruction, Motion Correction, Multi-Contrast, Brain, Total variation

\section{Introduction}

Magnetic resonance imaging (MRI) is fundamentally prone to motion artifacts, since the data acquisition process usually lasts several minutes for each acquired contrast, and the MR exam can be an uncomfortable experience for the patient. Motion corruption impedes a correct radiological assessment, which then may require a scan repetition, leading to considerable waste of resources for the hospital \citep{Andre2015}.

Motion reduction strategies are broadly classified as preventive, prospective, or retrospective techniques \citep{Zaitsev2015,godenschweger2016motion}. Preventive strategies include physical devices to limit the motion (e.g. head holders) or sedation, but their application is limited by ethical or health considerations, and are often ineffective in eliminating patient movement. Prospective and retrospective strategies, on the other hand, directly or indirectly estimate the motion that the object of interest undergoes inside the scanner, and remove its effect from the data or in the reconstruction phase. This correction step is said to be applied ``prospectively'' \citep{maclaren2013prospective} when the position of the patient is tracked in real time and the scan settings are adjusted accordingly on-the-fly. For example, the relative change of position can be estimated by acquiring additional $k$-space or image-space navigators \citep{ehman1989adaptive,welch2002spherical}, or with ``self-navigating'' sequences \citep{pipe1999motion,welch2004self,Bookwalter2010}. Alternatively, camera devices or markers \citep{zaitsev2006magnetic,forman2011self} can be used to estimate the imaging object position. However, most tracking modalities are often defective in terms of either precision, patient interaction, or sequence independence \citep{maclaren2013prospective}. Therefore, although effective in many respects, prospective methods have somewhat limited range of application.

Retrospective algorithms are characterized by the removal of motion artifacts in the final reconstruction phase, after the data acquisition. The main advantage of retrospective schemes is in their flexibility, since they do not necessarily require additional hardware, scanner modifications, MR navigators, markers, and so on. Note, however, that they may benefit from using prior information about the target imaging object and motion pattern. One main challenge for this class of methods is the need for time-intensive computations. The scientific literature on retrospective motion correction is quite rich: examples of retrospective techniques for rigid motion using navigators or markers can be found in \citet{Ehman1989,korin1995spatial,Mendes2009,Bookwalter2010,Vaillant2014}, while examples of ``blind'' techniques (in this context, meaning that they are not using navigators or markers) are presented in \citet{Atkinson1997,Atkinson1999,Manduca2000,Lin2006,Loktyushin2013}.

Retrospective correction schemes are typically formulated as a bi-level optimization problem, where two types of unknown are jointly estimated: the reconstructed (2D/3D) image and the motion parameters. Due to the ill-posedness of the problem here considered, the choice of the regularization method is crucial: see, for example, gradient-entropy regularization in \citet{Manduca2000,Lin2006,Loktyushin2013}, sparsity regularization in \citet{moller2015blind}, or iteratively re-weighted least-squares regularization in \citet{Cordero-Grande2020}. Another strategy to ease the ill-posedness is to resort to special acquisition patterns in $k$-space that are more robust in terms of motion correction, as described in the DISORDER method in \citet{Cordero-Grande2020}. Alternatively, many machine-learning approaches have been recently proposed for retrospective motion correction \citep{Pawar2018,Kustner2019,Haskell2019,lee2020deep,Ghaffari2021,MC2Net,Hossbach2022}.

Some previous work in \citet{rizzutiMOCO} introduced a retrospective motion correction scheme, whose novel aspect is the use of a contrast free of motion artifacts that can be leveraged as a reference to remove motion effects from any other contrast from the same patient, akin to a generalized rigid motion registration. The chief assumption of this work is the following: in a multi-contrast MR session, motion does not typically affect all the scans and some motion-free scans are generally available, so that we can exploit their anatomic similarity. Structural similarity is technically achieved via structure-guided total variation (TV), as originally proposed in \citet{Ehrhardt2016} and further developed in \citet{bungert2020robust} (see also \citet{Ehrhardt2015}).

The goal of this paper is to extend the scope of \citet{rizzutiMOCO}, limited to 2D synthetic results, to general 3D randomized acquisitions and 3D rigid-motion correction. We experimentally verify that a 3D extension is indeed feasible for brain imaging. We do not assume data-driven priors (so that machine learning is not available), any additional navigator data, nor consider motion-resilient acquisition schemes, in order to conform to more broadly available clinical protocols. Note that the proposed method can employ any acquisition scheme, in principle, but we stick to Cartesian acquisition, which are the standard encoding strategies of clinical protocols. Since we focus on brain imaging, rigid motion can be effectively assumed for our scope. The reference and the corrupted contrast do not need be co-registered or acquired with the same resolution.

We thoroughly validate the method with a prospective in-vivo study based on three volunteers and several motion types. The strength and limitations of the method are highlighted with the comparison of correction quality with varying degrees of motion artifacts and contrast type as a reference prior.

\section{Theory}\label{sec:theory}

In this section, we present the basic mathematical formulation underpinning the proposed motion correction method (further details can be found in \citet{rizzutiMOCO}).

The contrast volume, in the remainder of this section, will be denoted by $\bu\in\CC^{\nx}$, where $\nx$ is the number of voxels contained in a rectangular field of view. The 3D image undergoes a time-dependent rigid motion
\begin{equation}
    \bu_t=T_{\btheta_t}\bu,
\end{equation}
where $t$ is a time-related label. In practice, $t$ corresponds to the index of the $k$-space readout line in the phase-encoding plane. The corresponding rigid transformation is given by $T_{\btheta_t}$, and is parameterized by a time-dependent motion parameter $\btheta_t\in\RR^6$, which includes translations and rotations in 3D:
\begin{equation}
    \btheta=(\btau,\bphi),\qquad\btau=(\tau_x,\tau_y,\tau_z),\quad\bphi=(\varphi_{xy},\varphi_{xz},\varphi_{yz}).
\end{equation}
The rigid motion consists of a 3D rotation (defined by the 2D rotation angles $\varphi_{xy},\varphi_{xz},\varphi_{yz}$, performed in this order in the corresponding planes) followed by a translation (governed by the translation parameters $\tau_x,\tau_y,\tau_z$).

Without loss of generality, we are assuming a Cartesian acquisition. At each given time $t$, the MR acquisition process corresponds to the evaluation of the Fourier transform $\FF$ of $\bu_t$ in a particular subset $K_t$ of the $k$-space. In practice, the acquisition is structured in such a way that all the subsets $K_t$ consist of parallel lines in the $k$-space (the common direction being the readout direction). We refer to the Fourier transform of a rigidly moving object $\bu_{\btheta}:=T_{\btheta}\bu$ as the perturbed Fourier transform $\FF_{\btheta}\bu:=\FF\bu_{\btheta}$, and can be directly characterized as
\begin{equation}\label{eq:fouriermotion}
    \FF_{\btheta}\bu\,(\bk):=\exp{(-\ii\,\bk\cdot\btau)}\FF\bu\,(R_{\bphi}^{-1}\bk),
\end{equation}
where the rotational operator with respect to the 3D angle $\bphi$ is indicated by $R_{\bphi}$. This definition is motivated by classical Fourier identities that describe the action of rigid motion under the Fourier transform. Due to rotational effects, one must resort to the non-uniform discrete Fourier transform (NUFFT) to evaluate equation \eqref{eq:fouriermotion} \citep{Barnett2019, Barnett2021}.

Note that we implicitly assumed that no motion occurs while sampling the elements of $K_t$, since the state of the object at the time $t$ is associated to a single motion parameter $\btheta_t$. The assumption is motivated by the fact that $K_t$ will correspond, in practice, to a single Cartesian readout line, which lasts few milliseconds. Hence, the data acquisition at time $t$ is symbolized by the application of the selection operator $\SS_t$ to the Fourier-transformed volume:
\begin{equation}\label{eq:acq}
    \bd_t=\SS_t\FF_{\btheta_t}\bu=(\FF_{\btheta_t}\bu\,(\bk_1),\ldots,\FF_{\btheta_t}\bu\,(\bk_{\nr})),\qquad\bk_1,\ldots,\bk_{\nr}\in K_t.
\end{equation}
Here, $\nr$ is the number of $k$-space samples in a single readout.

The resulting inverse problem can be cast as an optimization problem over the reconstruction unknowns $\bu$ and the motion parameters $\btheta_t$, that is:
\begin{equation}\label{eq:opt}
    \min_{\bu,\btheta_{1:\nt}}f(\bu,\btheta_{1:\nt})+\lambda g_u(\bu)+\mu g_{\theta}(\btheta_{1:\nt}),
\end{equation}
where $\btheta_{1:\nt}=(\btheta_1,\ldots,\btheta_{\nt})$, and $\nt$ is the number of time steps. The weighting parameters $\lambda$, $\mu$ (both positive numbers) set the strength of the corresponding regularization terms. The first term of the objective functional in equation \eqref{eq:opt} corresponds to the data misfit:
\begin{equation}\label{eq:datamisfit}
    f(\bu,\btheta_{1:\nt})=\sum_{t=1}^{\nt}\dfrac{1}{2}\norm{\FF_{\btheta_t}\bu-\bd_t}^2.
\end{equation}
The least-squares norm is indicated here by $\norm{\cdot}$. The regularization terms $g_u$ and $g_{\theta}$ are crucial in ensuring the well-posedness of the problem. Indeed, the objective in equation \eqref{eq:datamisfit} will be sensitive to the relatively high signal-to-noise ratios (SNR) of the high-frequency components of the data. Moreover, the objective is highly non-convex as a function of $\btheta_{1:\nt}$. The motion-parameter regularization is designed to ensure some form of regularity in time (e.g. smoothness), this can be achieved for example by setting
\begin{equation}\label{eq:regpar}
    g_{\theta}(\btheta_{1:\nt})=\sum_{t=1}^{\nt-1}\dfrac{1}{2}\norm{\btheta_{t+1}-\btheta_t}^2.
\end{equation}
Alternatively, higher-order derivatives may be used. Another strategy, adopted in this paper, is to impose smoothness by setting hard constraints for the motion parameters, rather than via an additive penalty term as in equation \eqref{eq:regpar} \citep{rizzutiMOCO}. 

\subsection{Reference-guided total variation regularization}

The crux of the proposed method is related to the choice of the regularization term $g_u$ in equation \eqref{eq:opt}. We adopt the structure-guided total variation scheme proposed in \citet{Ehrhardt2016} in the context of multi-contrast imaging, that is:
\begin{equation}\label{eq:wtv}
    g_u(\bu)=\sum_{\bx}\norm{\Pi_{\bv}\rvert_{\bx}\nabla\bu\rvert_{\bx}},\quad\Pi_{\bv}\rvert_{\bx}=I_3-\xi_{\bv}\rvert_{\bx}\xi_{\bv}\rvert_{\bx}{}^{\tr},
\end{equation}
where $I_3$ is the $3\times3$ identity matrix, $\nabla\cdot\rvert_{\bx}$ is the discretized gradient operator evaluated at the voxel with center $\bx$, and $\Pi_{\bv}\rvert_{\bx}$ is the projection operator on the linear space that is orthogonal to the vector $\xi_{\bv}\rvert_{\bx}\in\CC^3$. The symbol $^{\tr}$ indicates the adjoint operation. The vector $\xi_{\bv}\rvert_{\bx}$ corresponds to the normalized gradient of a given motion-free contrast $\bv$, e.g. $\xi_{\bv}\rvert_{\bx}\approx\nabla\bv\rvert_{\bx}/\norm{\nabla\bv\rvert_{\bx}}$. The actual definition is
\begin{equation}\label{eq:wtv_projvec}
    \xi_{\bv}\rvert_{\bx}=\dfrac{\nabla\bv\rvert_{\bx}}{\sqrt{\norm{\nabla\bv\rvert_{\bx}}^2+\eta^2}},
\end{equation}
for some constant $\eta>0$. The regularization term in equation \eqref{eq:wtv} enforces the gradient structure of $\bv$ onto $\bu$, when $\bv$ and $\bu$ are anatomically compatible. It is important to observe that $\bv$ is not required to be registered with the target contrast $\bu$, since the estimation of the motion parameters in equation \eqref{eq:opt} will automatically compensate for the initial misalignment \citep[see also][]{bungert2020robust}. 
In this work, we actually adopt a constrained formulation of equation \eqref{eq:wtv}, meaning that structural similarity is imposed by forcing the solution to belong to the constraint set $C_u=\{\bu:g_u(\bu)\le\varepsilon\}$, where $\varepsilon>0$ is a prescribed regularization level \citep[see][for more details]{rizzutiMOCO,peters2018pmf}.

\subsection{Optimization}\label{sec:optimization}

In order to solve equation \eqref{eq:opt}, we adopt an alternating update scheme based on the proximal alternating minimization algorithm (PALM) described in \citet{bolte2014proximal}. The algorithm of the optimization strategy is exemplified in Algorithm \ref{alg:opt}. Each update requires the linearization of the smooth objective $f$ and the application of the proximal operators associated to $g_u$ and $g_{\theta}$. As it is commonly noted in the image registration literature, we will make use of multi-scale methods to ease the ill-posedness of the problem. Two types of scale are considered, here. One is traditionally associated to the reconstruction grid size, by considering a sequence of optimization problems defined on progressively finer grids. Note that spatial coarsening of the reconstructed image $\bu$ is intertwined with the temporal coarsening of the motion parameters $\btheta_{1:\nt}$, since they are associated to sample locations in the $k$-space. The other scale is related to the regularization strength $\lambda$, as defined in equation \eqref{eq:opt}. Hence, strongly weighted problems are solved first, and the regularization is gradually relaxed as in a continuation strategy. Overall, two nested sequences of optimization problems are considered here.
\begin{algorithm}
    \caption{Joint motion correction and reconstruction with alternating proximal operator evaluation}\label{alg:opt}
    \begin{algorithmic}
        \Require $\bd$, $\bu$(=0), $\btheta$(=0), $\alpha_u$, $\alpha_{\theta}$, $N$ \Comment{Data, starting guesses, steplengths, iters}
        \Ensure $\bu$, $\btheta$
        \For{scale = coarse,$\ldots$, fine}
            \State Downscaling of $\bd$, $\bu$, $\btheta$
            \For{$\lambda$ = high,$\ldots$, low}
                \For{$n=1:N$}
                    \State $\bu\gets\mathrm{prox}_{\alpha_u g_u}(\bu-\alpha_u\nabla_{\bu}f(\cdot,\btheta))$ \Comment{Reconstruction proxy}
                    \State $\btheta\gets\mathrm{prox}_{\alpha_{\theta}g_{\theta}}(\btheta-\alpha_{\theta}\nabla_{\btheta}f(\bu,\cdot))$ \Comment{Motion parameter proxy}
                \EndFor
            \EndFor
            \State Upscaling of $\bu$, $\btheta$
        \EndFor
    \end{algorithmic}
\end{algorithm}

\section{Experiments}\label{sec:exp}

In this section, we set up several experiments that demonstrate the capabilities of the retrospective motion correction algorithm detailed in Section \ref{sec:theory}, whose main novel aspect and strength is the use of a reference contrast to guide the correction. Our objective is to tackle motion correction for brain imaging, and we focus on acquisition protocols that are relevant for the clinical practice. All the imaging sequences considered in this study were taken from actual clinical brain protocols of the Radiology and Radiotherapy departments of the UMC Utrecht. The data considered in this section is based on 3D Cartesian acquisition. The sampling pattern used in these acquisitions typically utilizes pseudo-random undersampling. The main assumptions underlying the proposed method are related to the availability of a motion-free reference contrast and the motion artifacts being produced by rigid motion.

We consider several studies with volunteer data (three volunteers in total\footnote{We have informed written consent from the volunteers. The experiments were approved by the ethical review board of the UMC Utrecht.}), where motion artifacts are prospectively generated by instructing the volunteer to actively move during the scan (a certain number of times). While we did not track the type of rigid motion produced by the volunteers, we prompted them to maintain the same position in between our instructions. In this way, we have some fair qualitative expectations about the motion estimated by the correction algorithm (that is, a stepwise behavior, see also Appendix \ref{app:motionparameters} for the estimated motion unknowns associated to each of the experiments described in this section). The `ground-truth' acquisition and reconstruction is obtained by simply asking the volunteers not to move.

The volunteer studies aim at investigating several relevant questions related to the application of the proposed retrospective motion correction technique. The first study in Section \ref{exp:robust} is a qualitative assessment of the robustness of the motion correction with respect to motion complexity, here equated to the number of volunteer poses during the scan. In Section \ref{exp:prior}, we demonstrate that many combinations of corrupted-contrast and reference-contrast types are possible for adequate correction. In the experiment in Section \ref{exp:data}, we ascertain whether using the scanner reconstruction in the DICOM format, as opposed to the raw $k$-space data, is suited as input data for the algorithm after applying the Fourier transform (e.g., $\bd$ in equation \ref{eq:datamisfit}). We note that the proposed method assumes coil-resolved data as input for computational reasons, therefore it is sensitive to how the raw $k$-space data is post-processed, and, in particular, to the degree of which the post-processed data can be adequately corrected by rigid-motion estimation. Finally, further experimentation is deferred to the supplemental section in Appendix \ref{app:baseline}, where we demonstrate the effectiveness of the reference-based motion correction against a ``blind'' motion correction method, which does not use a reference contrast to eliminate the motion artifacts. 

All the following investigations use a 1.5 T Philips Ingenia scanner with a 15-channel head coil. We considered several contrast acquisition sequences with the specifications highlighted in Table \ref{tab:contrasts}. For all the experiments except the one described in Section \ref{exp:data}, the raw $k$-space data (pertaining to corrupted or ground-truth scans) was exported for off-line processing. A pre-processing step is dedicated to remove coil dependency, by performing a SENSE reconstruction and then applying the Fourier transform.
\begin{table}
    \centering
    \resizebox{\textwidth}{!}{%
    \begin{tabular}{|c|c|c|r|r|r|r|r|c|r|}
        \hline
        Experiment & Contrast & Sequence & Resolution & FOV & TR & TE & Flip angle & Phase-encoding pattern & Duration\\
        \hline
        Section \ref{exp:robust} & T2-FLAIR & 3D TSE & 1.2 mm$^3$ & 230$\times$230$\times$237.6 mm$^3$ & 4800 ms & 320 ms & 90$^{\circ}$ & Randomized & ~350 s\\
        & T1$^*$ & 3D TFE & 1 mm$^3$ & 230$\times$230$\times$238 mm$^3$ & 7.8 ms & 3.6 ms & 8$^{\circ}$ & Randomized & ~180 s\\
        \hline
        Section \ref{exp:prior} & T1$^{\phantom{*}}$ & 3D TFE & 1 mm$^3$ & 230$\times$230$\times$238 mm$^3$ & 7.9 ms & 3.6 ms & 8$^{\circ}$ & Randomized & ~180 s\\
        & T2$^*$ & 3D TSE & 1.1 mm$^3$ & 250$\times$250$\times$190.3 mm$^3$ & 3000 ms & 260 ms & 90$^{\circ}$ & Randomized & ~180 s\\
        \hline
        Section \ref{exp:data} & T2$^{\phantom{*}}$ & 3D TSE & 1.3 mm$^3$ & 250$\times$250$\times$183.26 mm$^3$ & 2000 ms & 318 ms & 90$^{\circ}$ & Regular (acc. 2$\times$2) & ~300 s\\
        & T1$^*$ & 3D TFE & 1 mm$^3$ & 250$\times$250$\times$183 mm$^3$ & 7.7 ms & 3.6 ms & 8$^{\circ}$ & Randomized & ~150 s\\
        \hline
        Section \ref{exp:data} & T2-FLAIR & 3D TSE & 1.2 mm$^3$ & 230$\times$230$\times$238 mm$^3$ & 4800 ms & 291 ms & 90$^{\circ}$ & Randomized & ~350 s\\
        & T1$^*$ & 3D TFE & 1 mm$^3$ & 230$\times$230$\times$238 mm$^3$ & 7.5 ms & 3.4 ms & 8$^{\circ}$ & Randomized & ~180 s\\
        \hline
    \end{tabular}%
    }%
    \caption{Specification of the acquisition sequences utilized in the experiments in Section \ref{sec:exp}. We use a 1.5 T Philips Ingenia scanner with a 15-channel head coil. For each experiment, the asterisk indicates the reference contrast. The ``randomized'' sampling pattern indicated in this table more specifically refers to variable density Cartesian randomized undersampling, while the ``regular'' pattern refers to classical accelerated linear filling undersampling.}\label{tab:contrasts}
\end{table}

\subsection{Experiment 1: robustness with respect to motion complexity}\label{exp:robust}

In order to test the robustness of the proposed motion correction scheme in terms of motion complexity, we instruct volunteer 1 to move multiple times during acquisition. With ``motion complexity'' we specifically refer to the number of position changes performed by the volunteer within one prospectively corrupted scan. The goal of this in-vivo study is to provide a qualitative assessment of the degradation of the reconstruction quality as a function of motion complexity.

We consider three levels of motion corruption: (\textit{i}) the volunteer moves once, (\textit{ii}) the volunteer moves twice, and (\textit{iii}) the volunteer moves five times. The volunteer is instructed to change its head position every time it is prompted to do so, and maintain that position in between instructions. We use T2-FLAIR-weighted contrasts as corrupted scans, with T1-weighted contrast as a reference (see Table \ref{tab:contrasts} for further details). The corrupted acquisition employs randomized sampling.

The results of this experiment are collected in Section \ref{res:robust}. Note that, in Appendix \ref{app:baseline}, we use the same settings detailed in this experiment to compare the proposed algorithm with a baseline method without a reference guide.

\subsection{Experiment 2: on the choice of the reference contrast}\label{exp:prior}

This in-vivo experiment tests the proposed correction scheme with respect to a different combination of corrupted and reference contrast, namely a T1-weighted corrupted contrast with a T2-weighted reference contrast (see Table \ref{tab:contrasts}). For this experiment, we prompt volunteer 2 to move five times during the acquisition. The corrupted acquisition employs randomized sampling.

In Section \ref{res:prior}, we gather the results for this experiment.

\subsection{Experiment 3: scanner reconstruction vs processed raw k-space data as input for retrospective motion correction}\label{exp:data}

With the in-vivo studies presented in this section, we investigate a question related to the nature of the input data $\bd$ (equation \ref{eq:datamisfit}) required by the algorithm. Due to the formulation of the problem directly in $k$-space (by means of the NUFFT), the method assumes coil-resolved data. One must then assess whether the scanner reconstruction (available in the DICOM format) is suitable for this purpose, since many different reconstruction methods are available depending on the acquisition protocol. In particular, the default reconstruction method for linear-filling patterns in $k$-space employs the SENSE framework \citep{Pruessmann1999}, while compressed-sensing reconstruction (via the wavelet transform) is used for randomized acquisitions \citep{Lustig2007}. Note that our experimentation suggests that without the phase map of the scanner reconstruction our motion correction scheme does not perform adequately. Therefore, with ``scanner reconstruction'', we will always refer to the complex-valued scanner reconstruction (comprising both the respective amplitude and phase).

In the first experiment, we asked volunteer 3 to change position once during the prospectively-corrupted acquisition. We consider a corrupted T2-weighted contrast and a reference T1-weighted contrast (see Table \ref{tab:contrasts}). One important aspect of this experiment is related to the acquisition protocol of the T2-weighted contrast, based on a linear-filling pattern in $k$-space. The corrupted data used as input for the proposed motion-correction algorithm is obtained by exporting the reconstructed volume directly from the scanner, followed by a simple Fourier transform. Note that this 3D image has been obtained by a SENSE reconstruction.

The second experiment is set up similarly to the previous one. We asked volunteer 3 to change position only once during the acquisition phase. We consider, now, a corrupted T2-FLAIR-weighted contrast with a reference T1-weighted contrast (see Table \ref{tab:contrasts}). The most important difference with the previous experiment, besides the type of contrast pair considered, is related to the randomized acquisition protocol. In this case, the scanner reconstruction employs a compressed-sensing reconstruction, and is not suited as input for the proposed motion-correction algorithm (see Appendix \ref{app:scanreconvsrawdata3}). Therefore, for adequate motion correction, we must set up an intermediate step for processing the raw $k$-space data via the SENSE reconstruction.

We further discuss the results of this experiment in Section \ref{res:data}.

\section{Results}\label{sec:results}

In this section, we display and briefly analyze the results of the experiments presented in the previous section. For ease of exposition, we organized the power signal-to-noise ratio (PSNR) and structural similarity index (SSIM) values of the reconstructions (with respect to a known ground truth) in Table \ref{tab:results}. %
\begin{table}[hbt]
    \centering
    \resizebox{\textwidth}{!}{%
    \begin{tabular}{|l|l|cc|cc|}
        \hline
        Experiment & Slice orientation & \multicolumn{2}{c|}{PSNR ($\uparrow$)} & \multicolumn{2}{c|}{SSIM ($\uparrow$)}\\
        & & Corrupted & Corrected & Corrupted & Corrected\\
        \hline
        Section \ref{exp:robust}, Figure \ref{fig:robustness1_vol2} & Sagittal & 23.94 & \textbf{27.95} & 0.7068 & \textbf{0.7936}\\
                                          & Coronal  & 26.66 & \textbf{29.82} & 0.7653 & \textbf{0.8332}\\
                                          & Axial    & 25.40 & \textbf{30.16} & 0.7616 & \textbf{0.8490}\\
        \hline
        Section \ref{exp:robust}, Figure \ref{fig:robustness2_vol2} & Sagittal & 25.78 & \textbf{27.76} & 0.7263 & \textbf{0.7816}\\
                                          & Coronal  & 28.19 & \textbf{29.73} & 0.7847 & \textbf{0.8244}\\
                                          & Axial    & 27.79 & \textbf{29.70} & 0.8104 & \textbf{0.8362}\\
        \hline
        Section \ref{exp:robust}, Figure \ref{fig:robustness3_vol2} & Sagittal & 22.45 & \textbf{25.28} & 0.6116 & \textbf{0.7661}\\
                                          & Coronal  & 24.54 & \textbf{27.40} & 0.6734 & \textbf{0.8060}\\
                                          & Axial    & 24.15 & \textbf{27.66} & 0.7086 & \textbf{0.8298}\\  
        \hline
        Section \ref{exp:prior}, Figure \ref{fig:prior} & Sagittal & 25.84 & \textbf{28.07} & 0.7032 & \textbf{0.8093}\\
                               & Coronal  & 26.35 & \textbf{30.40} & 0.7851 & \textbf{0.9021}\\
                               & Axial    & 28.11 & \textbf{30.54} & 0.8248 & \textbf{0.9012}\\
        \hline
        Section \ref{exp:data}, Figure \ref{fig:scanreconvsrawdata1} & Sagittal & 22.26 & \textbf{27.54} & 0.6963 & \textbf{0.8409}\\
                                             & Coronal  & 23.46 & \textbf{31.65} & 0.7321 & \textbf{0.8370}\\
                                             & Axial    & 24.55 & \textbf{32.33} & 0.7895 & \textbf{0.8144}\\
        \hline
        Section \ref{exp:data}, Figure \ref{fig:scanreconvsrawdata2} & Sagittal & 24.72 & \textbf{28.76} & 0.6762 & \textbf{0.7818}\\
                                             & Coronal  & 25.95 & \textbf{29.54} & 0.7238 & \textbf{0.8107}\\
                                             & Axial    & 25.08 & \textbf{29.59} & 0.7263 & \textbf{0.8407}\\
        \hline
    \end{tabular}%
    }%
    \caption{Summary of the motion-correction results shown in Section \ref{sec:results} in terms of PSNR and SSIM}\label{tab:results}
\end{table}

The motion-corrected full-volume scans were analyzed by a neuroradiologist with 16 years of experience. These were generally deemed of good radiological quality. The motion-related artifacts have been completely removed, and the results are quite close to the ground truth. In Table \ref{tab:radiol}, we organized a more detailed qualitative analysis of the 3D results, geared toward a radiological assessment of the corrected scans.
\begin{table}[!htb]
    \centering
    \resizebox{\textwidth}{!}{%
    \begin{tabular}{|l|l|l|l|l|l|}
        \hline
        Experiment & Contrast & Motion resolution & Blurring & Artifacts & Additional comments\\
        \hline
        Section \ref{exp:robust}, Figure \ref{fig:robustness1_vol2} & T2-FLAIR & Completely corrected & Some blurring & No additional artifacts & Good grey white matter differentiation\\
        \hline
        Section \ref{exp:robust}, Figure \ref{fig:robustness2_vol2} & T2-FLAIR & Completely corrected & Some blurring & No additional artifacts & Good grey white matter differentiation\\
        \hline
        Section \ref{exp:robust}, Figure \ref{fig:robustness3_vol2} & T2-FLAIR & Completely corrected & Some blurring & Darker areas within the white matter & Good grey white matter differentiation\\
        \hline
        Section \ref{exp:prior}, Figure \ref{fig:prior} & T1 & Completely corrected & Some blurring & No additional artifacts & Good grey white matter differentiation,\\
        & & & & & some loss of grey matter low signal\\
        \hline
        Section \ref{exp:data}, Figure \ref{fig:scanreconvsrawdata1} & T2 & Completely corrected & No blurring & No additional artifacts &\\
        \hline
        Section \ref{exp:data}, Figure \ref{fig:scanreconvsrawdata2} & T2-FLAIR & Completely corrected & Some blurring & No additional artifacts & Good grey white matter differentiation\\ 
        \hline
    \end{tabular}%
    }%
    \caption{Qualitative radiological analysis of the motion-corrected results shown in Section \ref{sec:results}. The corrected scans are radiologically equivalent to the ground truth.}\label{tab:radiol}
\end{table}

\subsection{Experiment 1: robustness test}\label{res:robust}

We gather the results for the robustness test described in Section \ref{exp:robust} (volunteer 1) in Figures \ref{fig:robustness1_vol2}, \ref{fig:robustness2_vol2}, and \ref{fig:robustness3_vol2} for motion corruption mechanisms associated to one, two, and five changes of position, respectively. Furthermore, we juxtapose the corrected images with varying degrees of corruption in Figure \ref{fig:robustness_comparison}. We observe that the proposed method consistently ameliorates the corrupted scan. The quality indexes based on PSNR and SSIM show only a modest decrease in correction quality as a function of motion complexity (Figure \ref{fig:robustness_comparison}).

\subsection{Experiment 2: choice of the reference contrast}\label{res:prior}

With the experiment described in Section \ref{exp:prior}, we demonstrate the flexibility of the correction scheme with respect to the choice of the reference contrast. The results are shown in Figure \ref{fig:prior}. Contrary to the experiments detailed in the previous section, we are now considering a T2-weighted reference contrast to guide the correction of a T1-weighted corrupted contrast. The quality of the correction indicates that the proposed technique is rather flexible in terms of reference contrast.

\subsection{Experiment 3: scanner reconstruction vs raw k-space data}\label{res:data}

The results of the two experiments described in Section \ref{exp:data} are depicted in Figures \ref{fig:scanreconvsrawdata1} and \ref{fig:scanreconvsrawdata2}. The main difference between the two experiments is related to the input data for the proposed motion-correction algorithm.

In the first experiment, the corrupted contrast has been acquired with a protocol based on a linear filling pattern in $k$-space. Note that, in this particular case, the scanner reconstruction implements the SENSE method. We then extracted the DICOM of both amplitude and phase produced by the scanner, and used it as input data (after a Fourier transform) for the algorithm. The proposed scheme is able to successfully remove the motion artifacts in Figure \ref{fig:scanreconvsrawdata1}.

In the case of randomized sampling, the scanner reconstruction is not adequate as input data for the proposed  motion-correction algorithm, because it employs a compressed-sensing algorithm. We speculate that compressed-sensing reconstructions degrade the information contained in the corrupted volume, and the corrected contrast cannot be effectively recovered by simply removing rigid-motion artifacts (we defer the degraded results when using scanner reconstruction data in Appendix \ref{app:scanreconvsrawdata3}). However, when the input data is obtained by directly processing the raw $k$-space data via the SENSE reconstruction, the motion-correction scheme is able to successfully remove the motion artifacts (Figure \ref{fig:scanreconvsrawdata2}).

\section{Discussion}

Reference-guided TV regularization substantially improves the motion correction quality, both visually and in terms of quality metrics based on PSNR and SSIM, when compared to basic Fourier reconstruction without motion correction. The comparison is also substantially favorable with standard ``blind'' motion correction techniques, for example based on conventional regularization such as TV, which do not employ a reference to guide the correction (see Appendix \ref{app:baseline}). In fact, for randomized sampling patterns that are now common in the clinical practice, we verified that blind retrospective techniques are wholly inadequate for motion correction of radiological quality (cf. the comparison in Appendix \ref{app:baseline}, Figure \ref{fig:baseline}).

Our experimentation based on volunteer data aimed at assessing the robustness of the correction quality with respect to motion artifacts of increasing complexity. In this study, we equated this complexity to the number of volunteer changes of pose during the acquisition phase. Clearly, this does not fully describe the complexity of motion encountered in practice in the clinic, but it only constitutes a preliminary step in that direction. Nevertheless, the results described in Section \ref{res:robust} support the indication that the retrospective motion correction of T2-FLAIR weighted images based on a T1 reference contrast is quite robust in terms of reconstruction quality, with only minor degradations in terms of contrast and resolution.

Furthermore, the flexibility of the proposed motion-correction method is demonstrated with different combinations of motion-corrupted and reference contrasts (Section \ref{res:prior}). Our experience suggests that an important factor in assessing the effectiveness of the reference contrast as a guide for motion correction lies in the similarity of the $k$-space distribution of the two contrasts. Good reconstruction quality can be expected when the reference contrast has similar or higher frequency content when compared to the corrupted contrast, regardless of the type of contrast considered.

A significant part of our experimentation was devoted to assess whether the scanner reconstruction (available as DICOM format) can be directly used as input data for the proposed correction method (Section \ref{res:data}). We established that the scanner reconstruction is not suitable for this purpose when it is obtained via compressed-sensing algorithms (Appendix \ref{app:scanreconvsrawdata3}), which is the case for randomized sampling on the 1.5 T Philips Ingenia scanner utilized in this work. In this case, we must resort to the raw $k$-space data and perform an intermediate SENSE reconstruction for effective motion correction.

The computational times of the motion correction are, generally speaking, problem dependent, since complex motion artifacts require an increasing number of iterations as a function of motion complexity (Section \ref{sec:optimization}). The examples illustrated in this study, where a fixed number of iterations was consider irrespectively of motion complexity, are completed within 1 h 30 min for 3D images of approximately 256$\times$256$\times$256 voxels. The current CPU implementation was run on a consumer-grade laptop with the following processor specifications: Intel Core i7-10750H CPU@2.60GHz$\times$12. An effective implementation in a clinical scenario for on-line reconstructions will likely require GPUs.

The basic assumption of the proposed retrospective correction method is related to the availability of a motion-free contrast. While we believe that it is a realistic possibility within an MR session, we note that the reference contrast may come from previous MR sessions (or even different imaging modalities altogether, such as CT). In this particular case, the bias introduced by the structural prior may have an adverse effect in case of an evolving pathology. However, when structural changes involve a limited pathological region, the adverse bias can be easily mitigated by masking the affected zone.

Note that the motion-free reference can be exploited differently than the reference-guided TV regularization introduced in \citet{Ehrhardt2016}, and adopted in this work. For example, one may consider several competing techniques advanced for multi-contrast MRI, such as Bayesian compressed sensing \citep{bilgic2011multi}, group sparsity \citep{huang2014fast}, reference-based MRI \citep{weizman2015fast}, or multi-contrast graph-based sparsity \citep{lai2017sparse,lai2018joint}.

The method here presented is limited to rigid motion. Indeed, some decrease in correction quality is noticeable in Figure \ref{fig:scanreconvsrawdata1} in the neck region (which is not supposed to behave rigidly). However, our technique may be extended to non-rigid motion and, hence, different body regions other than the brain \citep[see, for example,][]{huttinga2020mr}. A major challenge for such extension is a computationally effective parameterization of the motion effects, and the resulting ill-posedness of the inverse problem. Note that a significant computational advantage of rigid motion over non-rigid motion is related to the direct implementation of the rigid motion in $k$-space, via equation \eqref{eq:fouriermotion}, which results in a data model that requires a single NUFFT evaluation, regardless of the number of time samples considered. Other interesting extensions of the method are related to the integration of specialized motion-resilient acquisition patterns, e.g. as described in \citet{Cordero-Grande2020}.

\section{Conclusions}

We assessed the performance of the proposed retrospective motion correction method based on a reference contrast not affected by motion artifacts. The current prospective in-vivo study targets 3D clinical protocols conventionally used in brain imaging.

The method is tested with several degrees of motion artifacts, by instructing the volunteers to change position during the scan multiple times. While, we observe that the corrupted images are severely degraded as a function of motion complexity, the corrected images are generally robustly estimated. We also verified that the proposed technique is agnostic with respect to the choice of the reference contrast, as long as the frequency content of the reference and target contrasts is comparable.

Further assessment of the proposed method will be devoted to patient data.

\section*{Data}

The 3D results of the experiment described in Sections \ref{sec:exp}, \ref{sec:results} are freely available online in the DICOM format at the following link:
\begin{center}
    \href{github.com/grizzuti/ReferenceGuidedMotionCorrection_Supplementary_DICOM}{{\small github.com/grizzuti/ReferenceGuidedMotionCorrection$\_$Supplementary$\_$DICOM}}.
\end{center}

\section*{Acknowledgments}

This publication is part of the project ``Reducing re-scans in clinical MRI exams'' (with project numbers 0104022007, 01040222210001 of the research program ``IMDI, Technologie voor bemensbare zorg: Doorbraakprojecten'') which is financed by The Netherlands Organization for Health Research and Development (ZonMW). The project is also supported by Philips Medical Systems Netherlands BV.

\begin{figure}[!htb]
    \centering
    \begin{subfigure}{\textwidth}
        \centering
        \rotatebox{90}{\hspace{2.5em}Sagittal} %
        \begin{subfigure}[b]{0.24\textwidth}
            \caption*{Corrupted}\vspace{-0.5em}%
            \begin{overpic}[width=\textwidth]{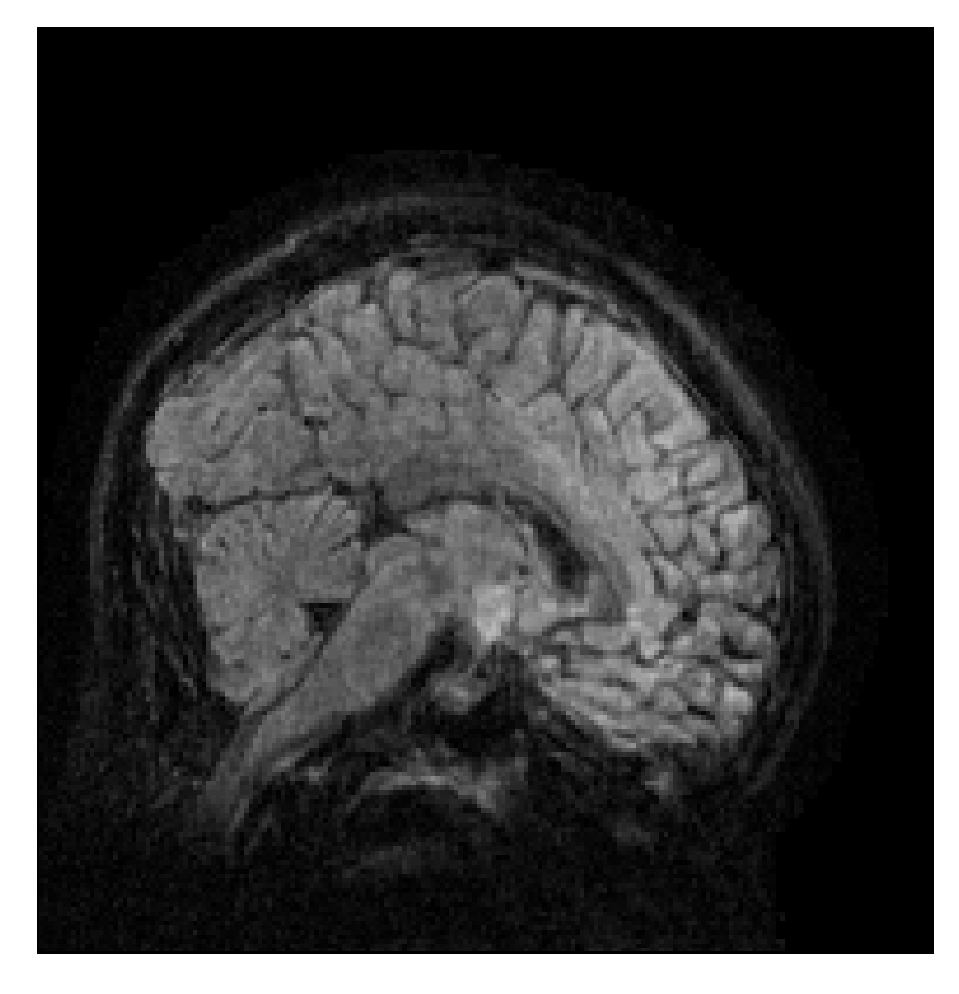}%
                \linethickness{2pt}
                \put(5,13){{\tiny\color{white}PSNR: 22.93}}
                \put(5,6){{\tiny\color{white}SSIM: 0.6655}}
            \end{overpic}%
        \end{subfigure}%
        \begin{subfigure}[b]{0.24\textwidth}
            \caption*{Corrected}\vspace{-0.5em}%
            \begin{overpic}[width=\textwidth]{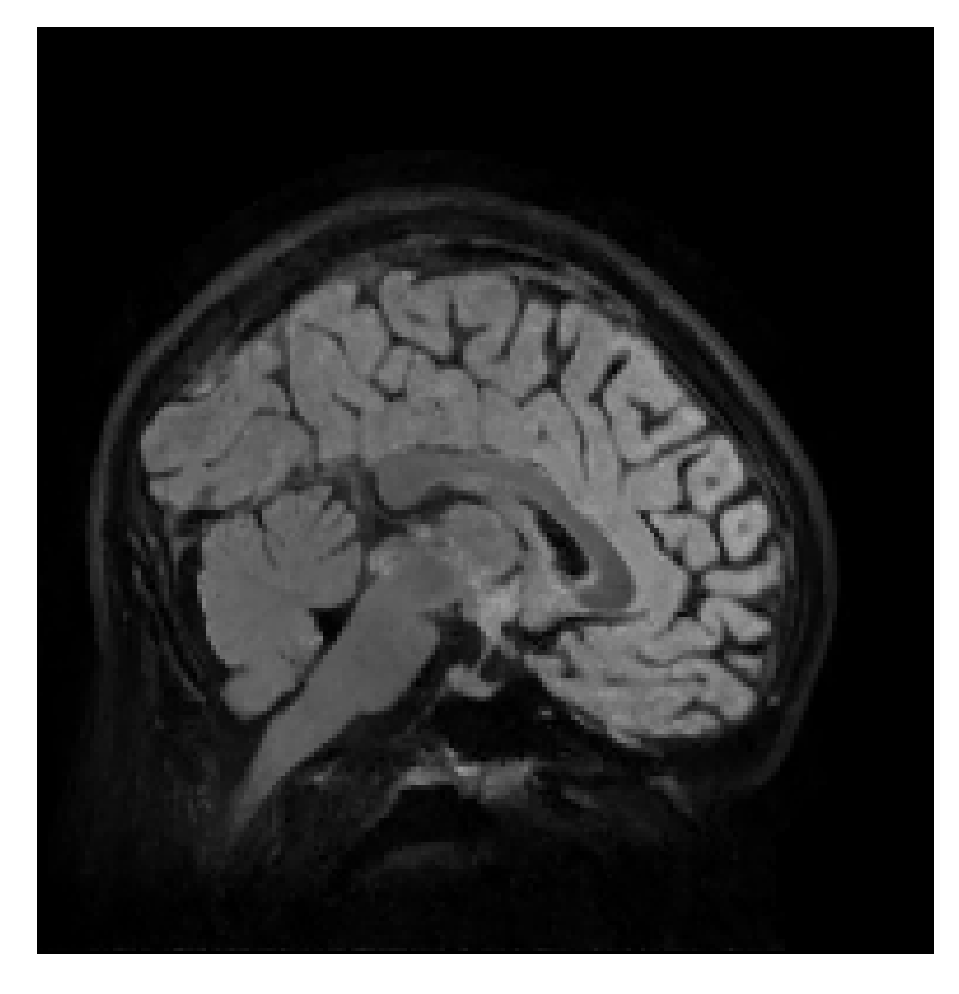}%
                \linethickness{2pt}
                \put(5,13){{\tiny\color{white}PSNR: 27.95}}
                \put(5,6){{\tiny\color{white}SSIM: 0.7936}}
            \end{overpic}%
        \end{subfigure}%
        \begin{subfigure}[b]{0.24\textwidth}
            \caption*{Ground truth}\vspace{-0.5em}%
            \includegraphics[width=\textwidth]{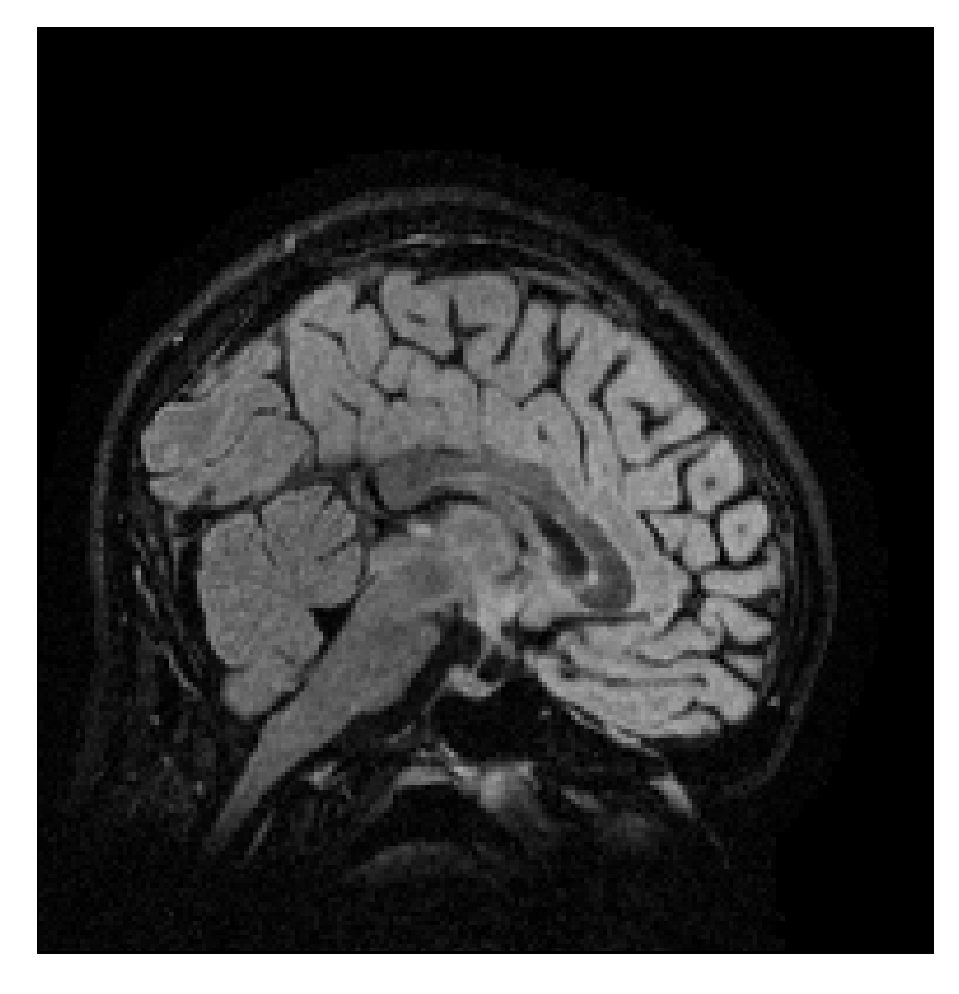}%
        \end{subfigure}%
        \begin{subfigure}[b]{0.24\textwidth}
            \caption*{Reference}\vspace{-0.5em}%
            \includegraphics[width=\textwidth]{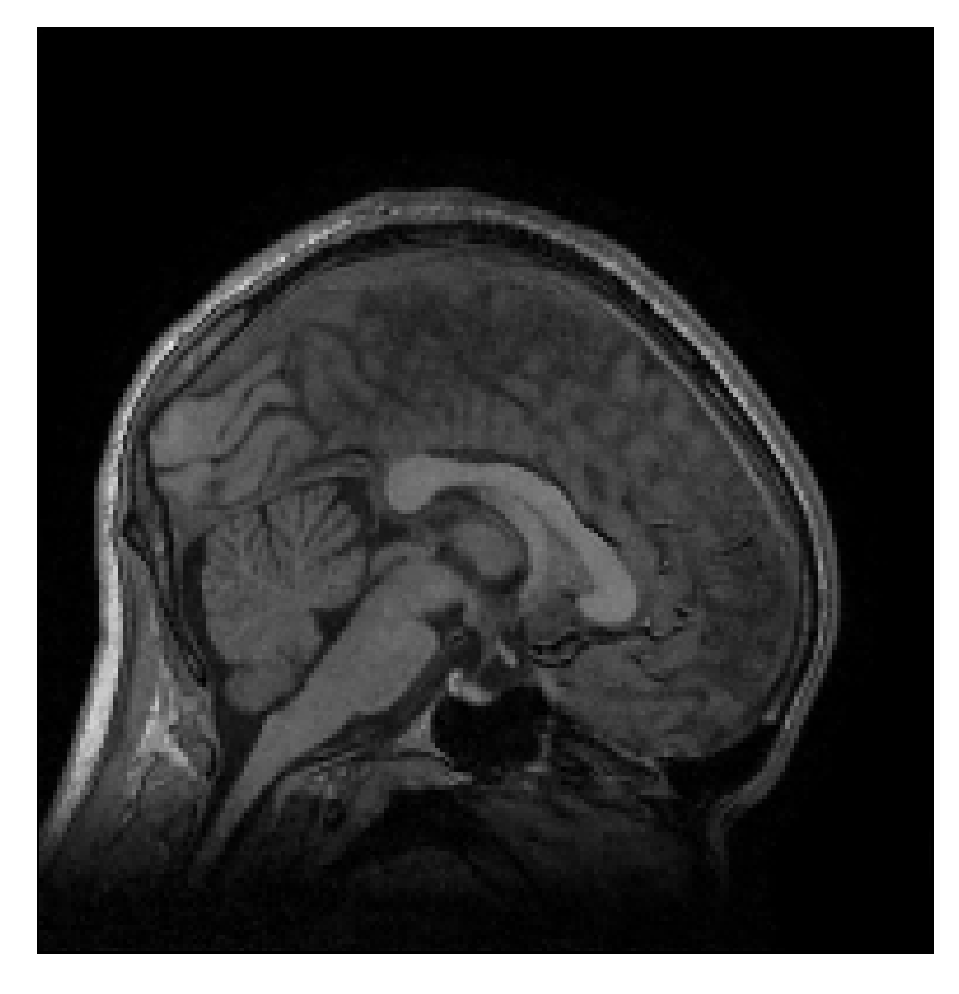}%
        \end{subfigure}%
    \end{subfigure}
    \begin{subfigure}{\textwidth}
        \centering
        \rotatebox{90}{\hspace{2.5em}Coronal} %
        \begin{subfigure}[b]{0.24\textwidth}
            \begin{overpic}[width=\textwidth]{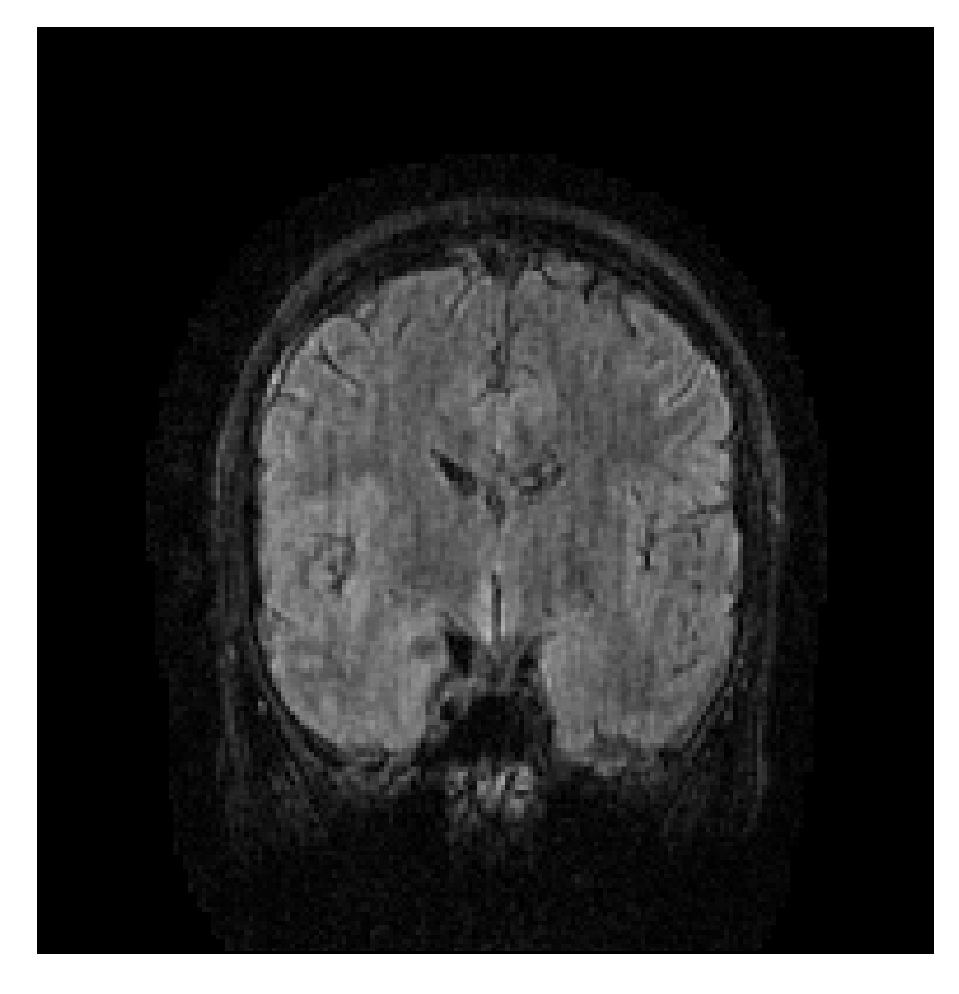}%
                \linethickness{2pt}
                \put(5,13){{\tiny\color{white}PSNR: 25.91}}
                \put(5,6){{\tiny\color{white}SSIM: 0.7461}}
            \end{overpic}%
        \end{subfigure}%
        \begin{subfigure}[b]{0.24\textwidth}
            \begin{overpic}[width=\textwidth]{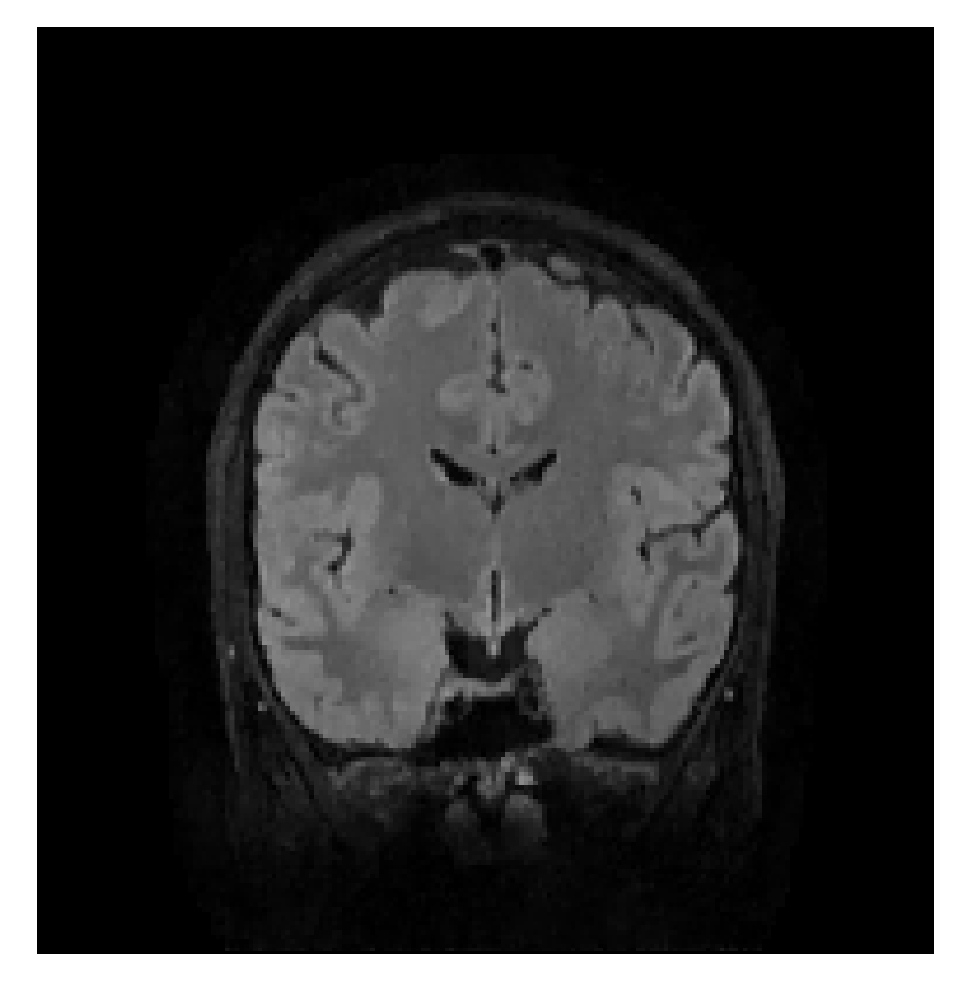}%
                \linethickness{2pt}
                \put(5,13){{\tiny\color{white}PSNR: 29.82}}
                \put(5,6){{\tiny\color{white}SSIM: 0.8332}}
            \end{overpic}%
        \end{subfigure}%
        \begin{subfigure}[b]{0.24\textwidth}
            \includegraphics[width=\textwidth]{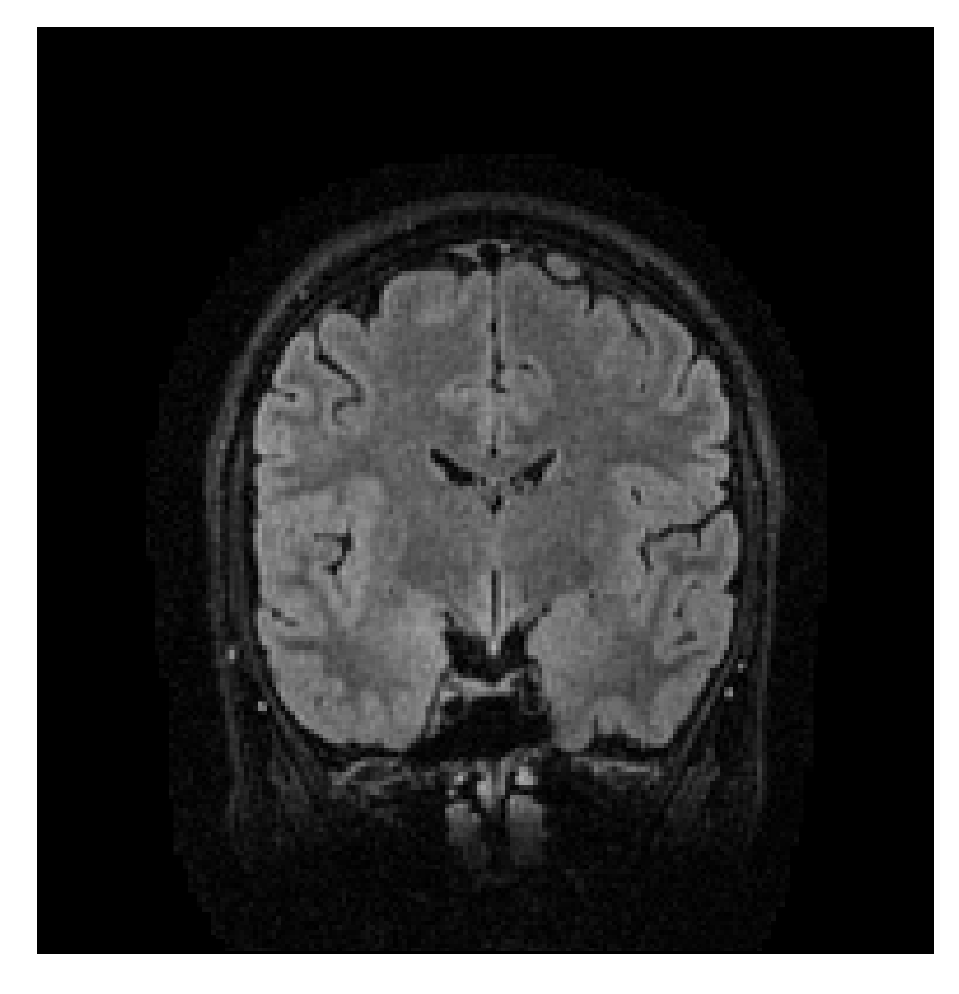}%
        \end{subfigure}%
        \begin{subfigure}[b]{0.24\textwidth}
            \includegraphics[width=\textwidth]{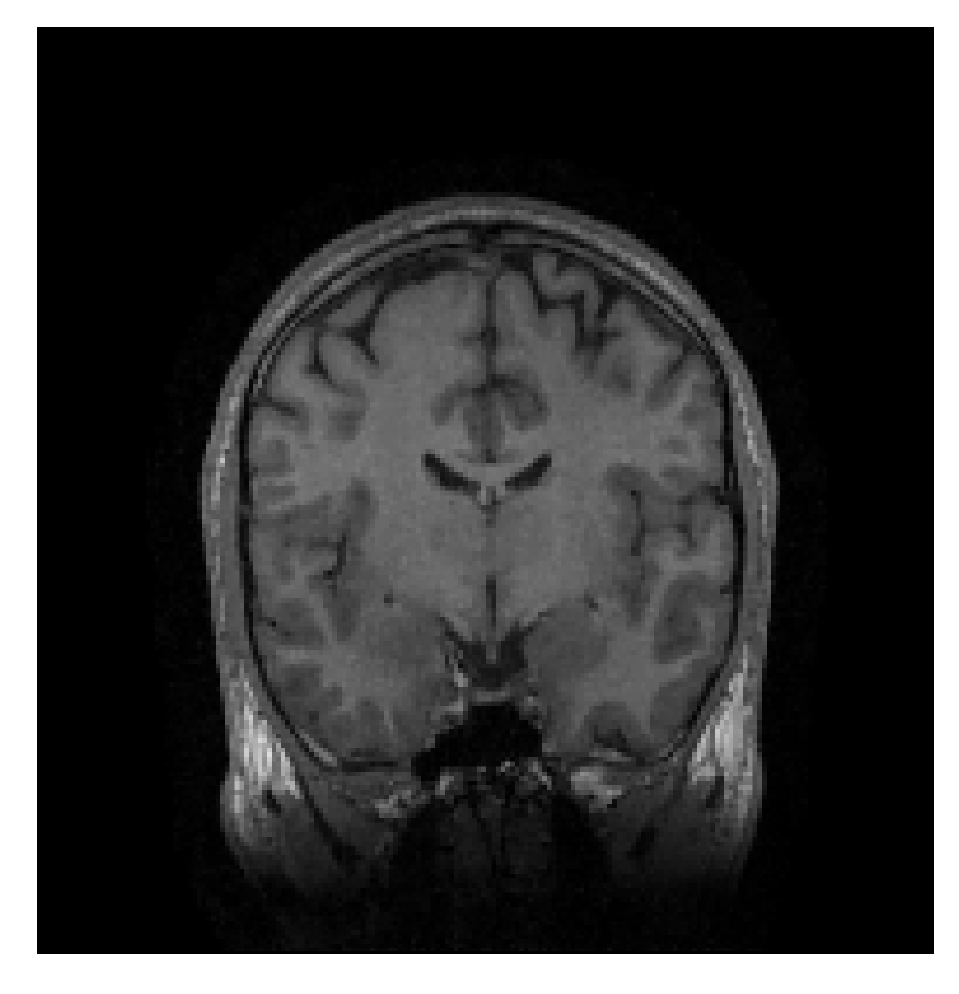}%
        \end{subfigure}%
    \end{subfigure}
    \begin{subfigure}{\textwidth}
        \centering
        \rotatebox{90}{\hspace{3em}Axial} %
        \begin{subfigure}[b]{0.24\textwidth}
            \begin{overpic}[width=\textwidth]{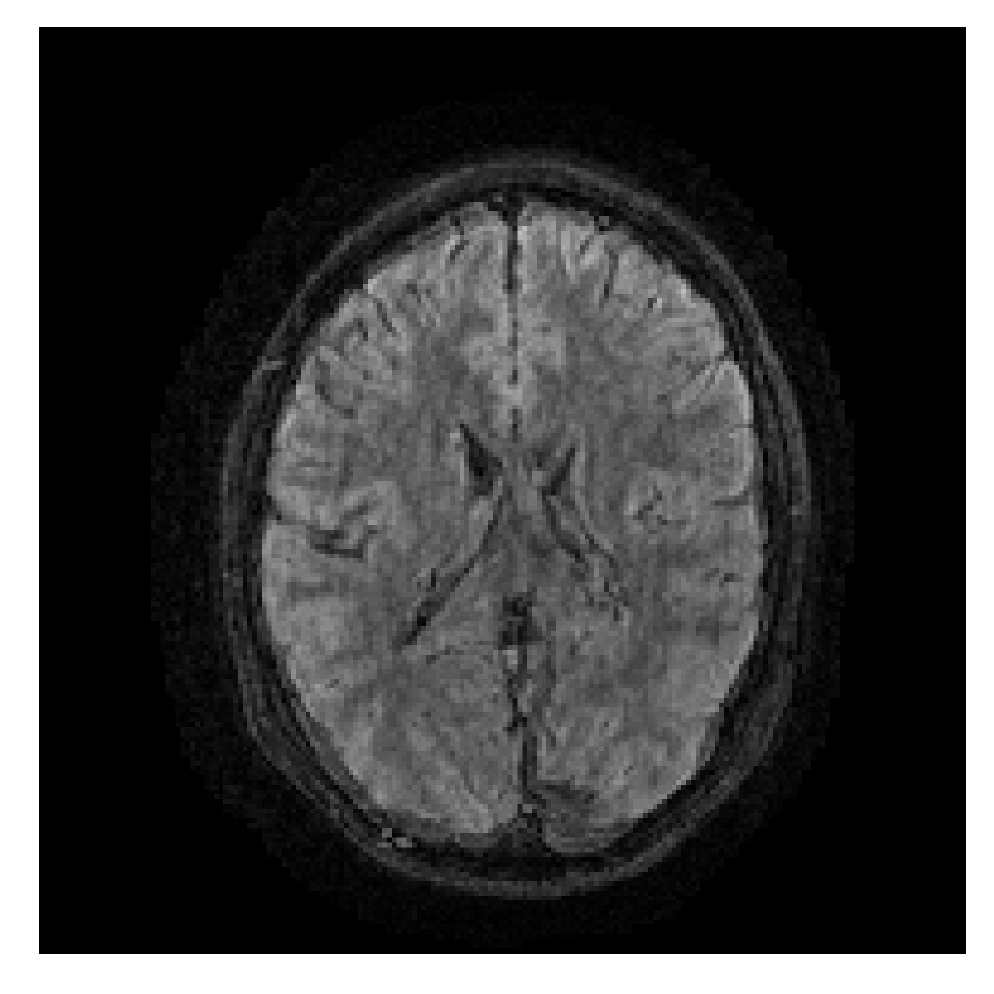}%
                \linethickness{2pt}
                \put(5,13){{\tiny\color{white}PSNR: 24.46}}
                \put(5,6){{\tiny\color{white}SSIM: 0.7371}}
            \end{overpic}%
        \end{subfigure}%
        \begin{subfigure}[b]{0.24\textwidth}
            \begin{overpic}[width=\textwidth]{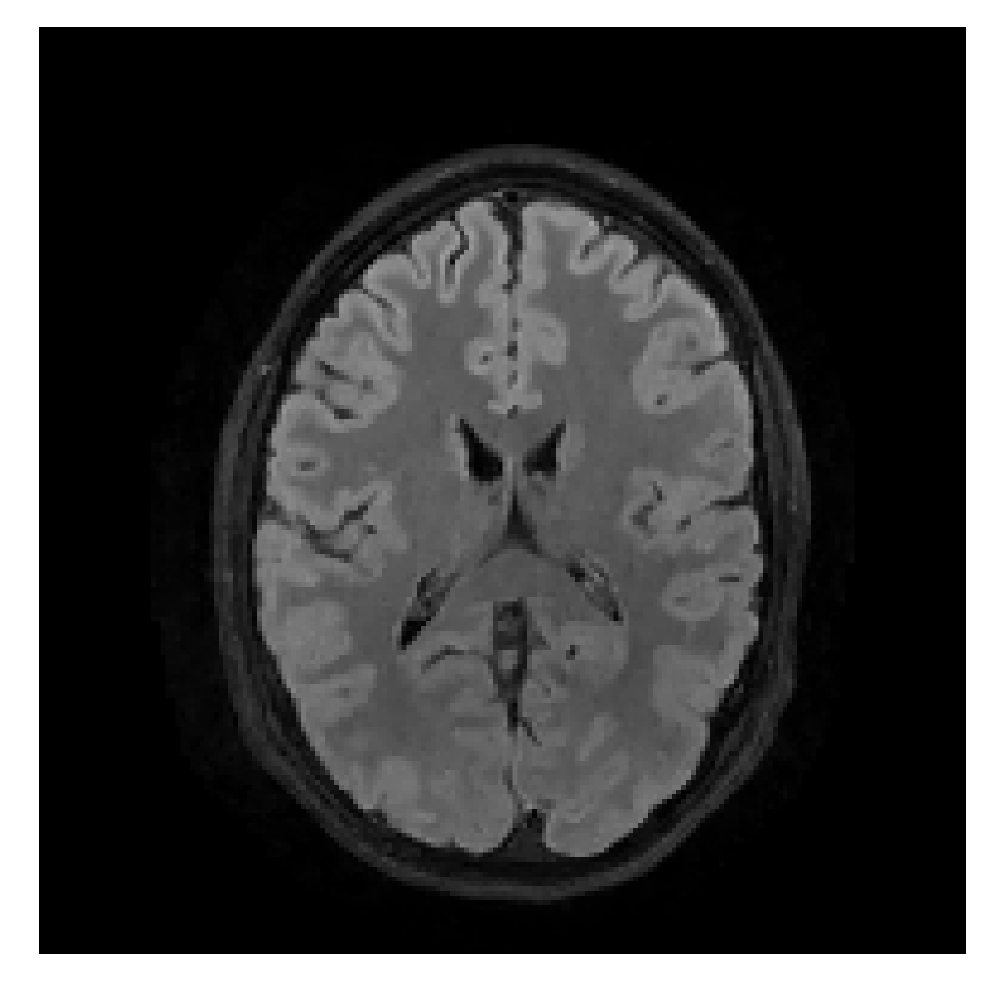}%
                \linethickness{2pt}
                \put(5,13){{\tiny\color{white}PSNR: 30.16}}
                \put(5,6){{\tiny\color{white}SSIM: 0.8490}}
            \end{overpic}
        \end{subfigure}%
        \begin{subfigure}[b]{0.24\textwidth}
            \includegraphics[width=\textwidth]{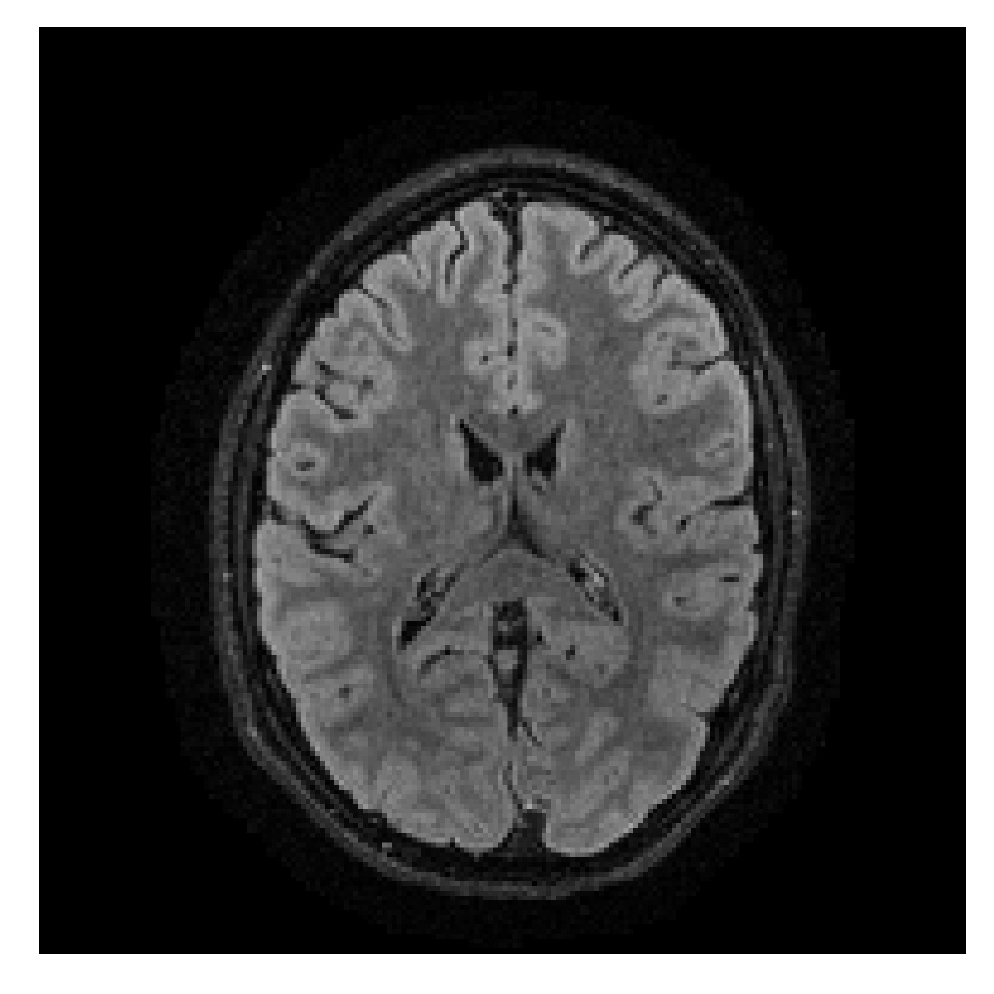}%
        \end{subfigure}%
        \begin{subfigure}[b]{0.24\textwidth}
            \includegraphics[width=\textwidth]{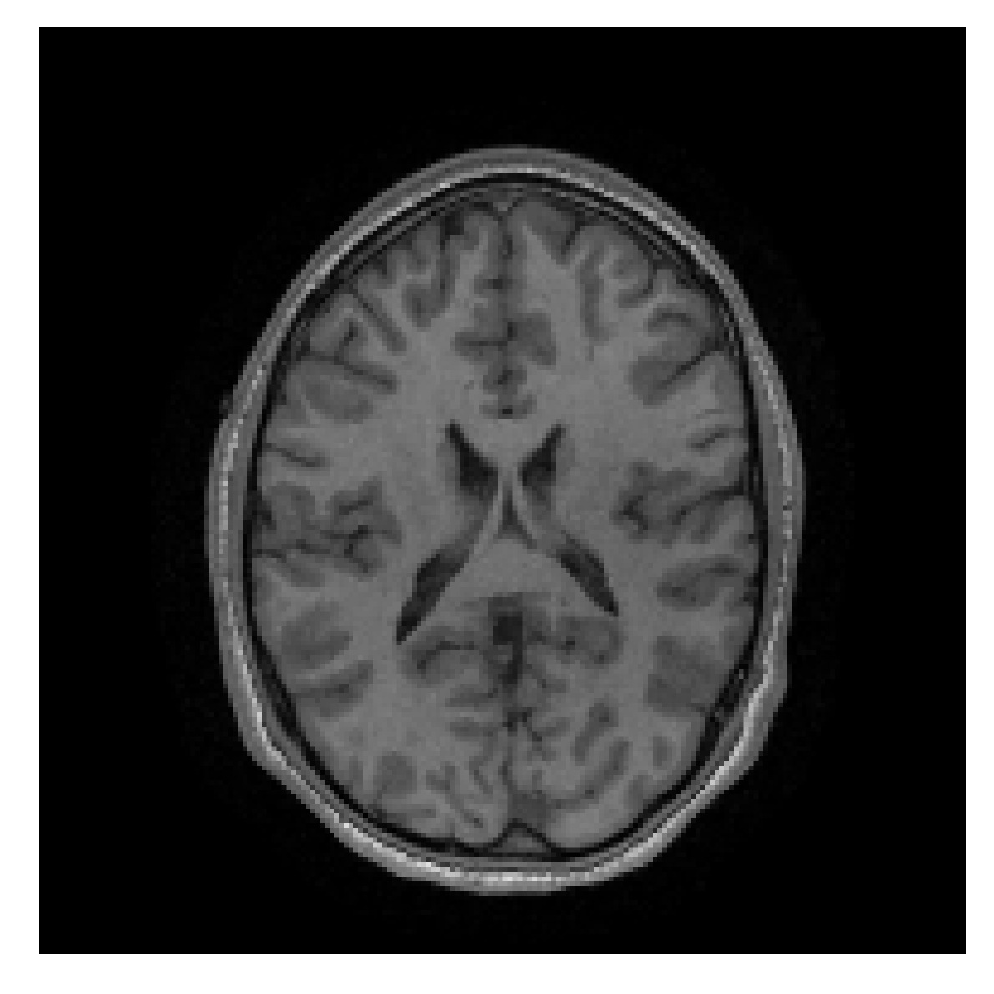}%
        \end{subfigure}%
    \end{subfigure}
    \begin{subfigure}{\textwidth}
        \centering
        \rotatebox{90}{\hspace{1.3em}Axial detail} %
        \begin{subfigure}[b]{0.24\textwidth}
            \includegraphics[width=\textwidth]{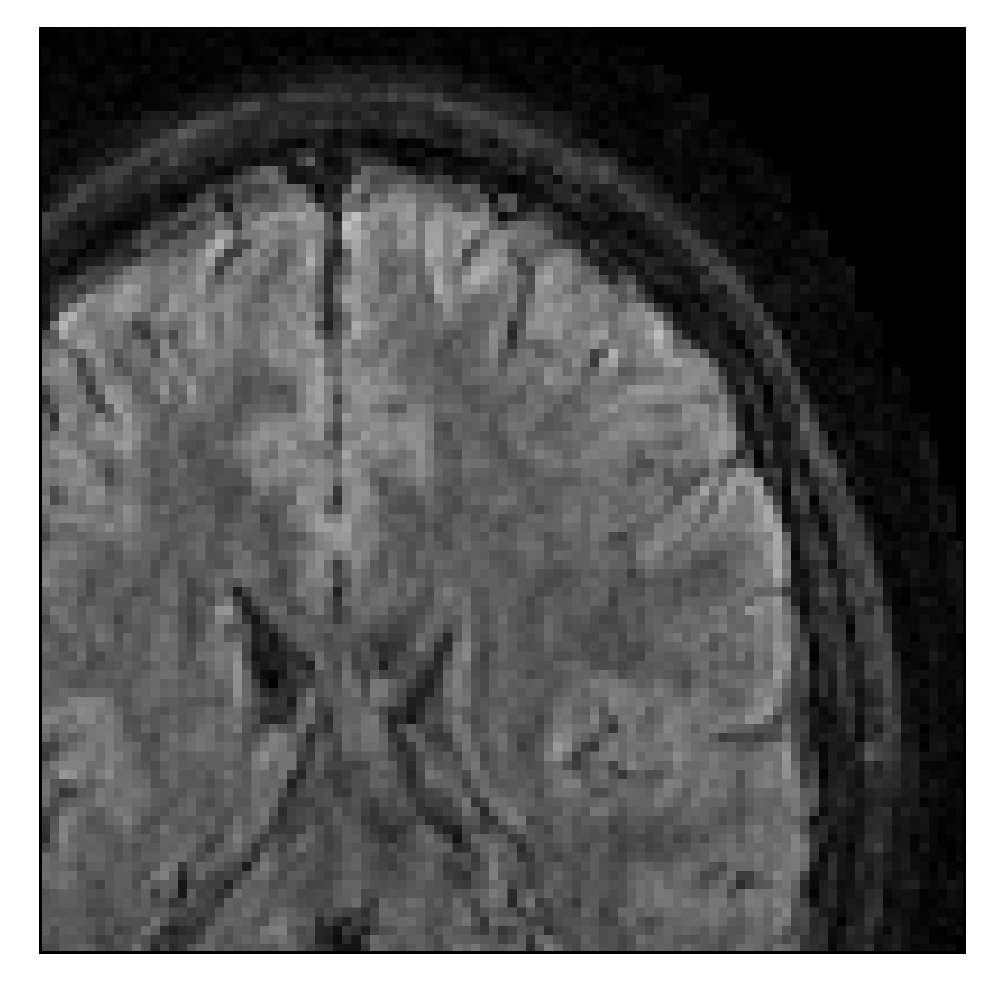}%
        \end{subfigure}%
        \begin{subfigure}[b]{0.24\textwidth}
            \includegraphics[width=\textwidth]{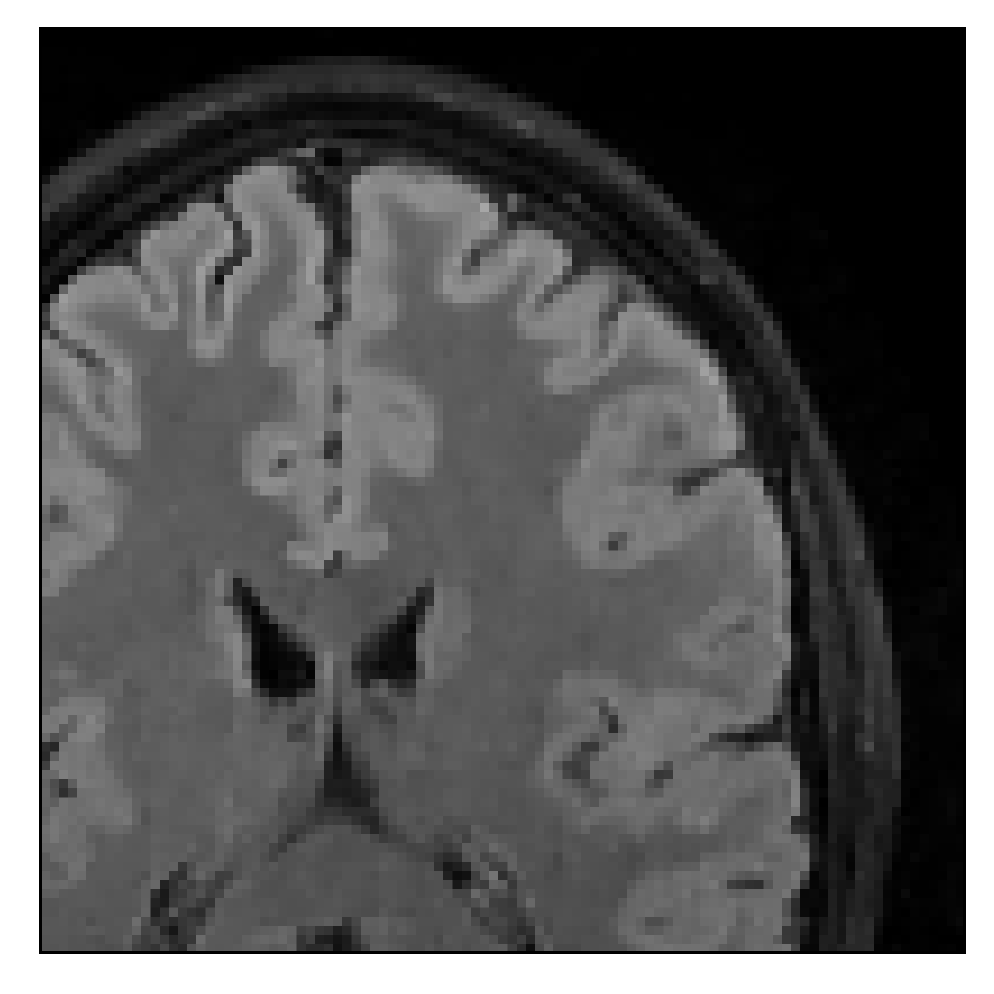}%
        \end{subfigure}%
        \begin{subfigure}[b]{0.24\textwidth}
            \includegraphics[width=\textwidth]{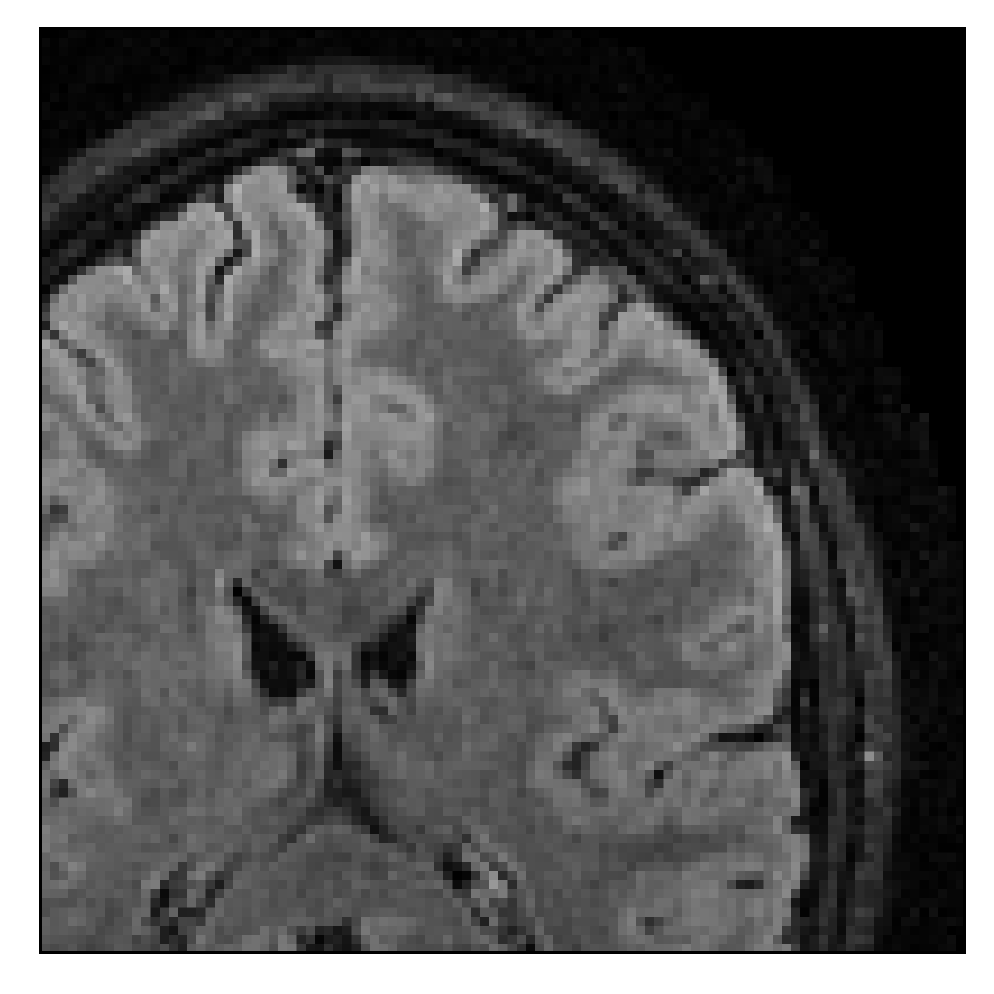}%
        \end{subfigure}%
        \begin{subfigure}[b]{0.24\textwidth}
            \includegraphics[width=\textwidth]{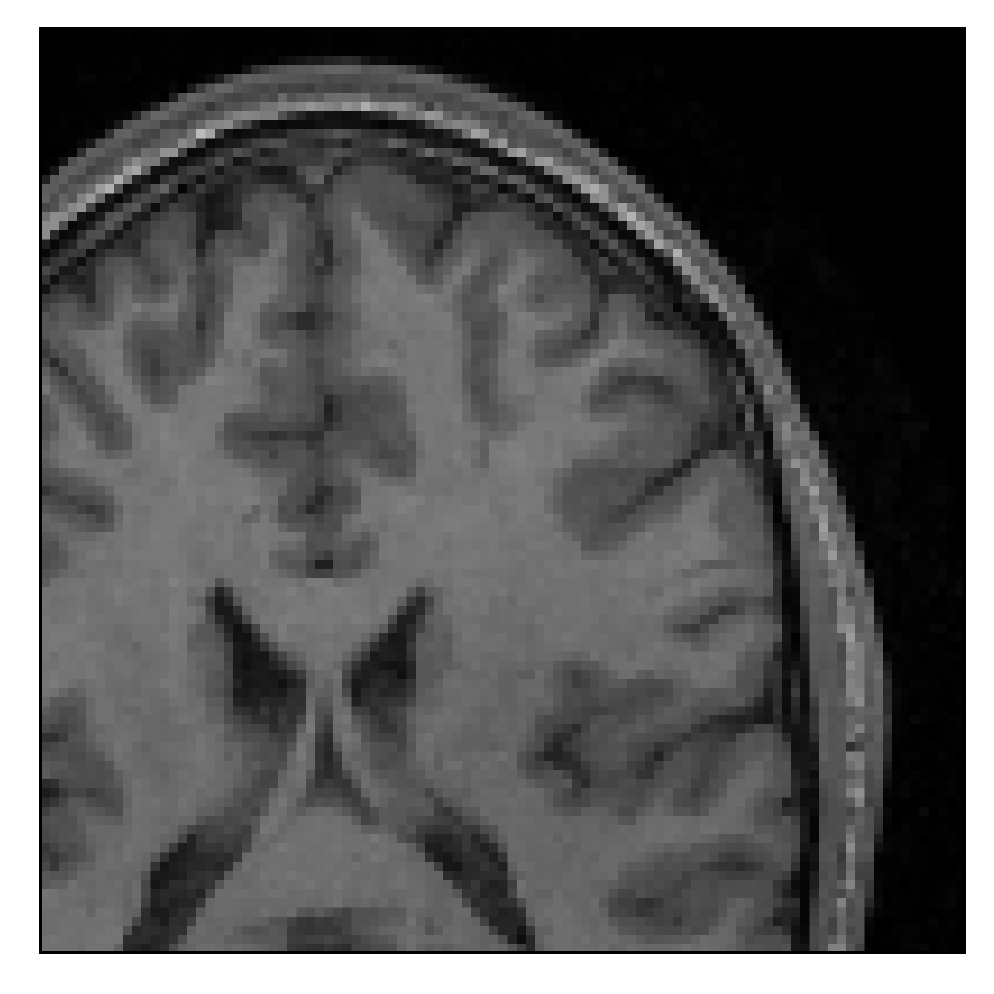}%
        \end{subfigure}%
    \end{subfigure}
    \caption{Reconstruction results for volunteer 1. The volunteer is instructed to move once during the scan. The corrupted contrast is T2-FLAIR-weighted, while the reference contrast is T1-weighted. Compare these results with the one obtained with different motion complexity in Figures \ref{fig:robustness2_vol2}, \ref{fig:robustness3_vol2}.}\label{fig:robustness1_vol2}
\end{figure}
\begin{figure}[!htb]
    \centering
    \begin{subfigure}{\textwidth}
        \centering
        \rotatebox{90}{\hspace{2.5em}Sagittal} %
        \begin{subfigure}[b]{0.24\textwidth}
            \caption*{Corrupted}\vspace{-0.5em}%
            \begin{overpic}[width=\textwidth]{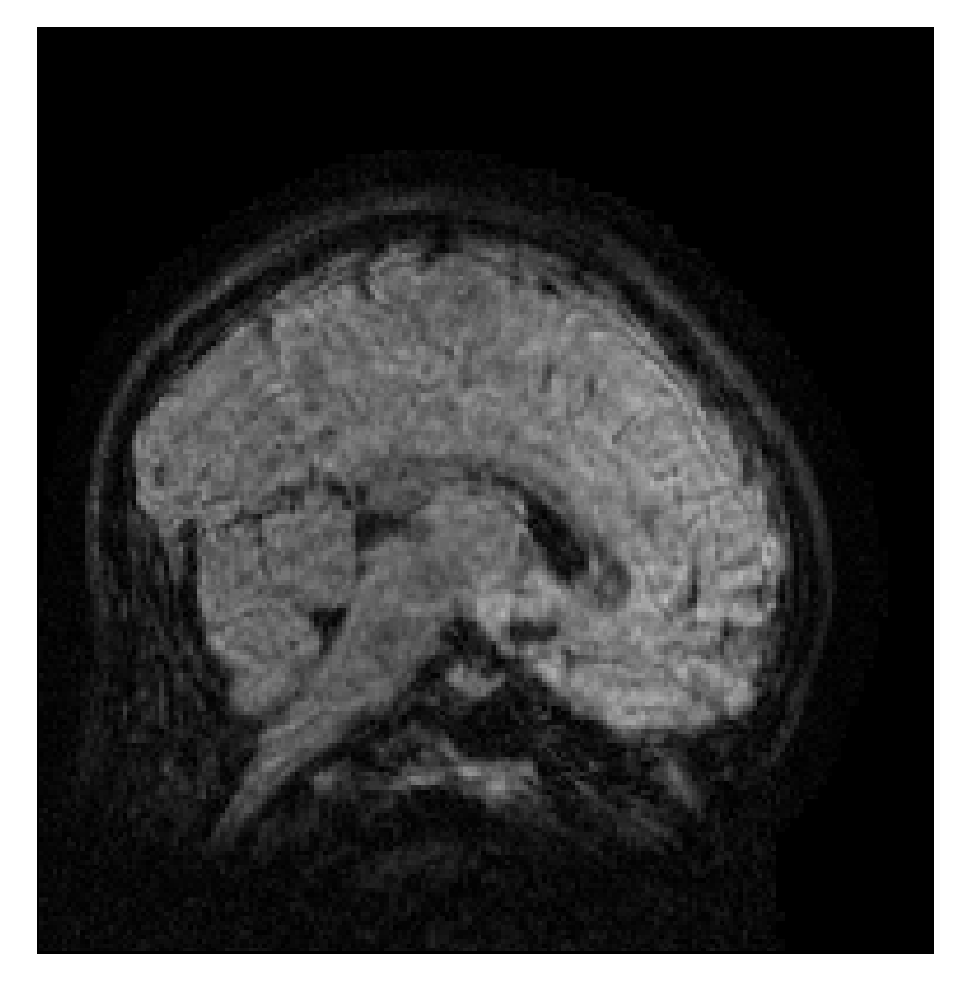}%
                \linethickness{2pt}
                \put(5,13){{\tiny\color{white}PSNR: 21.82}}
                \put(5,6){{\tiny\color{white}SSIM: 0.6071}}
            \end{overpic}%
        \end{subfigure}%
        \begin{subfigure}[b]{0.24\textwidth}
            \caption*{Corrected}\vspace{-0.5em}%
            \begin{overpic}[width=\textwidth]{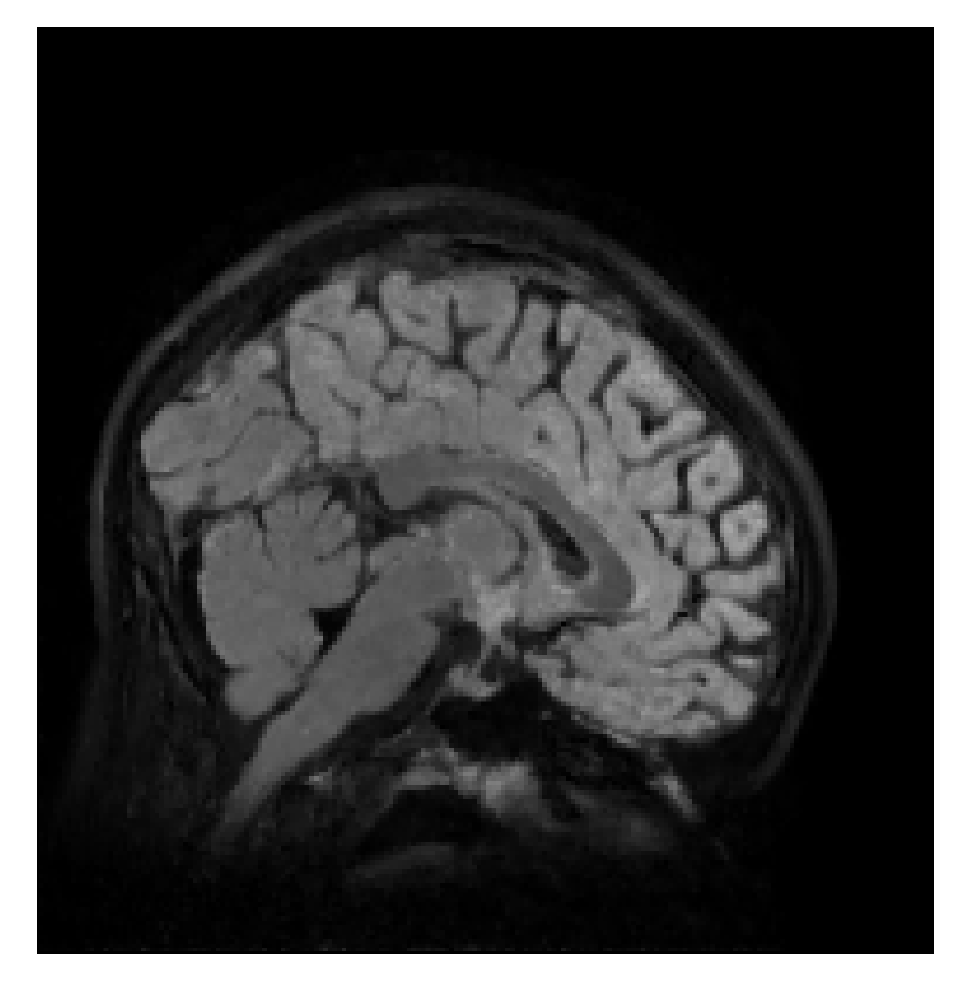}%
                \linethickness{2pt}
                \put(5,13){{\tiny\color{white}PSNR: 27.76}}
                \put(5,6){{\tiny\color{white}SSIM: 0.7816}}
            \end{overpic}%
        \end{subfigure}%
        \begin{subfigure}[b]{0.24\textwidth}
            \caption*{Ground truth}\vspace{-0.5em}%
            \includegraphics[width=\textwidth]{./figs/robustness/vol1/ground_truth_vol1_robustness_slice1-eps-converted-to.pdf}%
        \end{subfigure}%
        \begin{subfigure}[b]{0.24\textwidth}
            \caption*{Reference}\vspace{-0.5em}%
            \includegraphics[width=\textwidth]{./figs/robustness/vol1/reference_vol1_robustness_slice1-eps-converted-to.pdf}%
        \end{subfigure}%
    \end{subfigure}
    \begin{subfigure}{\textwidth}
        \centering
        \rotatebox{90}{\hspace{2.5em}Coronal} %
        \begin{subfigure}[b]{0.24\textwidth}
            \begin{overpic}[width=\textwidth]{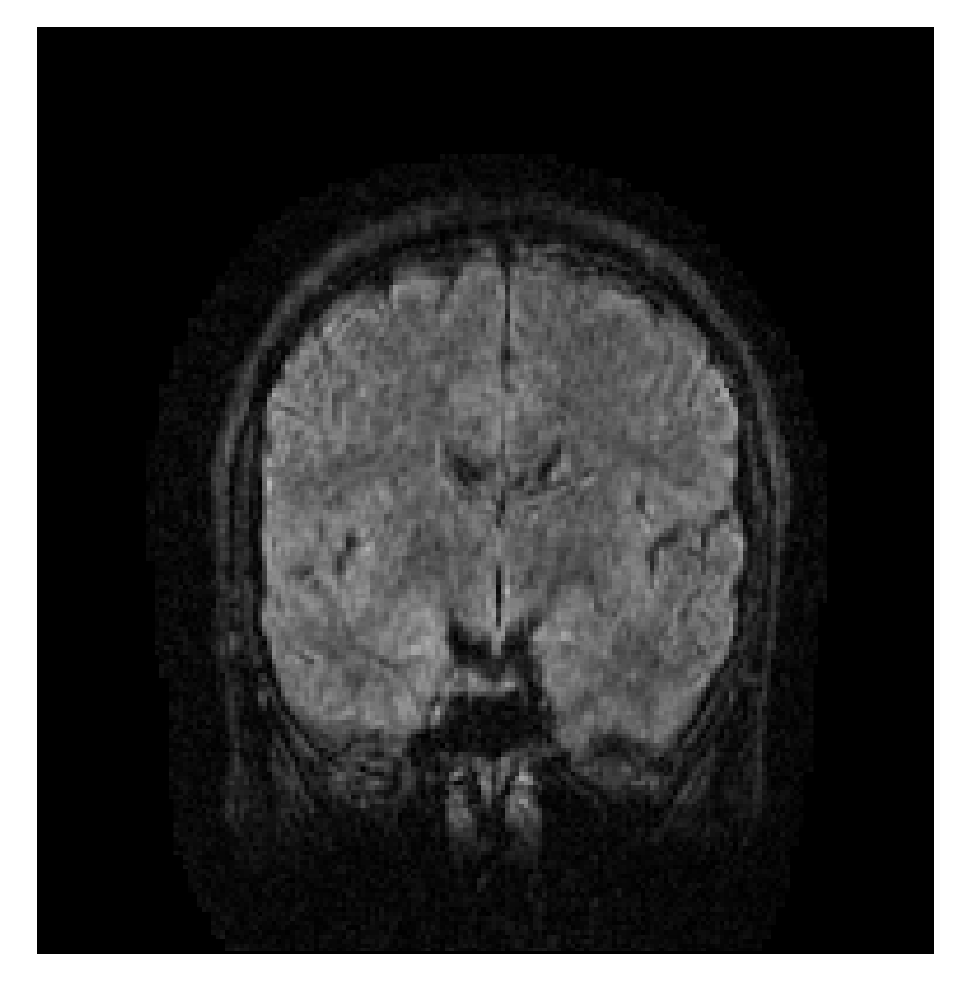}%
                \linethickness{2pt}
                \put(5,13){{\tiny\color{white}PSNR: 24.08}}
                \put(5,6){{\tiny\color{white}SSIM: 0.6751}}
            \end{overpic}%
        \end{subfigure}%
        \begin{subfigure}[b]{0.24\textwidth}
            \begin{overpic}[width=\textwidth]{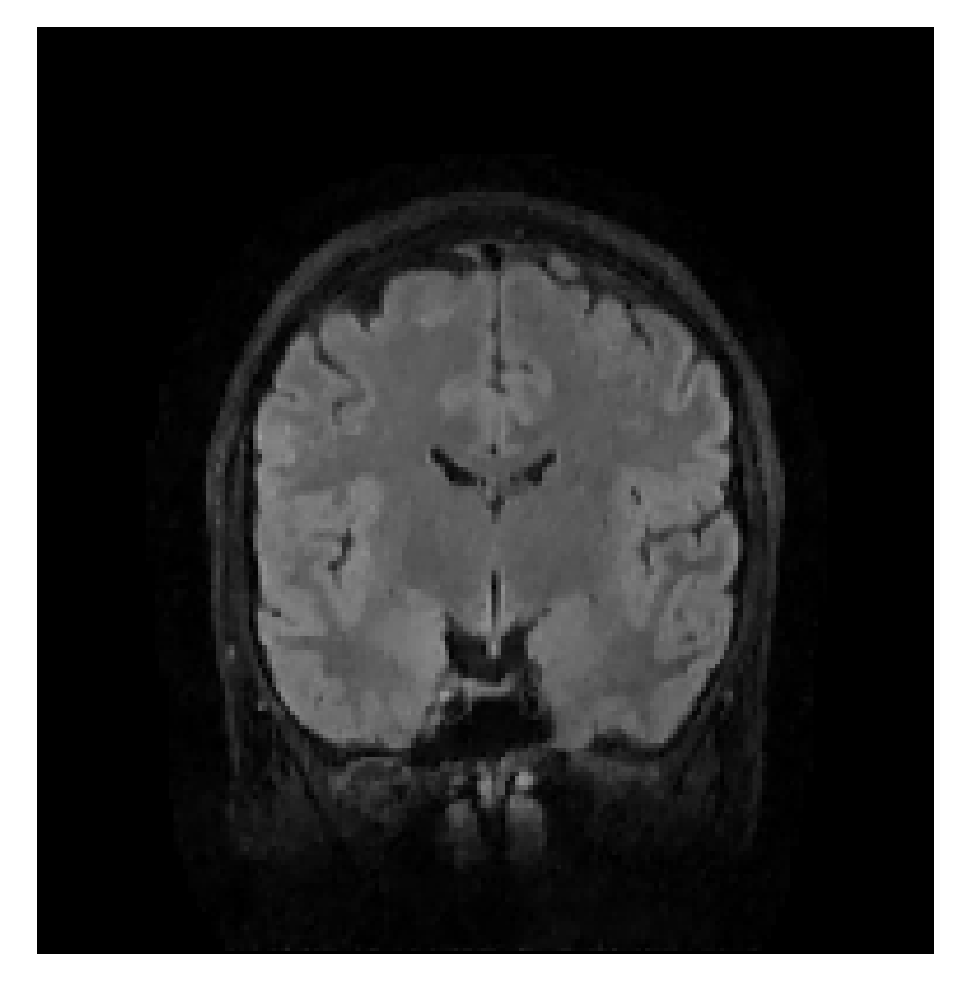}%
                \linethickness{2pt}
                \put(5,13){{\tiny\color{white}PSNR: 29.73}}
                \put(5,6){{\tiny\color{white}SSIM: 0.8244}}
            \end{overpic}%
        \end{subfigure}%
        \begin{subfigure}[b]{0.24\textwidth}
            \includegraphics[width=\textwidth]{./figs/robustness/vol1/ground_truth_vol1_robustness_slice2-eps-converted-to.pdf}%
        \end{subfigure}%
        \begin{subfigure}[b]{0.24\textwidth}
            \includegraphics[width=\textwidth]{./figs/robustness/vol1/reference_vol1_robustness_slice2-eps-converted-to.pdf}%
        \end{subfigure}%
    \end{subfigure}
    \begin{subfigure}{\textwidth}
        \centering
        \rotatebox{90}{\hspace{3em}Axial} %
        \begin{subfigure}[b]{0.24\textwidth}
            \begin{overpic}[width=\textwidth]{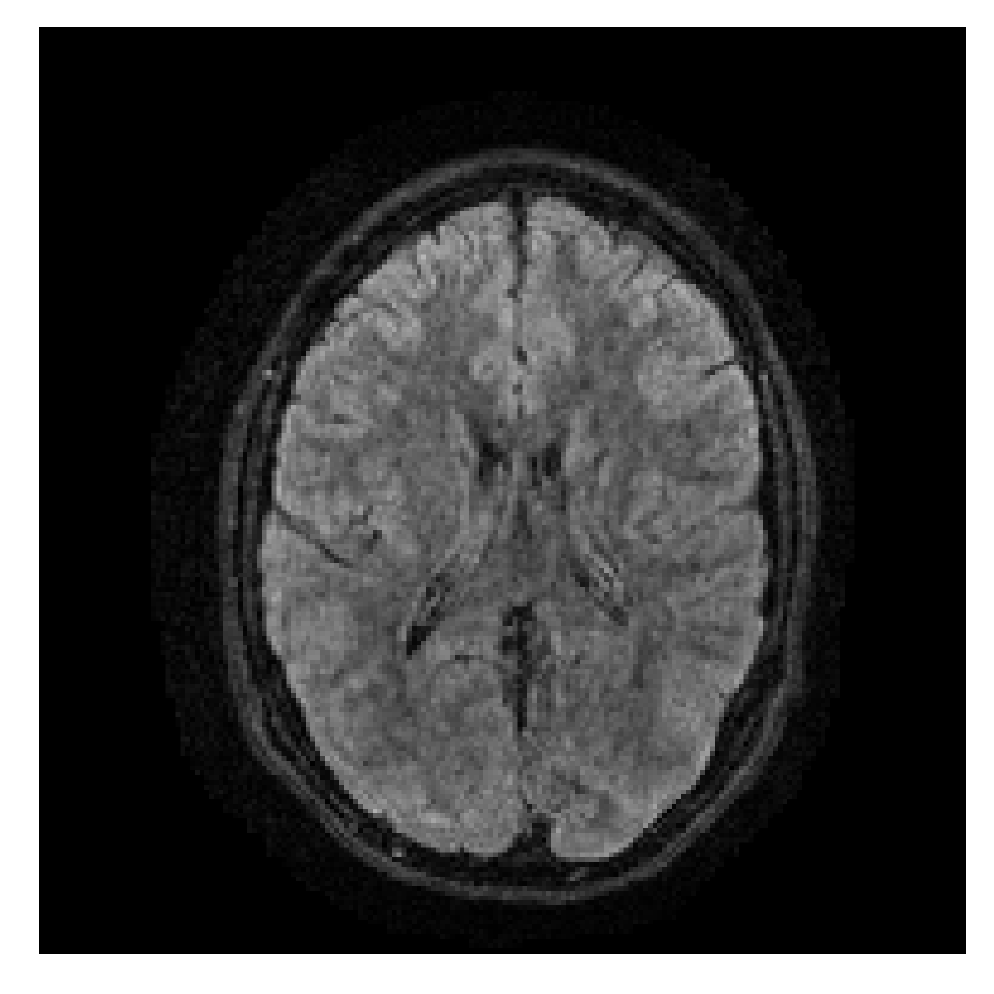}%
                \linethickness{2pt}
                \put(5,13){{\tiny\color{white}PSNR: 22.72}}
                \put(5,6){{\tiny\color{white}SSIM: 0.6440}}
            \end{overpic}%
        \end{subfigure}%
        \begin{subfigure}[b]{0.24\textwidth}
            \begin{overpic}[width=\textwidth]{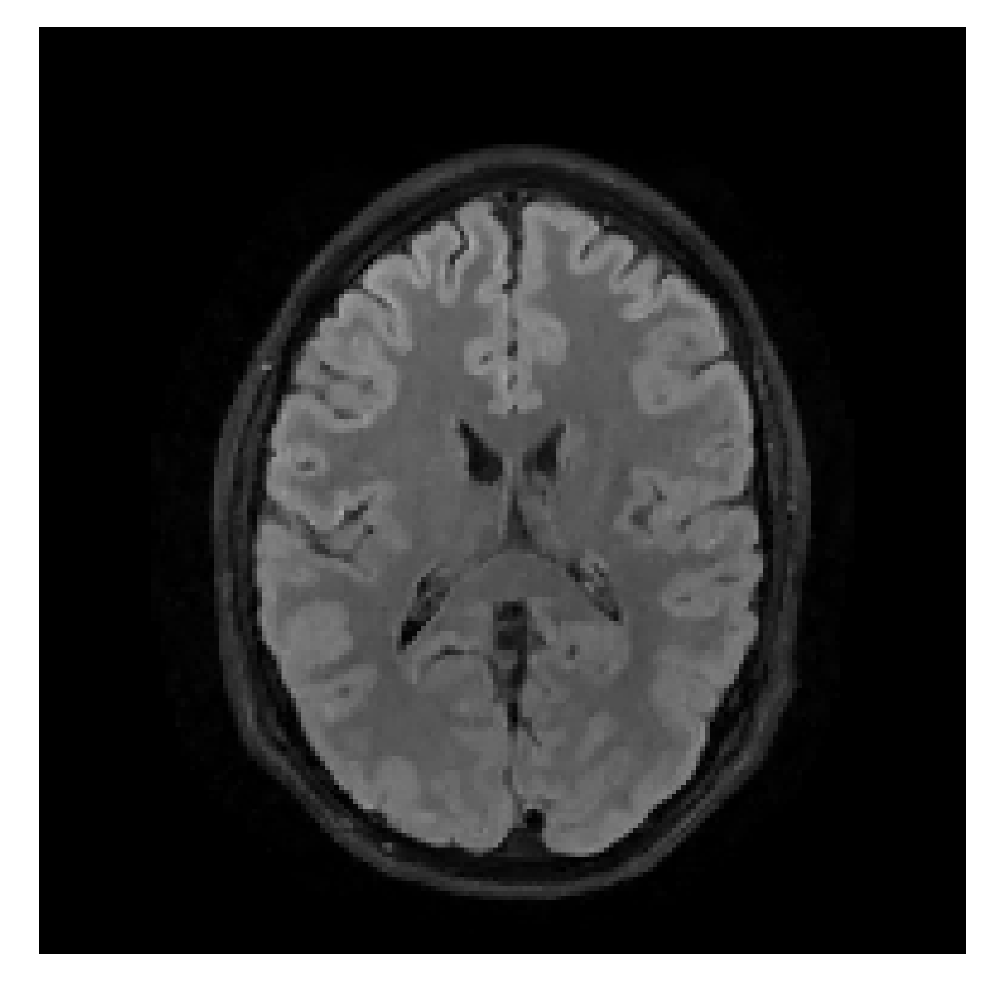}%
                \linethickness{2pt}
                \put(5,13){{\tiny\color{white}PSNR: 29.70}}
                \put(5,6){{\tiny\color{white}SSIM: 0.8362}}
            \end{overpic}
        \end{subfigure}%
        \begin{subfigure}[b]{0.24\textwidth}
            \includegraphics[width=\textwidth]{./figs/robustness/vol1/ground_truth_vol1_robustness_slice3-eps-converted-to.pdf}%
        \end{subfigure}%
        \begin{subfigure}[b]{0.24\textwidth}
            \includegraphics[width=\textwidth]{./figs/robustness/vol1/reference_vol1_robustness_slice3-eps-converted-to.pdf}%
        \end{subfigure}%
    \end{subfigure}
    \begin{subfigure}{\textwidth}
        \centering
        \rotatebox{90}{\hspace{1.3em}Axial detail} %
        \begin{subfigure}[b]{0.24\textwidth}
            \includegraphics[width=\textwidth]{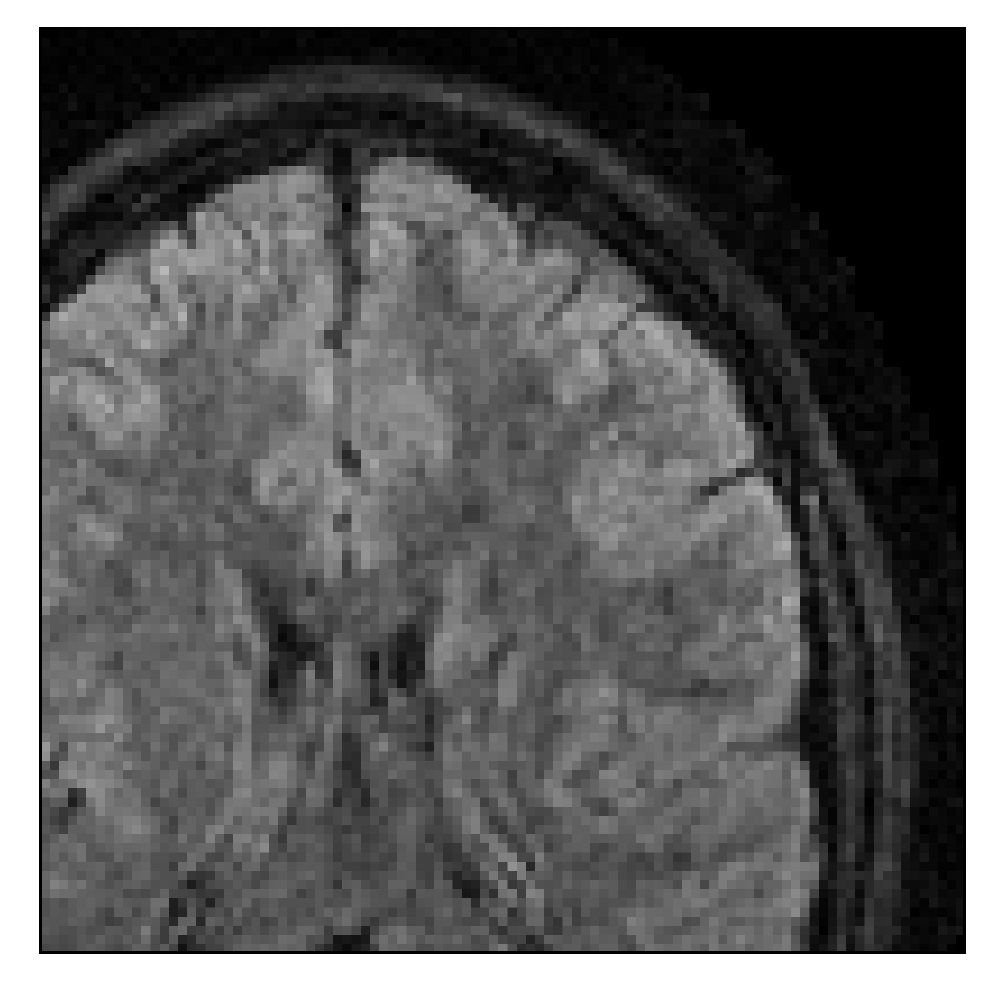}%
        \end{subfigure}%
        \begin{subfigure}[b]{0.24\textwidth}
            \includegraphics[width=\textwidth]{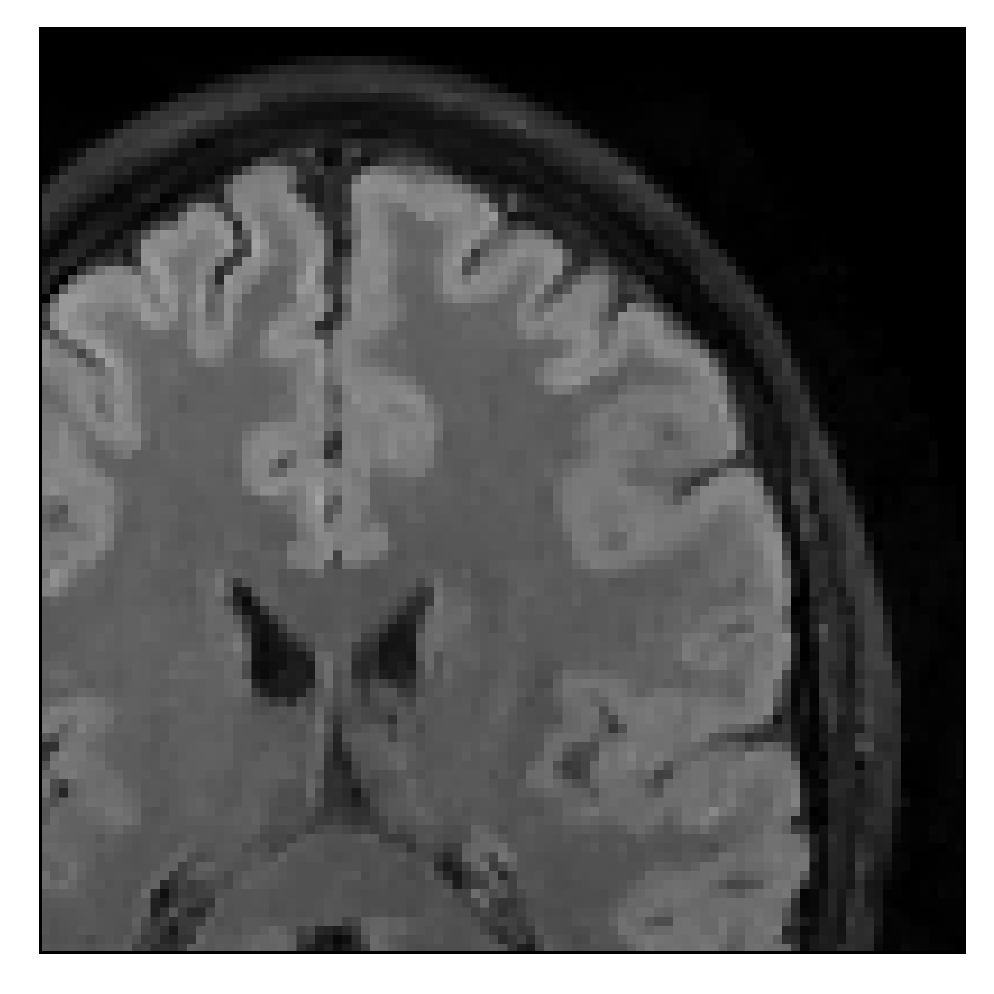}%
        \end{subfigure}%
        \begin{subfigure}[b]{0.24\textwidth}
            \includegraphics[width=\textwidth]{./figs/robustness/vol1/ground_truth_vol1_robustness_detail_slice1-eps-converted-to.pdf}%
        \end{subfigure}%
        \begin{subfigure}[b]{0.24\textwidth}
            \includegraphics[width=\textwidth]{./figs/robustness/vol1/reference_vol1_robustness_detail_slice1-eps-converted-to.pdf}%
        \end{subfigure}%
    \end{subfigure}
    \caption{Reconstruction results for volunteer 1. The volunteer is instructed to move twice during the scan. The corrupted contrast is T2-FLAIR-weighted, while the reference contrast is T1-weighted. Compare these results with the one obtained with different motion complexity in Figures \ref{fig:robustness1_vol2}, \ref{fig:robustness3_vol2}.}\label{fig:robustness2_vol2}
\end{figure}
\begin{figure}[!htb]
    \centering
    \begin{subfigure}{\textwidth}
        \centering
        \rotatebox{90}{\hspace{2.5em}Sagittal} %
        \begin{subfigure}[b]{0.24\textwidth}
            \caption*{Corrupted}\vspace{-0.5em}%
            \begin{overpic}[width=\textwidth]{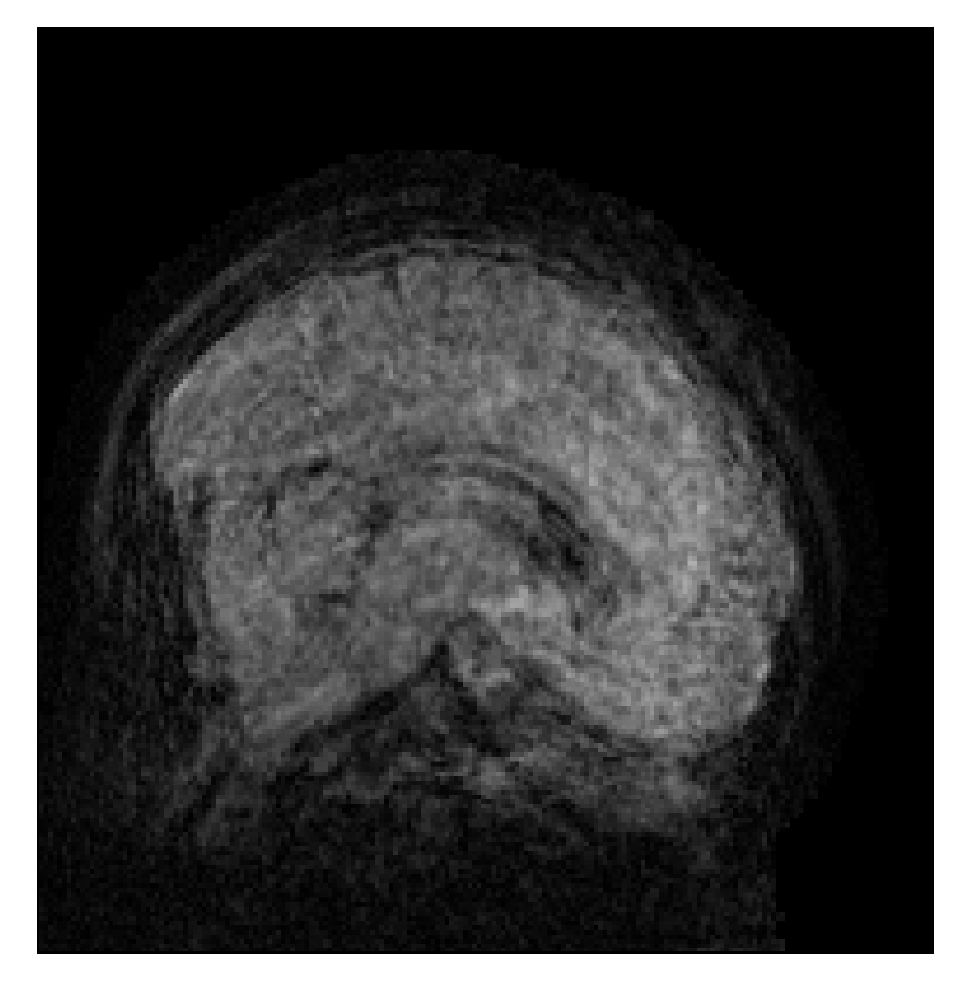}%
                \linethickness{2pt}
                \put(5,13){{\tiny\color{white}PSNR: 20.65}}
                \put(5,6){{\tiny\color{white}SSIM: 0.5141}}
            \end{overpic}%
        \end{subfigure}%
        \begin{subfigure}[b]{0.24\textwidth}
            \caption*{Corrected}\vspace{-0.5em}%
            \begin{overpic}[width=\textwidth]{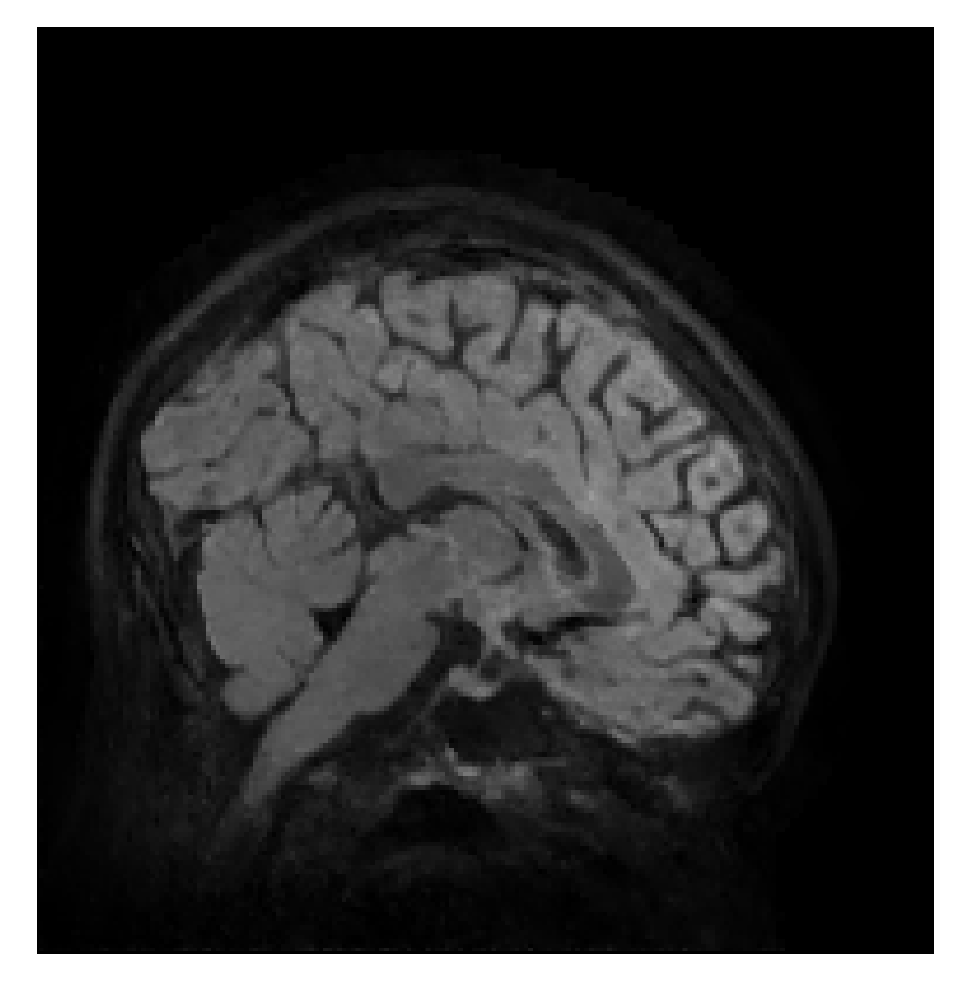}%
                \linethickness{2pt}
                \put(5,13){{\tiny\color{white}PSNR: 25.28}}
                \put(5,6){{\tiny\color{white}SSIM: 0.7661}}
            \end{overpic}%
        \end{subfigure}%
        \begin{subfigure}[b]{0.24\textwidth}
            \caption*{Ground truth}\vspace{-0.5em}%
            \includegraphics[width=\textwidth]{./figs/robustness/vol1/ground_truth_vol1_robustness_slice1-eps-converted-to.pdf}%
        \end{subfigure}%
        \begin{subfigure}[b]{0.24\textwidth}
            \caption*{Reference}\vspace{-0.5em}%
            \includegraphics[width=\textwidth]{./figs/robustness/vol1/reference_vol1_robustness_slice1-eps-converted-to.pdf}%
        \end{subfigure}%
    \end{subfigure}
    \begin{subfigure}{\textwidth}
        \centering
        \rotatebox{90}{\hspace{2.5em}Coronal} %
        \begin{subfigure}[b]{0.24\textwidth}
            \begin{overpic}[width=\textwidth]{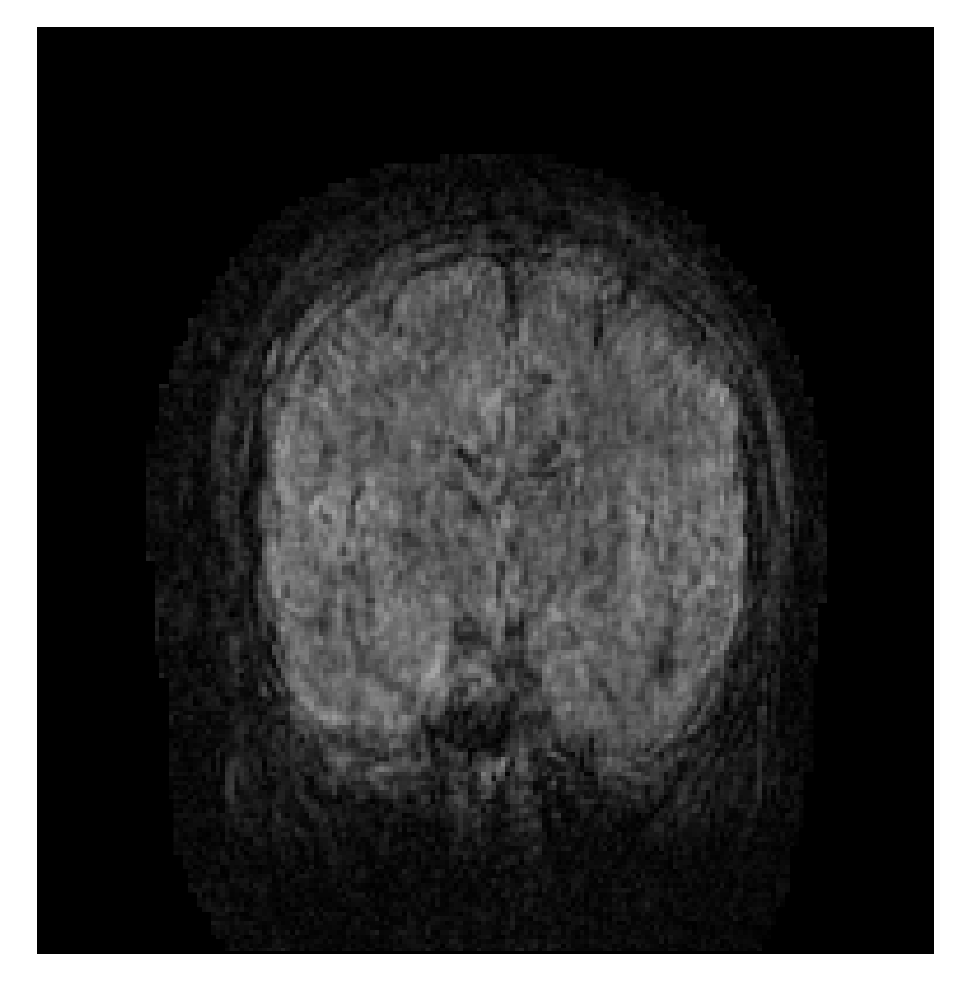}%
                \linethickness{2pt}
                \put(5,13){{\tiny\color{white}PSNR: 22.77}}
                \put(5,6){{\tiny\color{white}SSIM: 0.5998}}
            \end{overpic}%
        \end{subfigure}%
        \begin{subfigure}[b]{0.24\textwidth}
            \begin{overpic}[width=\textwidth]{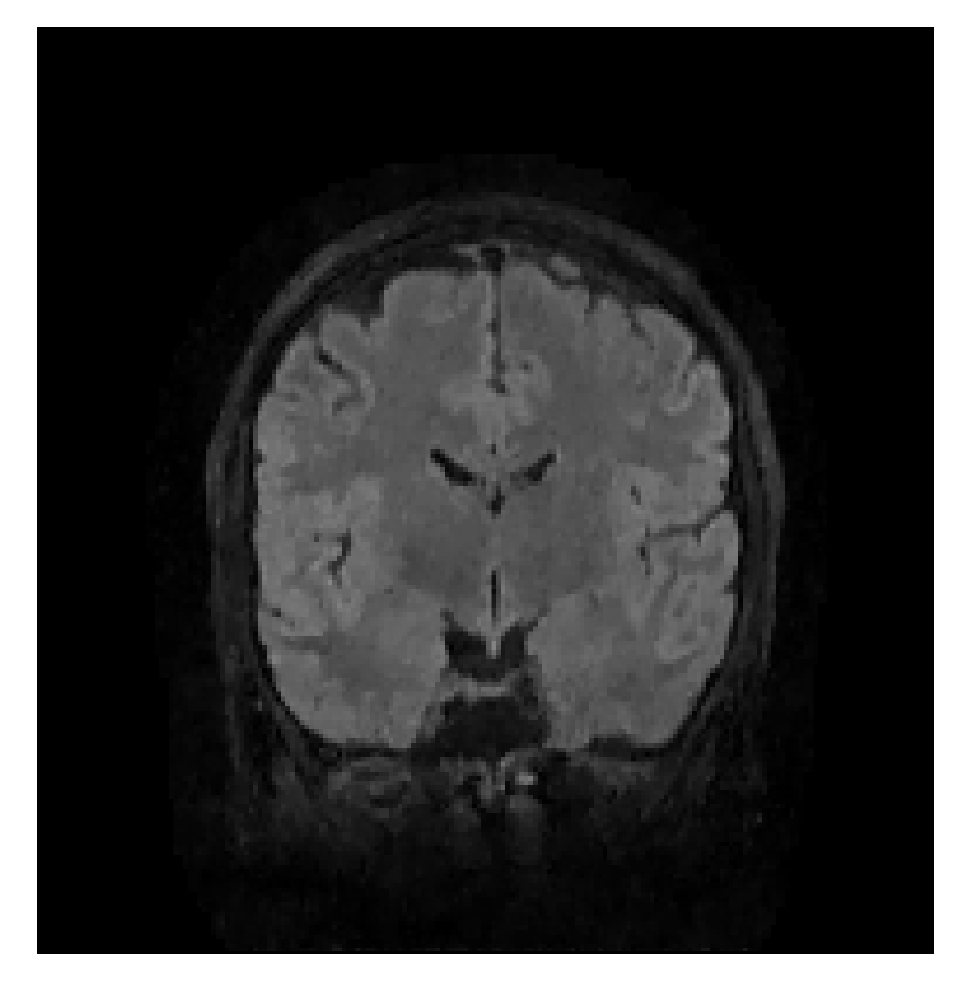}%
                \linethickness{2pt}
                \put(5,13){{\tiny\color{white}PSNR: 27.40}}
                \put(5,6){{\tiny\color{white}SSIM: 0.8060}}
            \end{overpic}%
        \end{subfigure}%
        \begin{subfigure}[b]{0.24\textwidth}
            \includegraphics[width=\textwidth]{./figs/robustness/vol1/ground_truth_vol1_robustness_slice2-eps-converted-to.pdf}%
        \end{subfigure}%
        \begin{subfigure}[b]{0.24\textwidth}
            \includegraphics[width=\textwidth]{./figs/robustness/vol1/reference_vol1_robustness_slice2-eps-converted-to.pdf}%
        \end{subfigure}%
    \end{subfigure}
    \begin{subfigure}{\textwidth}
        \centering
        \rotatebox{90}{\hspace{3em}Axial} %
        \begin{subfigure}[b]{0.24\textwidth}
            \begin{overpic}[width=\textwidth]{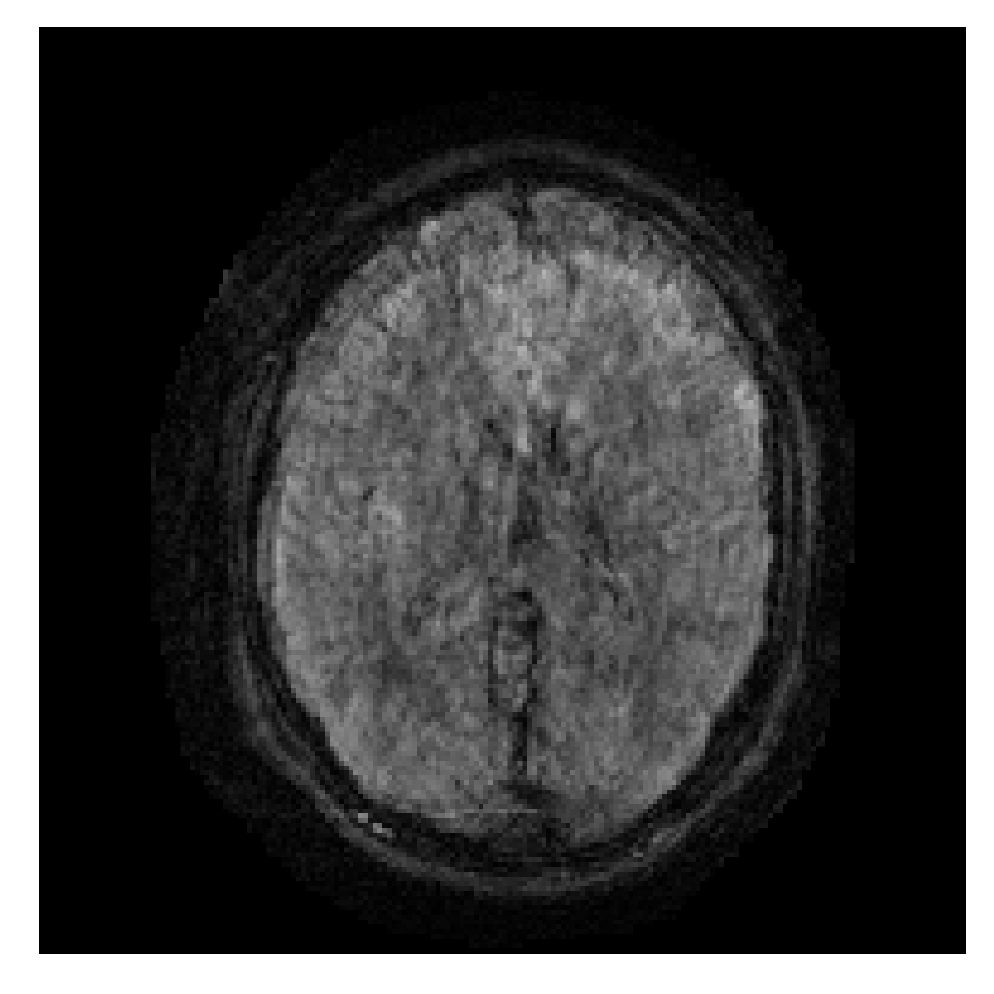}%
                \linethickness{2pt}
                \put(5,13){{\tiny\color{white}PSNR: 21.23}}
                \put(5,6){{\tiny\color{white}SSIM: 0.5727}}
            \end{overpic}%
        \end{subfigure}%
        \begin{subfigure}[b]{0.24\textwidth}
            \begin{overpic}[width=\textwidth]{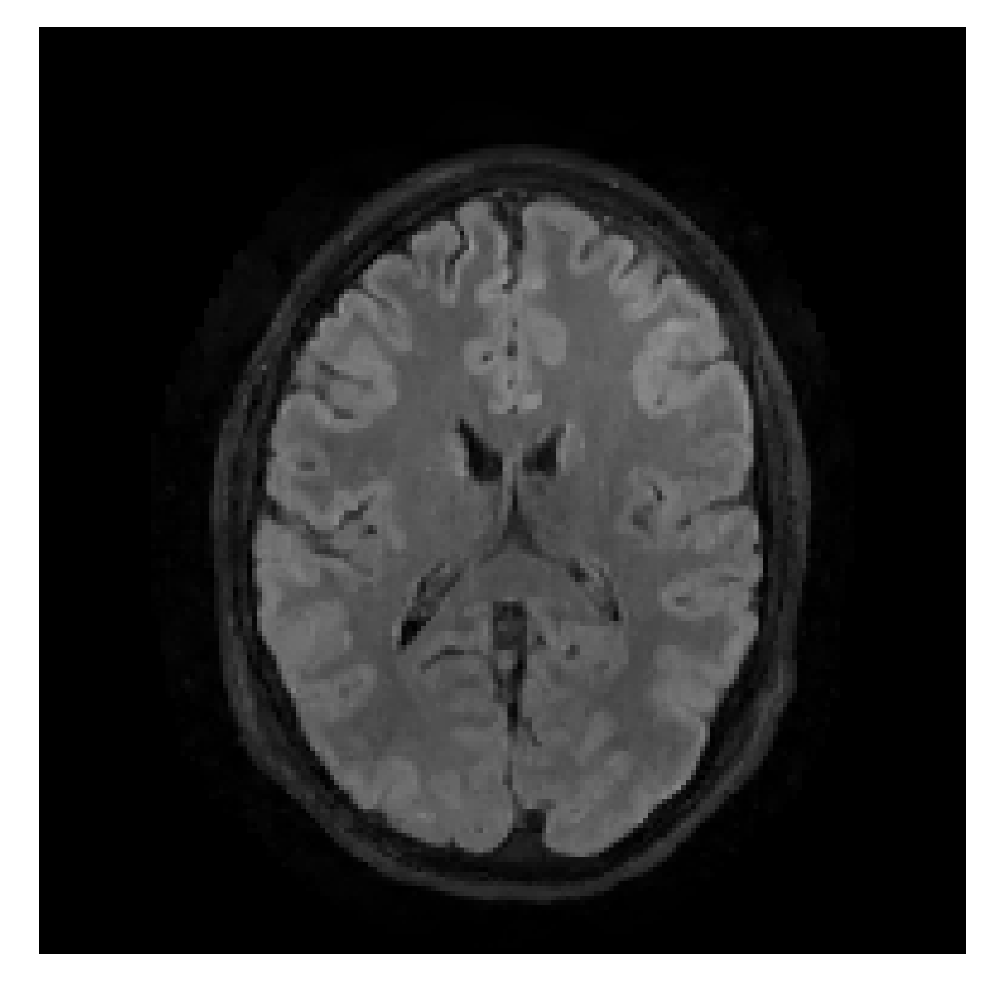}%
                \linethickness{2pt}
                \put(5,13){{\tiny\color{white}PSNR: 27.66}}
                \put(5,6){{\tiny\color{white}SSIM: 0.8298}}
            \end{overpic}
        \end{subfigure}%
        \begin{subfigure}[b]{0.24\textwidth}
            \includegraphics[width=\textwidth]{./figs/robustness/vol1/ground_truth_vol1_robustness_slice3-eps-converted-to.pdf}%
        \end{subfigure}%
        \begin{subfigure}[b]{0.24\textwidth}
            \includegraphics[width=\textwidth]{./figs/robustness/vol1/reference_vol1_robustness_slice3-eps-converted-to.pdf}%
        \end{subfigure}%
    \end{subfigure}
    \begin{subfigure}{\textwidth}
        \centering
        \rotatebox{90}{\hspace{1.3em}Axial detail} %
        \begin{subfigure}[b]{0.24\textwidth}
            \includegraphics[width=\textwidth]{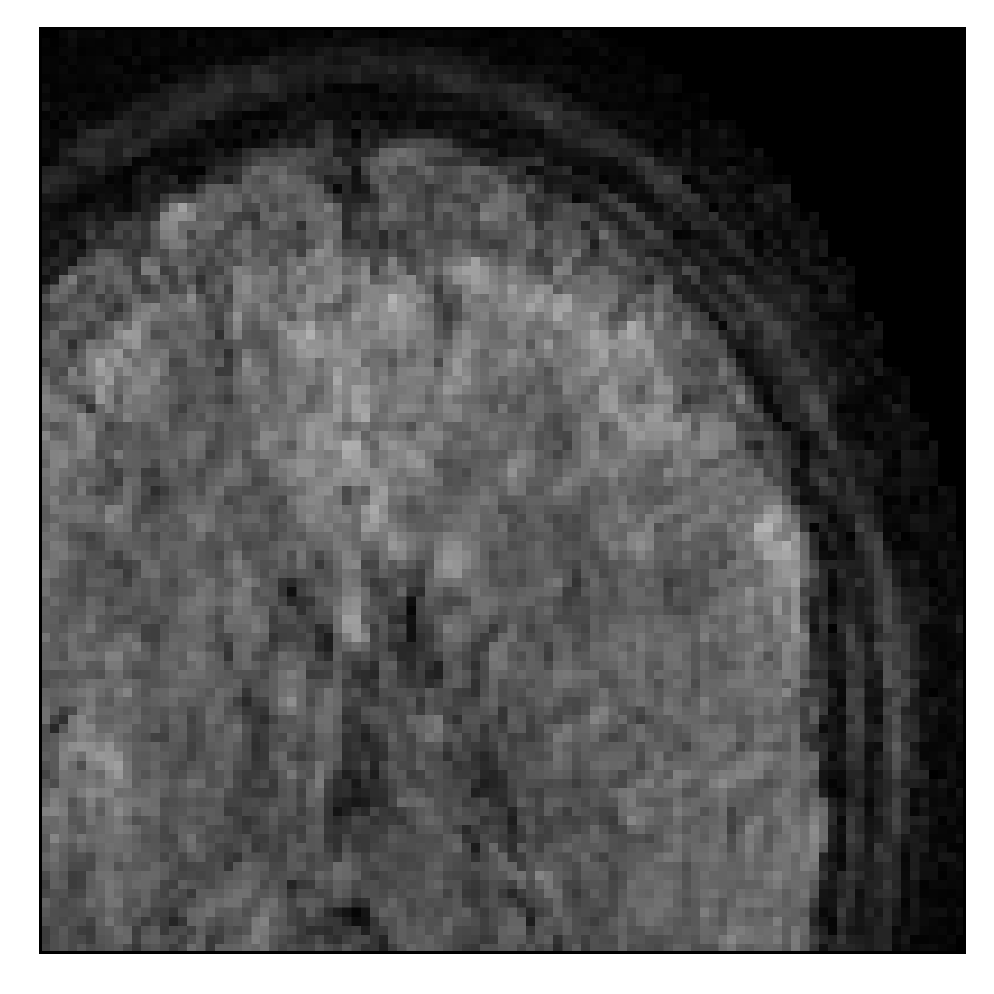}%
        \end{subfigure}%
        \begin{subfigure}[b]{0.24\textwidth}
            \includegraphics[width=\textwidth]{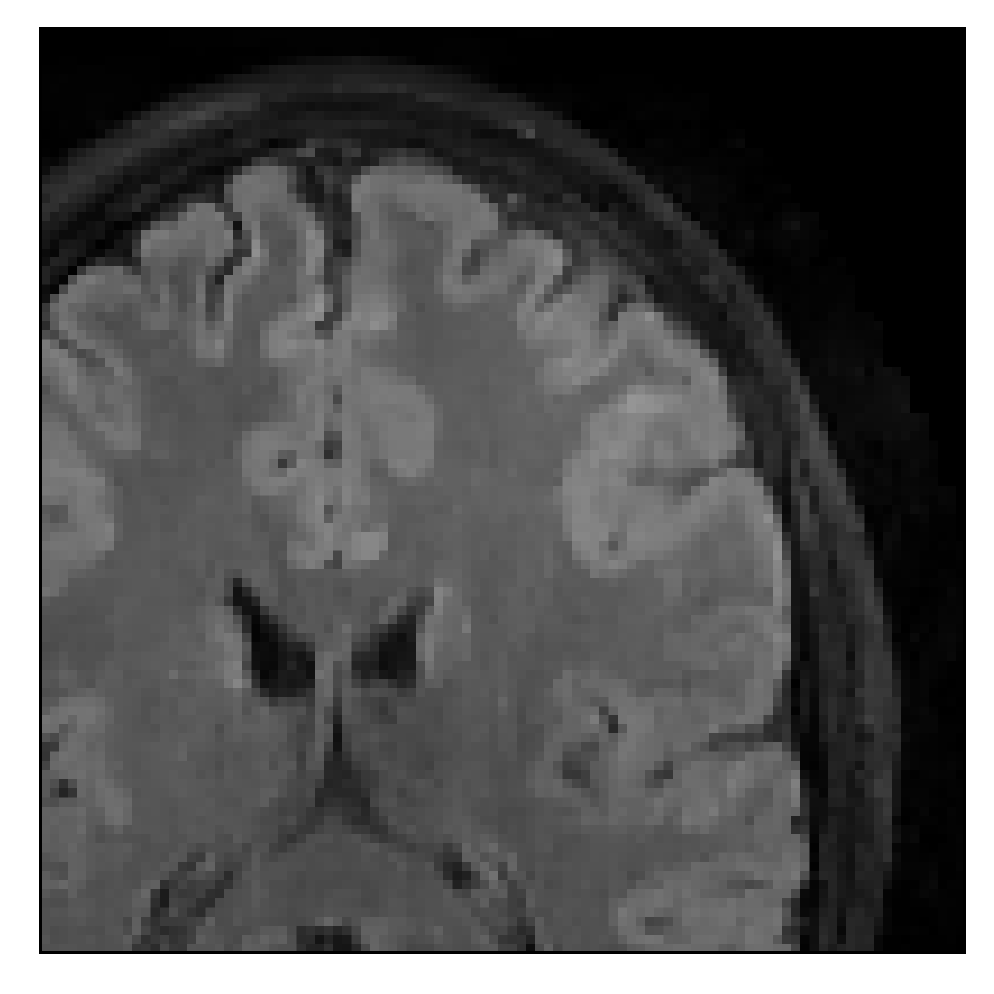}%
        \end{subfigure}%
        \begin{subfigure}[b]{0.24\textwidth}
            \includegraphics[width=\textwidth]{./figs/robustness/vol1/ground_truth_vol1_robustness_detail_slice1-eps-converted-to.pdf}%
        \end{subfigure}%
        \begin{subfigure}[b]{0.24\textwidth}
            \includegraphics[width=\textwidth]{./figs/robustness/vol1/reference_vol1_robustness_detail_slice1-eps-converted-to.pdf}%
        \end{subfigure}%
    \end{subfigure}
    \caption{Reconstruction results for volunteer 1. The volunteer is instructed to move five times during the scan. The corrupted contrast is T2-FLAIR-weighted, while the reference contrast is T1-weighted. Compare these results with the one obtained with different motion complexity in Figures \ref{fig:robustness1_vol2}, \ref{fig:robustness2_vol2}.}\label{fig:robustness3_vol2}
\end{figure}
\begin{figure}[!htb]
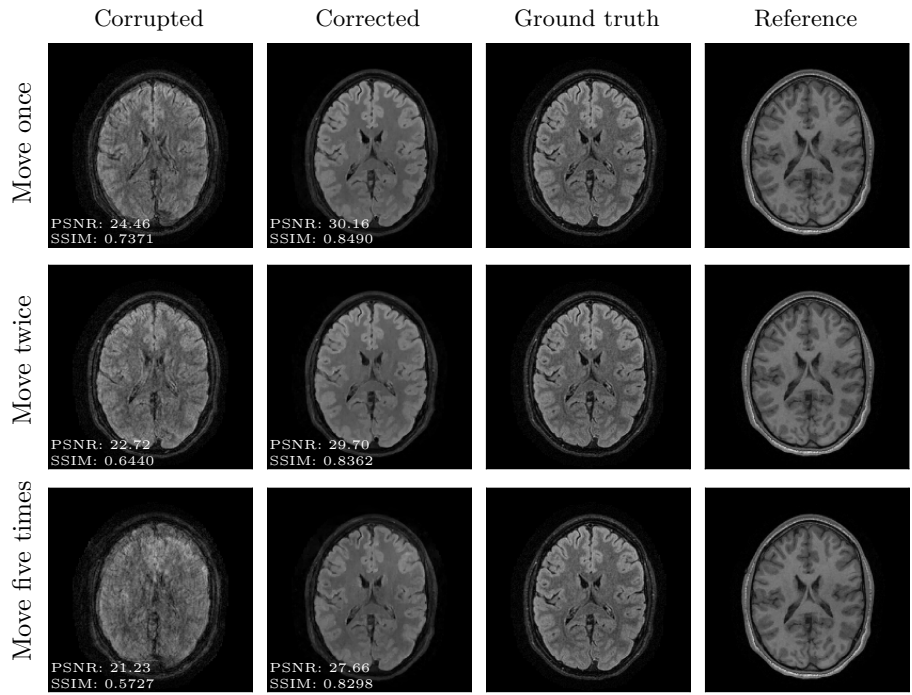

    \centering
    \begin{subfigure}{\textwidth}
        \centering
        \rotatebox{90}{\hspace{2em}Move once} %
        \begin{subfigure}[b]{0.24\textwidth}
            \caption*{Corrupted}\vspace{-0.5em}%
            \begin{overpic}[width=\textwidth]{./figs/robustness/vol1/corrupted1_vol1_robustness_slice3-eps-converted-to.pdf}%
                \linethickness{2pt}
                \put(5,13){{\tiny\color{white}PSNR: 24.46}}
                \put(5,6){{\tiny\color{white}SSIM: 0.7371}}
            \end{overpic}%
        \end{subfigure}%
        \begin{subfigure}[b]{0.24\textwidth}
            \caption*{Corrected}\vspace{-0.5em}%
            \begin{overpic}[width=\textwidth]{./figs/robustness/vol1/corrected1_vol1_robustness_slice3-eps-converted-to.pdf}%
                \linethickness{2pt}
                \put(5,13){{\tiny\color{white}PSNR: 30.16}}
                \put(5,6){{\tiny\color{white}SSIM: 0.8490}}
            \end{overpic}
        \end{subfigure}%
        \begin{subfigure}[b]{0.24\textwidth}
            \caption*{Ground truth}\vspace{-0.5em}%
            \includegraphics[width=\textwidth]{./figs/robustness/vol1/ground_truth_vol1_robustness_slice3-eps-converted-to.pdf}%
        \end{subfigure}%
        \begin{subfigure}[b]{0.24\textwidth}
            \caption*{Reference}\vspace{-0.5em}%
            \includegraphics[width=\textwidth]{./figs/robustness/vol1/reference_vol1_robustness_slice3-eps-converted-to.pdf}%
        \end{subfigure}%
    \end{subfigure}
    \begin{subfigure}{\textwidth}
        \centering
        \rotatebox{90}{\hspace{2em}Move twice} %
        \begin{subfigure}[b]{0.24\textwidth}
            \begin{overpic}[width=\textwidth]{./figs/robustness/vol1/corrupted2_vol1_robustness_slice3-eps-converted-to.pdf}%
                \linethickness{2pt}
                \put(5,13){{\tiny\color{white}PSNR: 22.72}}
                \put(5,6){{\tiny\color{white}SSIM: 0.6440}}
            \end{overpic}%
        \end{subfigure}%
        \begin{subfigure}[b]{0.24\textwidth}
            \begin{overpic}[width=\textwidth]{./figs/robustness/vol1/corrected2_vol1_robustness_slice3-eps-converted-to.pdf}%
                \linethickness{2pt}
                \put(5,13){{\tiny\color{white}PSNR: 29.70}}
                \put(5,6){{\tiny\color{white}SSIM: 0.8362}}
            \end{overpic}
        \end{subfigure}%
        \begin{subfigure}[b]{0.24\textwidth}
            \includegraphics[width=\textwidth]{./figs/robustness/vol1/ground_truth_vol1_robustness_slice3-eps-converted-to.pdf}%
        \end{subfigure}%
        \begin{subfigure}[b]{0.24\textwidth}
            \includegraphics[width=\textwidth]{./figs/robustness/vol1/reference_vol1_robustness_slice3-eps-converted-to.pdf}%
        \end{subfigure}%
    \end{subfigure}
    \begin{subfigure}{\textwidth}
        \centering
        \rotatebox{90}{\hspace{1.5em}Move five times} %
        \begin{subfigure}[b]{0.24\textwidth}
            \begin{overpic}[width=\textwidth]{./figs/robustness/vol1/corrupted3_vol1_robustness_slice3-eps-converted-to.pdf}%
                \linethickness{2pt}
                \put(5,13){{\tiny\color{white}PSNR: 21.23}}
                \put(5,6){{\tiny\color{white}SSIM: 0.5727}}
            \end{overpic}%
        \end{subfigure}%
        \begin{subfigure}[b]{0.24\textwidth}
            \begin{overpic}[width=\textwidth]{./figs/robustness/vol1/corrected3_vol1_robustness_slice3-eps-converted-to.pdf}%
                \linethickness{2pt}
                \put(5,13){{\tiny\color{white}PSNR: 27.66}}
                \put(5,6){{\tiny\color{white}SSIM: 0.8298}}
            \end{overpic}
        \end{subfigure}%
        \begin{subfigure}[b]{0.24\textwidth}
            \includegraphics[width=\textwidth]{./figs/robustness/vol1/ground_truth_vol1_robustness_slice3-eps-converted-to.pdf}%
        \end{subfigure}%
        \begin{subfigure}[b]{0.24\textwidth}
            \includegraphics[width=\textwidth]{./figs/robustness/vol1/reference_vol1_robustness_slice3-eps-converted-to.pdf}%
        \end{subfigure}%
    \end{subfigure}
    \caption{Summary of the reconstruction results for volunteer 1 (see Figures \ref{fig:robustness1_vol2}, \ref{fig:robustness2_vol2}, \ref{fig:robustness3_vol2}). The volunteer is instructed to move a variable number of times during the scan in order to test the robustness of the proposed correction scheme with respect the motion complexity. The corrupted images are increasingly affected by motion artifacts, however only modest decrease in reconstruction quality can be observed for the corrected images (here axial slices).}\label{fig:robustness_comparison}
\end{figure}
\begin{figure}[!htb]
    \centering
    \begin{subfigure}{\textwidth}
        \centering
        \rotatebox{90}{\hspace{2.5em}Sagittal} %
        \begin{subfigure}[b]{0.24\textwidth}
            \caption*{Corrupted}\vspace{-0.5em}%
            \begin{overpic}[width=\textwidth]{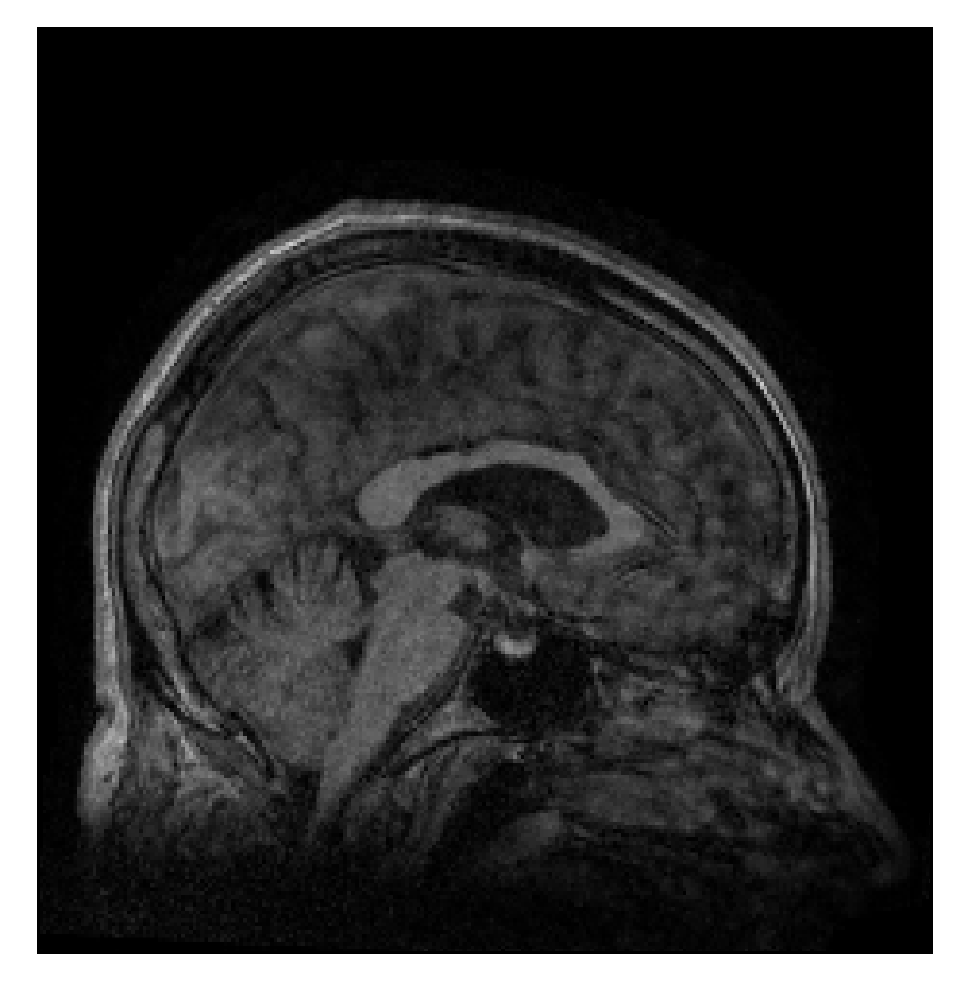}%
                \linethickness{2pt}
                \put(5,13){{\tiny\color{white}PSNR: 25.84}}
                \put(5,6){{\tiny\color{white}SSIM: 0.7032}}
            \end{overpic}%
        \end{subfigure}%
        \begin{subfigure}[b]{0.24\textwidth}
            \caption*{Corrected}\vspace{-0.5em}%
            \begin{overpic}[width=\textwidth]{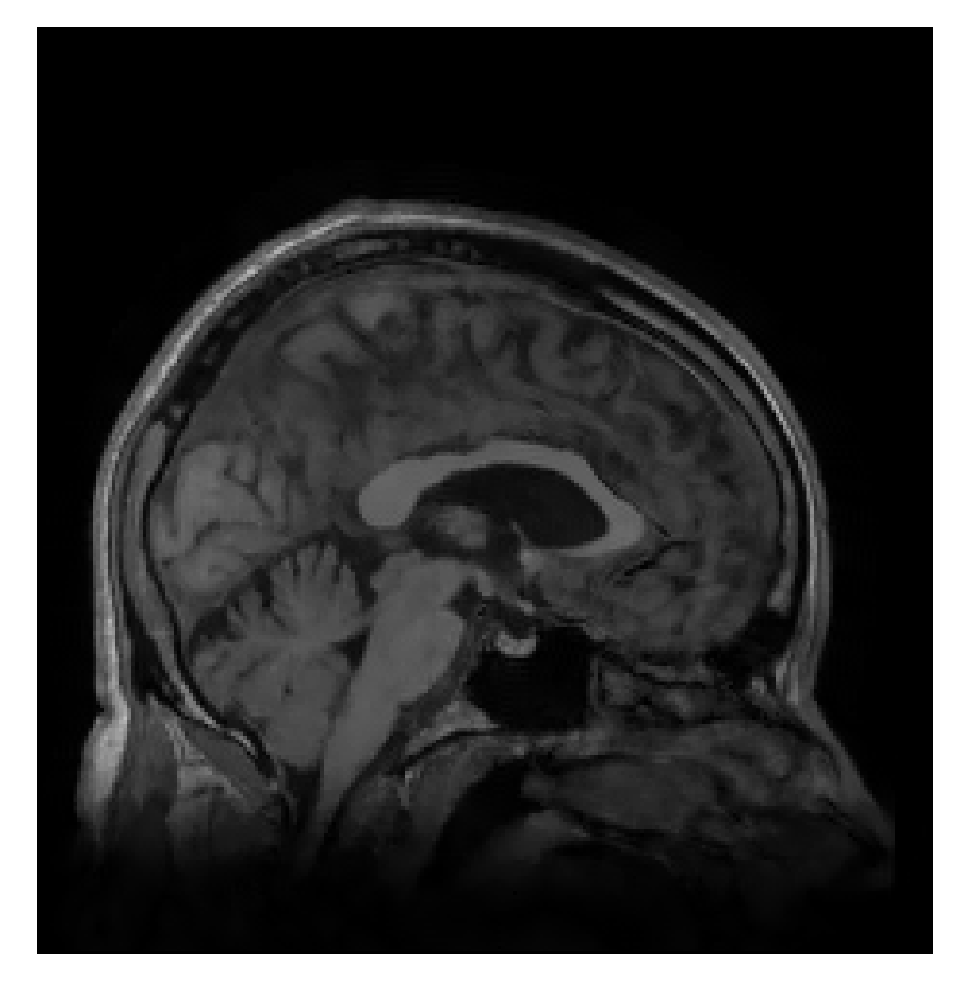}%
                \linethickness{2pt}
                \put(5,13){{\tiny\color{white}PSNR: 28.07}}
                \put(5,6){{\tiny\color{white}SSIM: 0.8093}}
            \end{overpic}%
        \end{subfigure}%
        \begin{subfigure}[b]{0.24\textwidth}
            \caption*{Ground truth}\vspace{-0.5em}%
            \includegraphics[width=\textwidth]{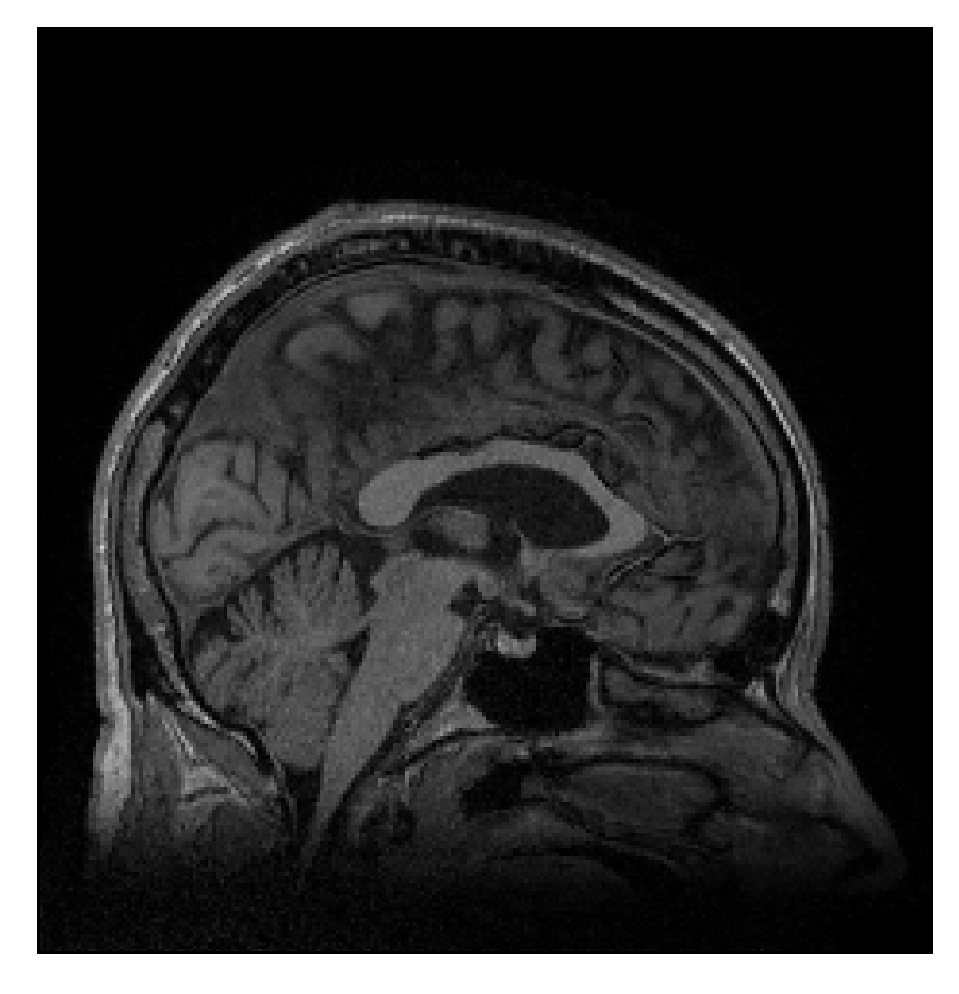}%
        \end{subfigure}%
        \begin{subfigure}[b]{0.24\textwidth}
            \caption*{Reference}\vspace{-0.5em}%
            \includegraphics[width=\textwidth]{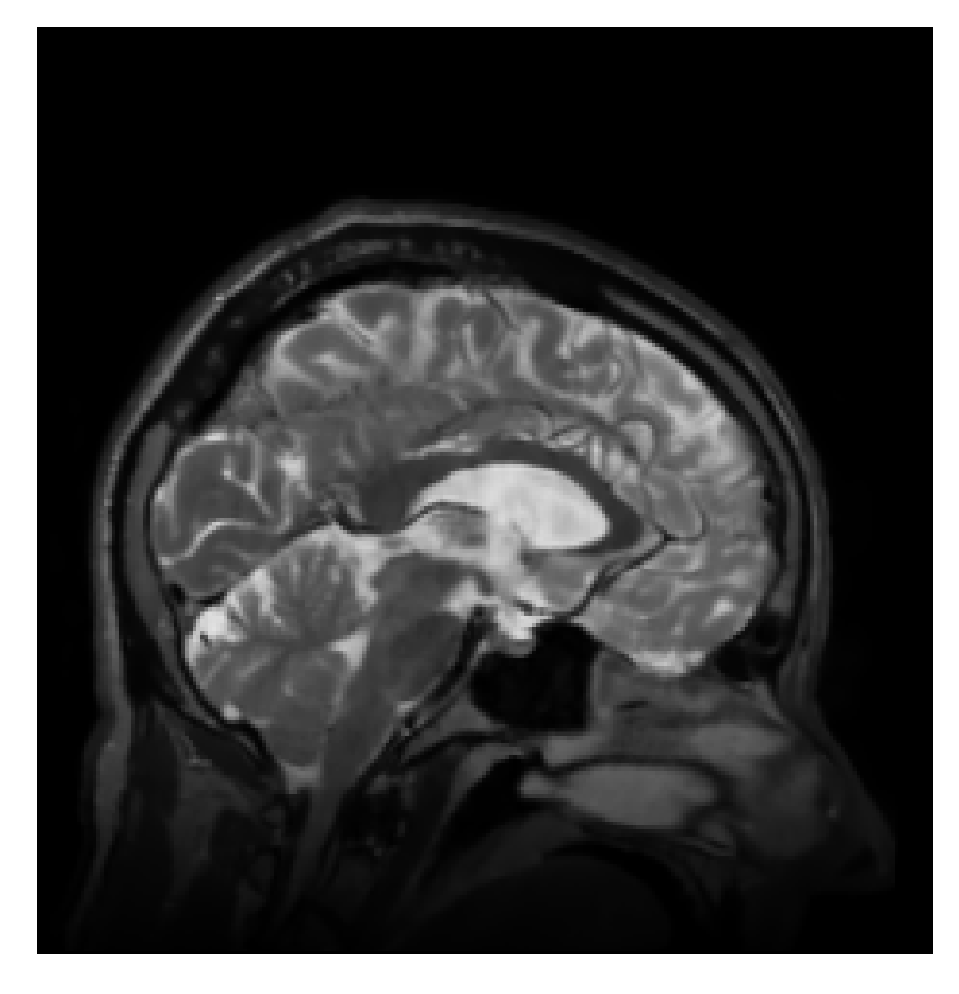}%
        \end{subfigure}%
    \end{subfigure}
    \begin{subfigure}{\textwidth}
        \centering
        \rotatebox{90}{\hspace{2.5em}Coronal} %
        \begin{subfigure}[b]{0.24\textwidth}
            \begin{overpic}[width=\textwidth]{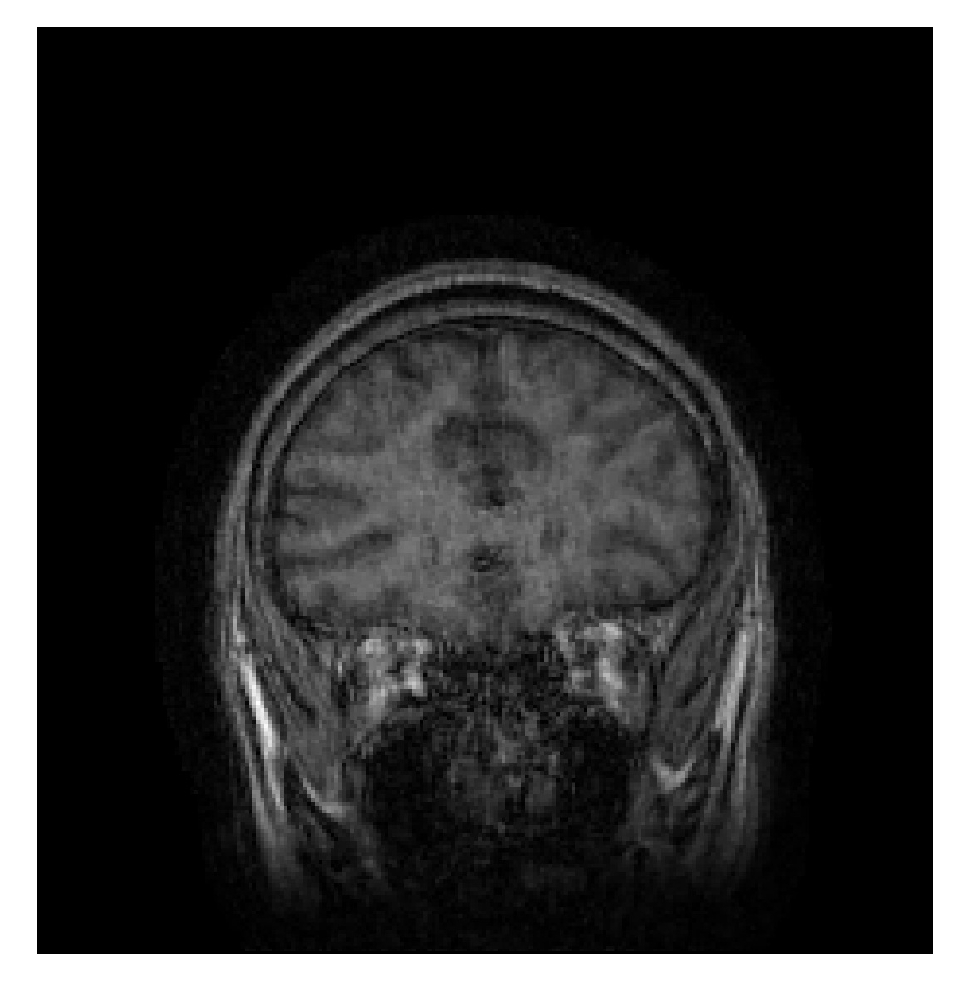}%
                \linethickness{2pt}
                \put(5,13){{\tiny\color{white}PSNR: 26.35}}
                \put(5,6){{\tiny\color{white}SSIM: 0.7851}}
            \end{overpic}%
        \end{subfigure}%
        \begin{subfigure}[b]{0.24\textwidth}
            \begin{overpic}[width=\textwidth]{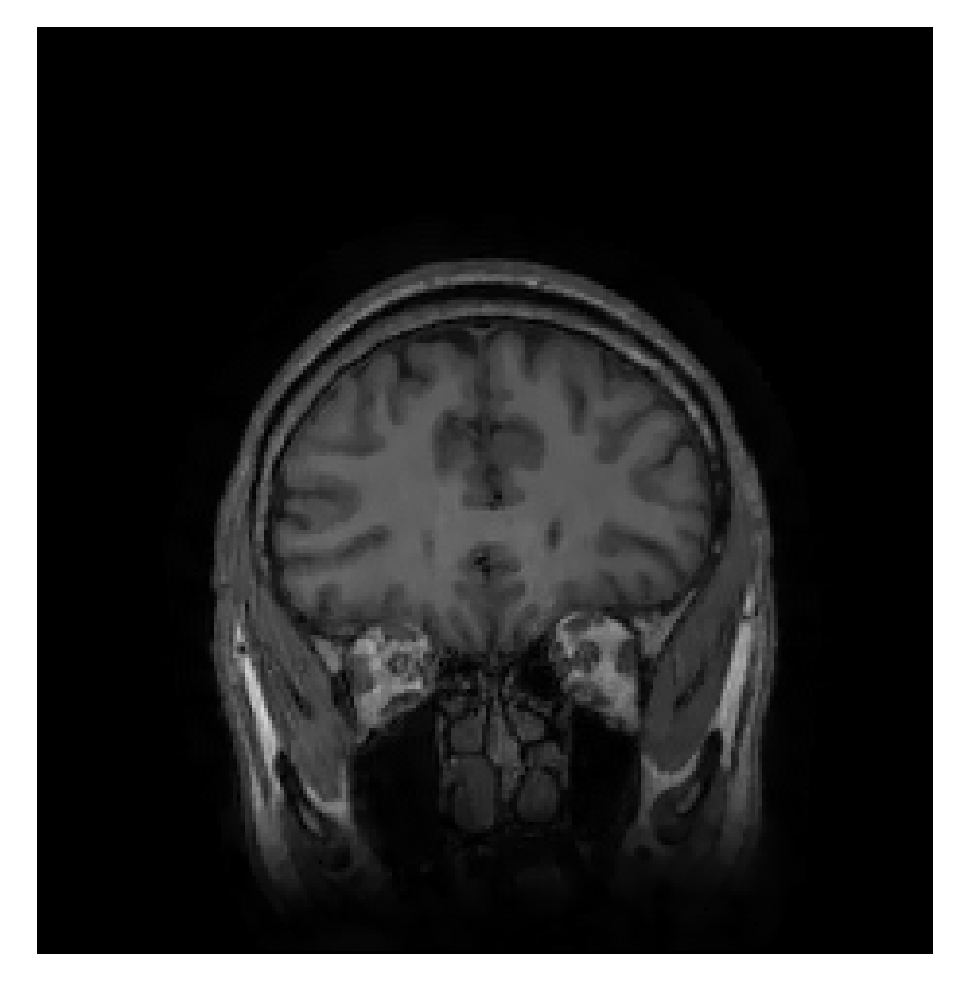}%
                \linethickness{2pt}
                \put(5,13){{\tiny\color{white}PSNR: 30.40}}
                \put(5,6){{\tiny\color{white}SSIM: 0.9021}}
            \end{overpic}%
        \end{subfigure}%
        \begin{subfigure}[b]{0.24\textwidth}
            \includegraphics[width=\textwidth]{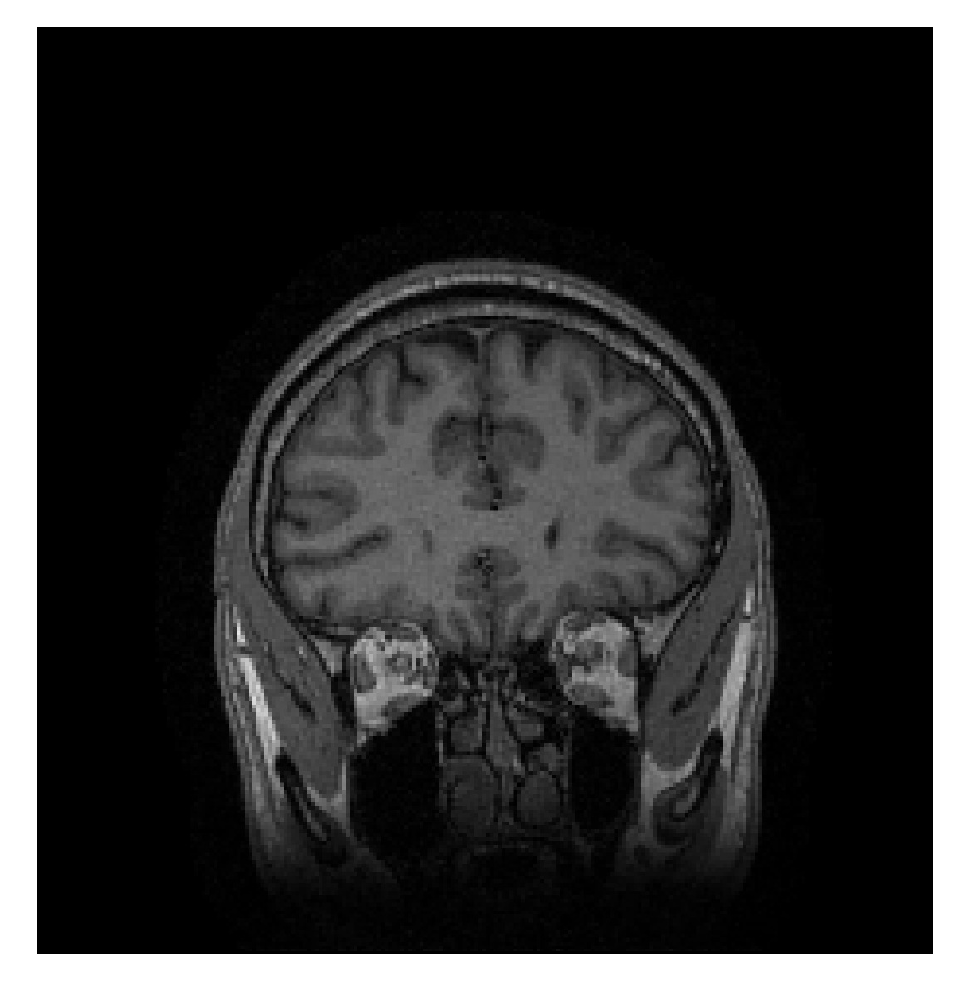}%
        \end{subfigure}%
        \begin{subfigure}[b]{0.24\textwidth}
            \includegraphics[width=\textwidth]{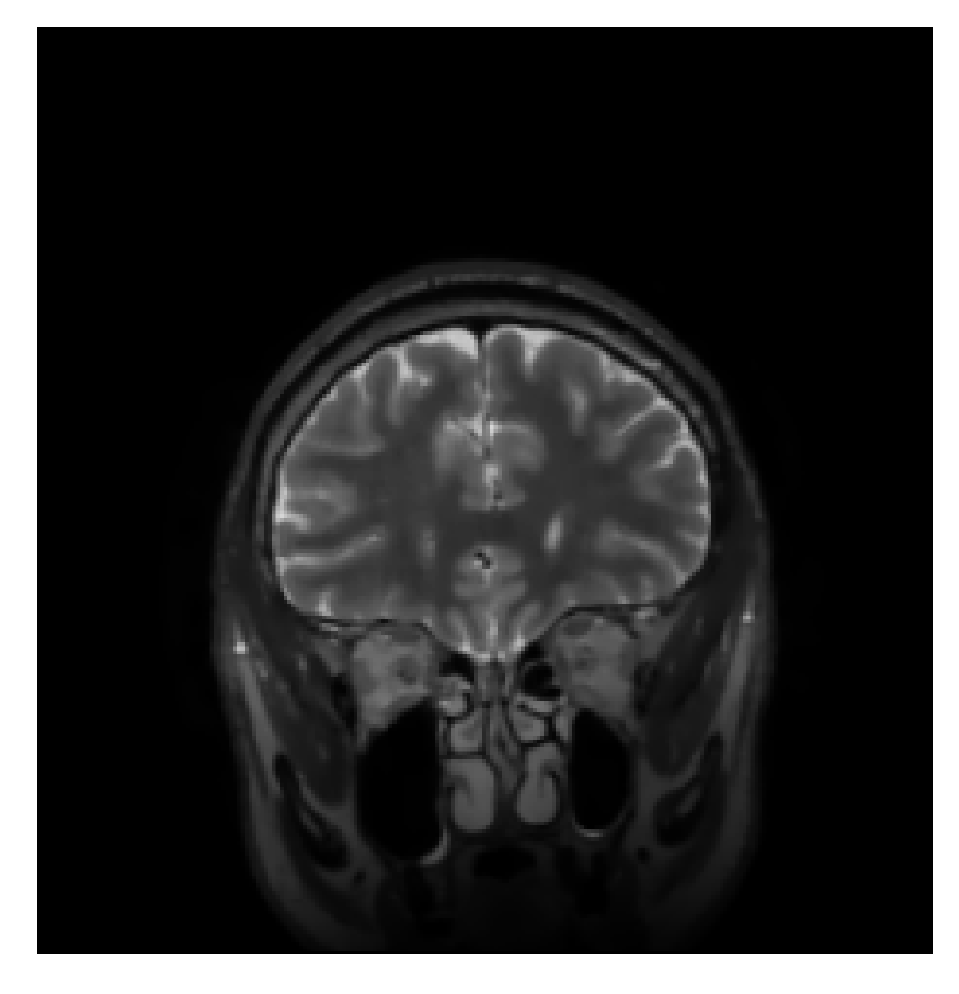}%
        \end{subfigure}%
    \end{subfigure}
    \begin{subfigure}{\textwidth}
        \centering
        \rotatebox{90}{\hspace{3em}Axial} %
        \begin{subfigure}[b]{0.24\textwidth}
            \begin{overpic}[width=\textwidth]{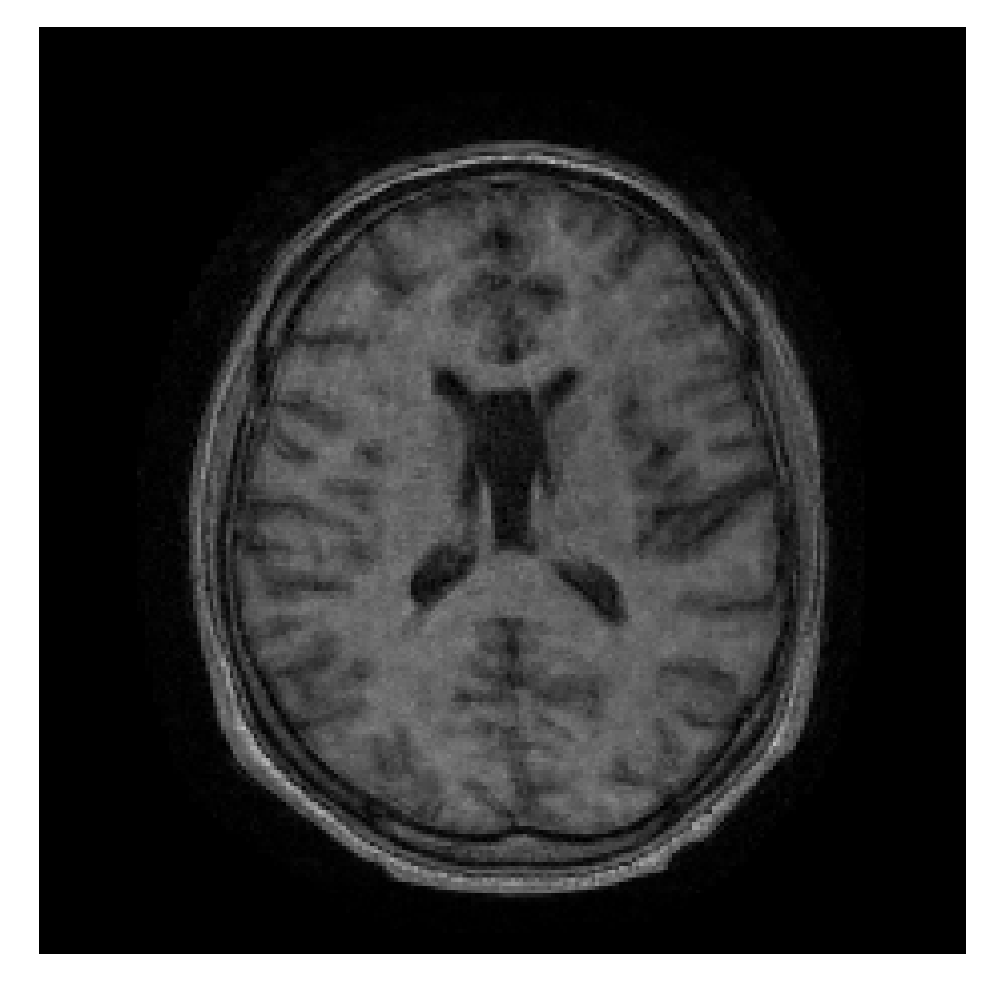}%
                \linethickness{2pt}
                \put(5,13){{\tiny\color{white}PSNR: 28.11}}
                \put(5,6){{\tiny\color{white}SSIM: 0.8248}}
            \end{overpic}%
        \end{subfigure}%
        \begin{subfigure}[b]{0.24\textwidth}
            \begin{overpic}[width=\textwidth]{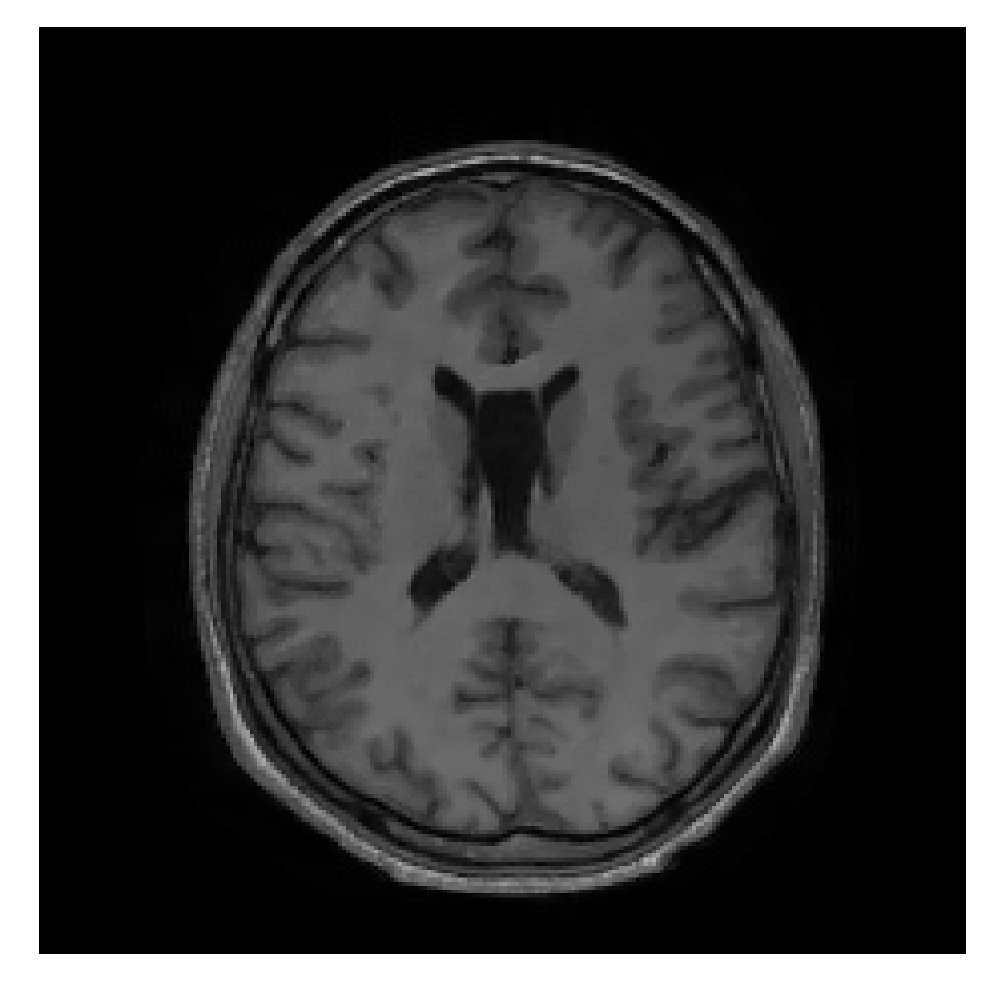}%
                \linethickness{2pt}
                \put(5,13){{\tiny\color{white}PSNR: 30.55}}
                \put(5,6){{\tiny\color{white}SSIM: 0.9012}}
            \end{overpic}
        \end{subfigure}%
        \begin{subfigure}[b]{0.24\textwidth}
            \includegraphics[width=\textwidth]{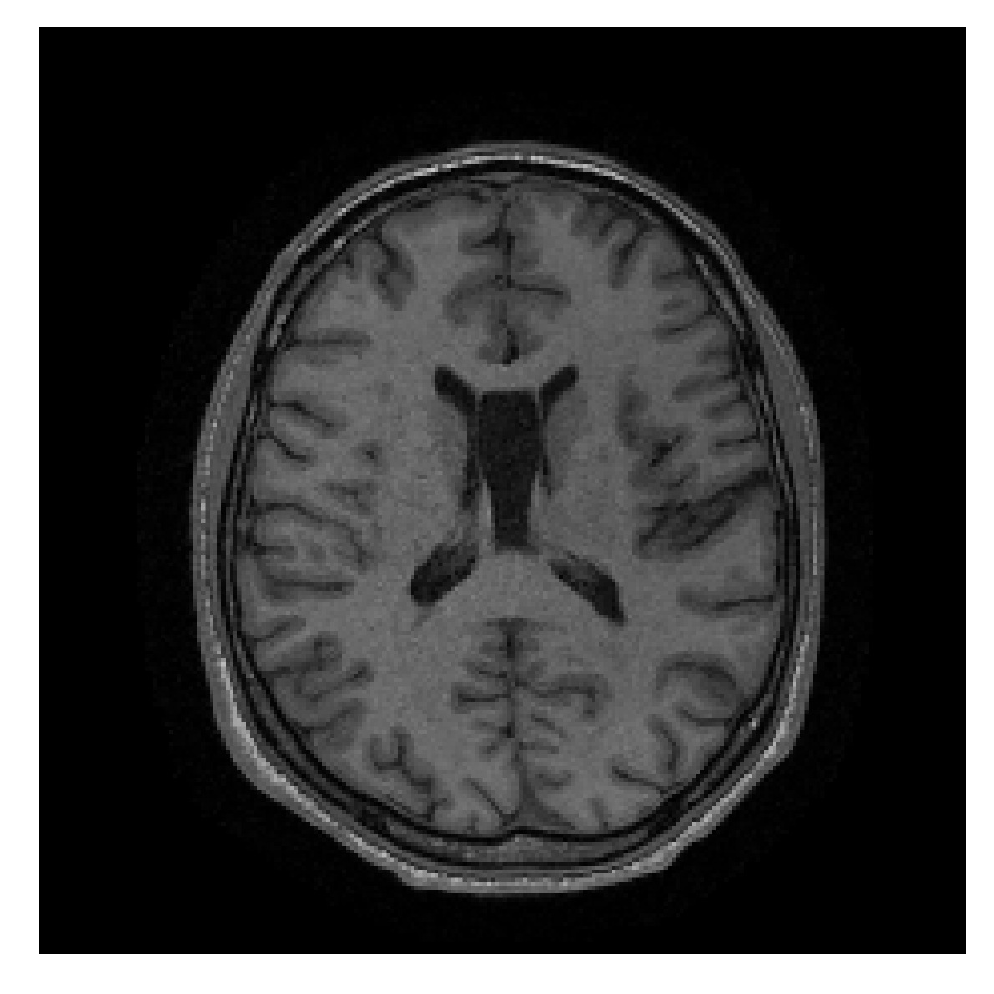}%
        \end{subfigure}%
        \begin{subfigure}[b]{0.24\textwidth}
            \includegraphics[width=\textwidth]{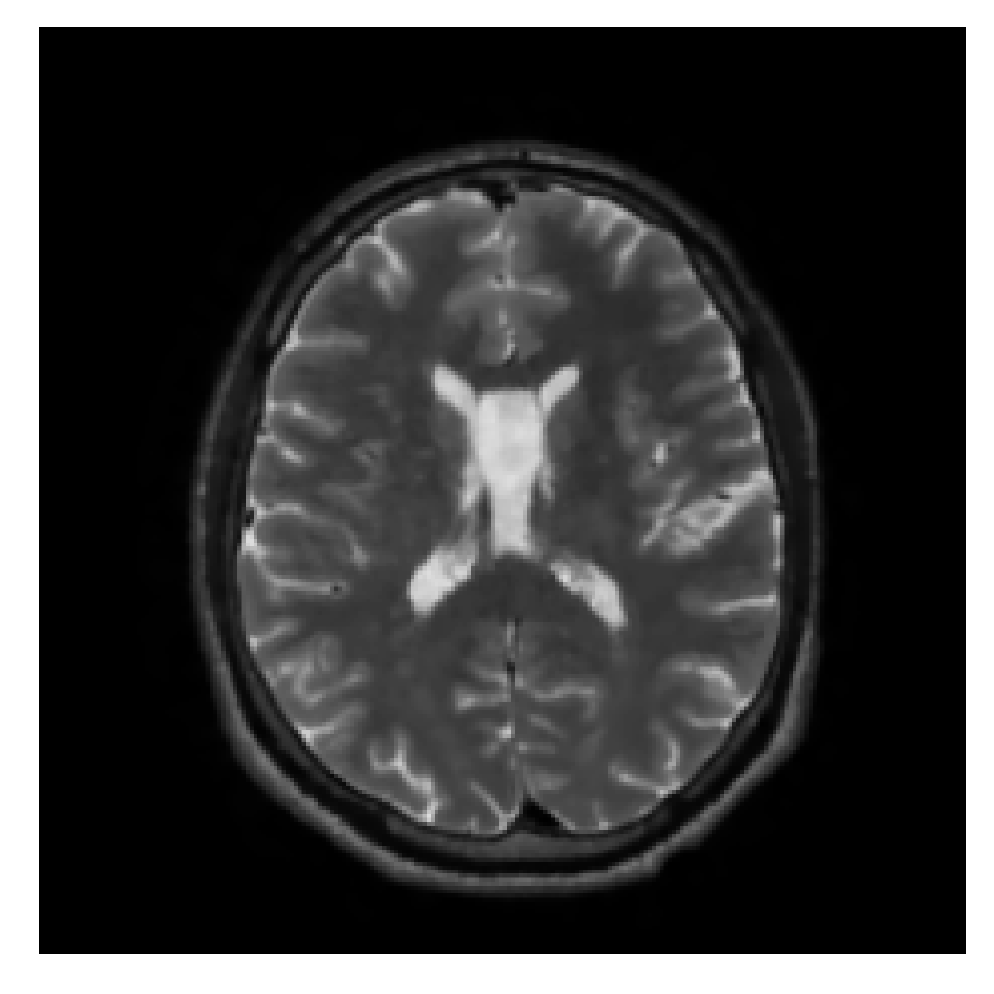}%
        \end{subfigure}%
    \end{subfigure}
    \begin{subfigure}{\textwidth}
        \centering
        \rotatebox{90}{\hspace{0.8em}Coronal detail} %
        \begin{subfigure}[b]{0.24\textwidth}
            \includegraphics[width=\textwidth]{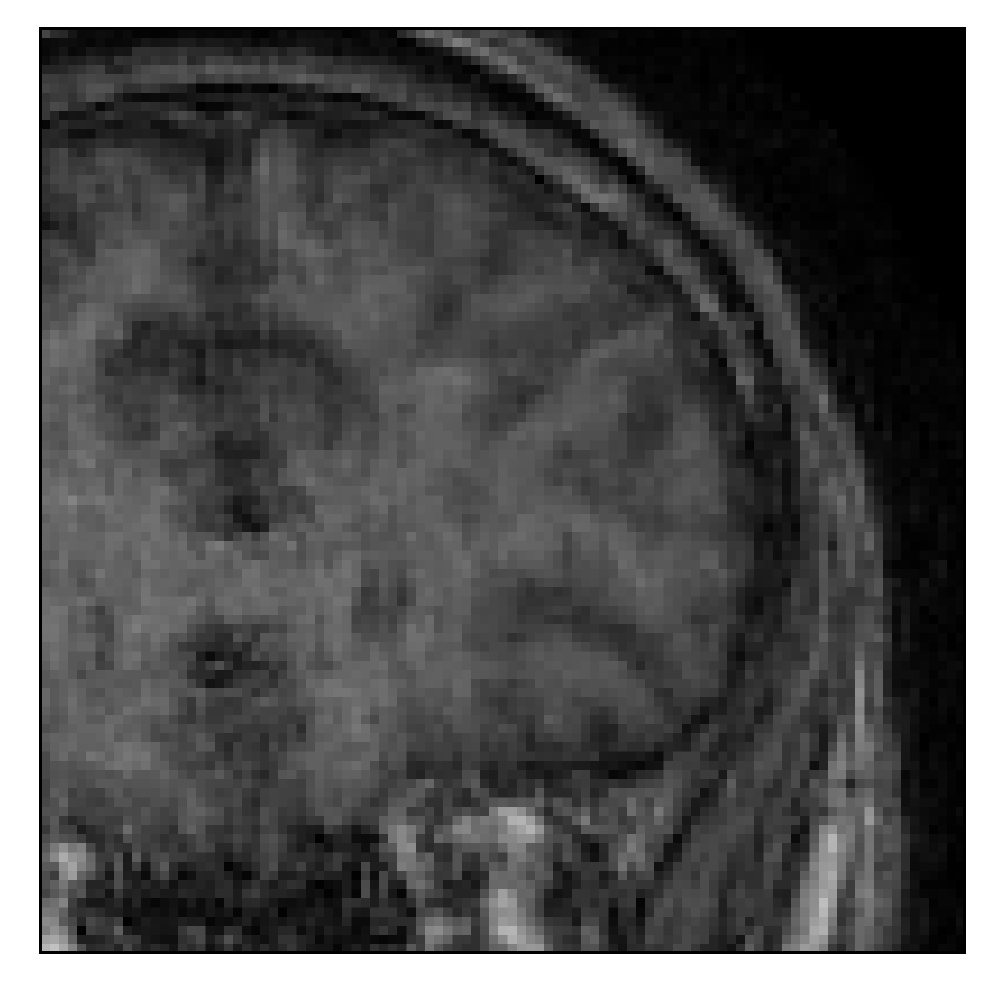}%
        \end{subfigure}%
        \begin{subfigure}[b]{0.24\textwidth}
            \includegraphics[width=\textwidth]{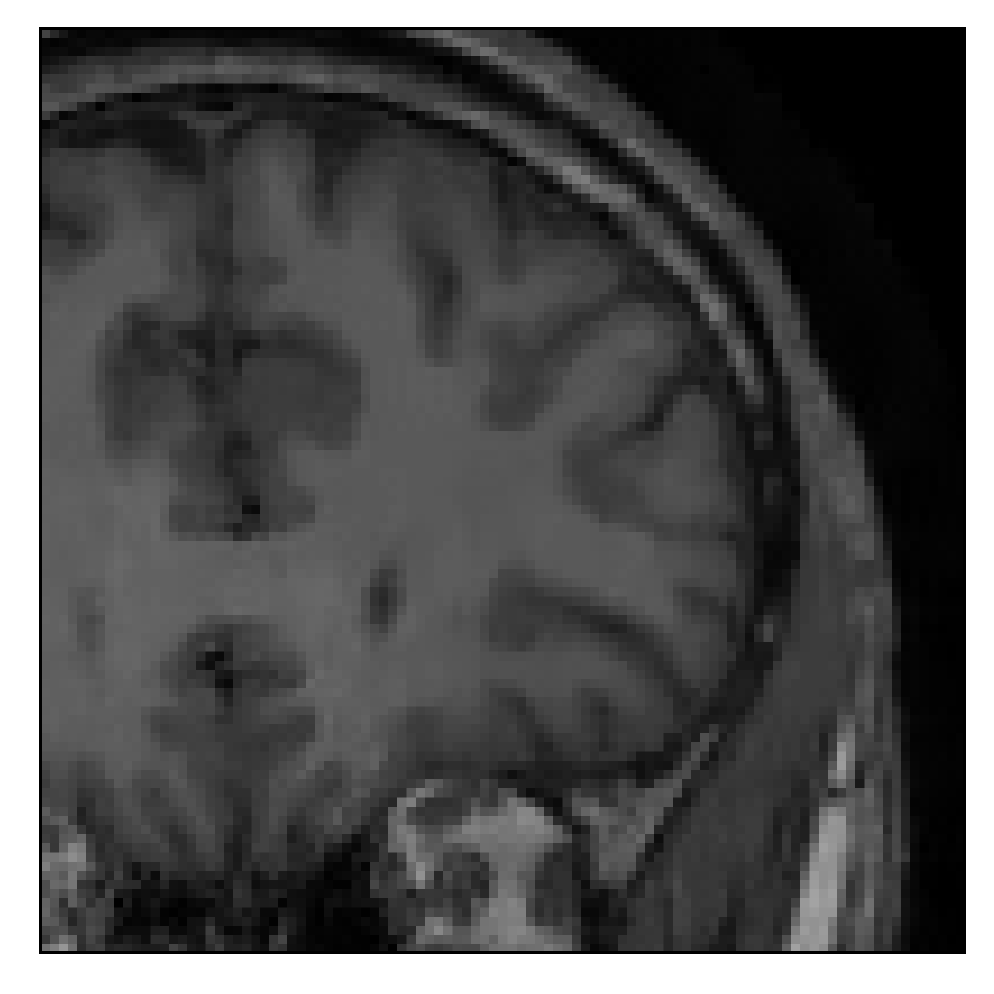}%
        \end{subfigure}%
        \begin{subfigure}[b]{0.24\textwidth}
            \includegraphics[width=\textwidth]{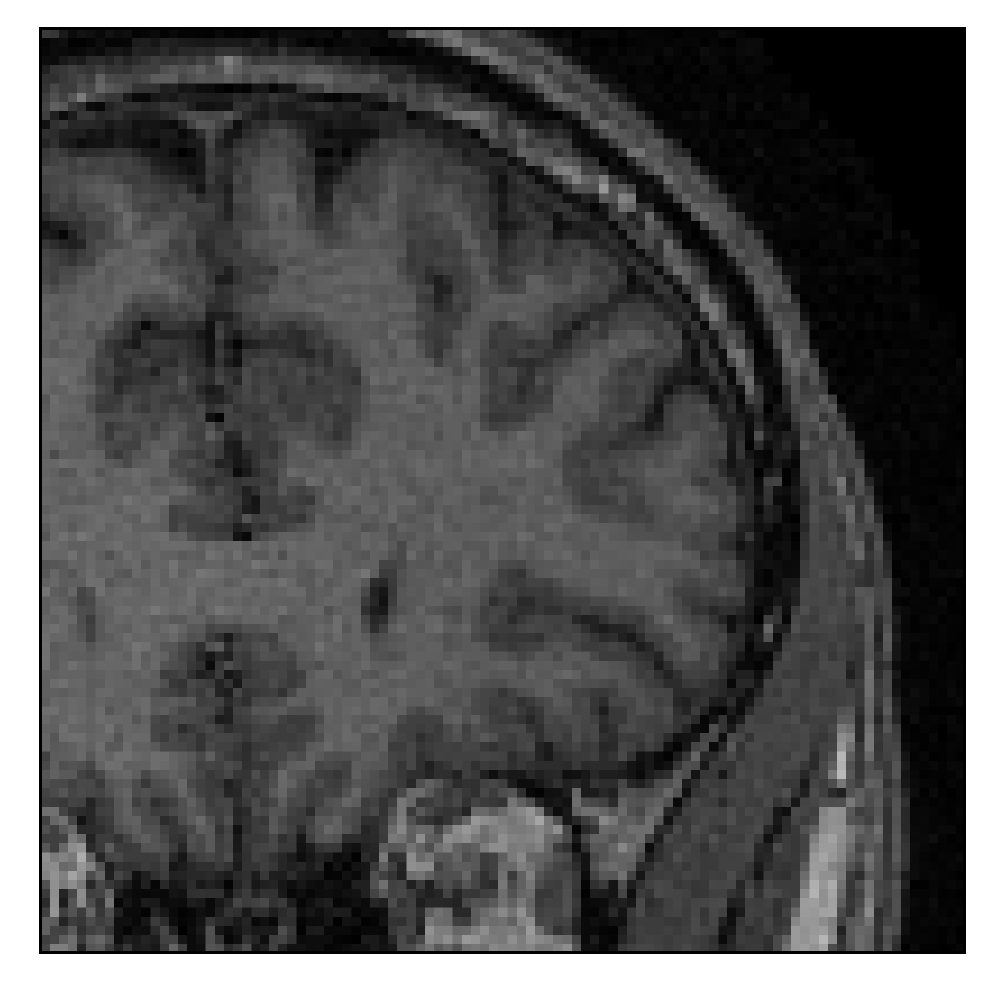}%
        \end{subfigure}%
        \begin{subfigure}[b]{0.24\textwidth}
            \includegraphics[width=\textwidth]{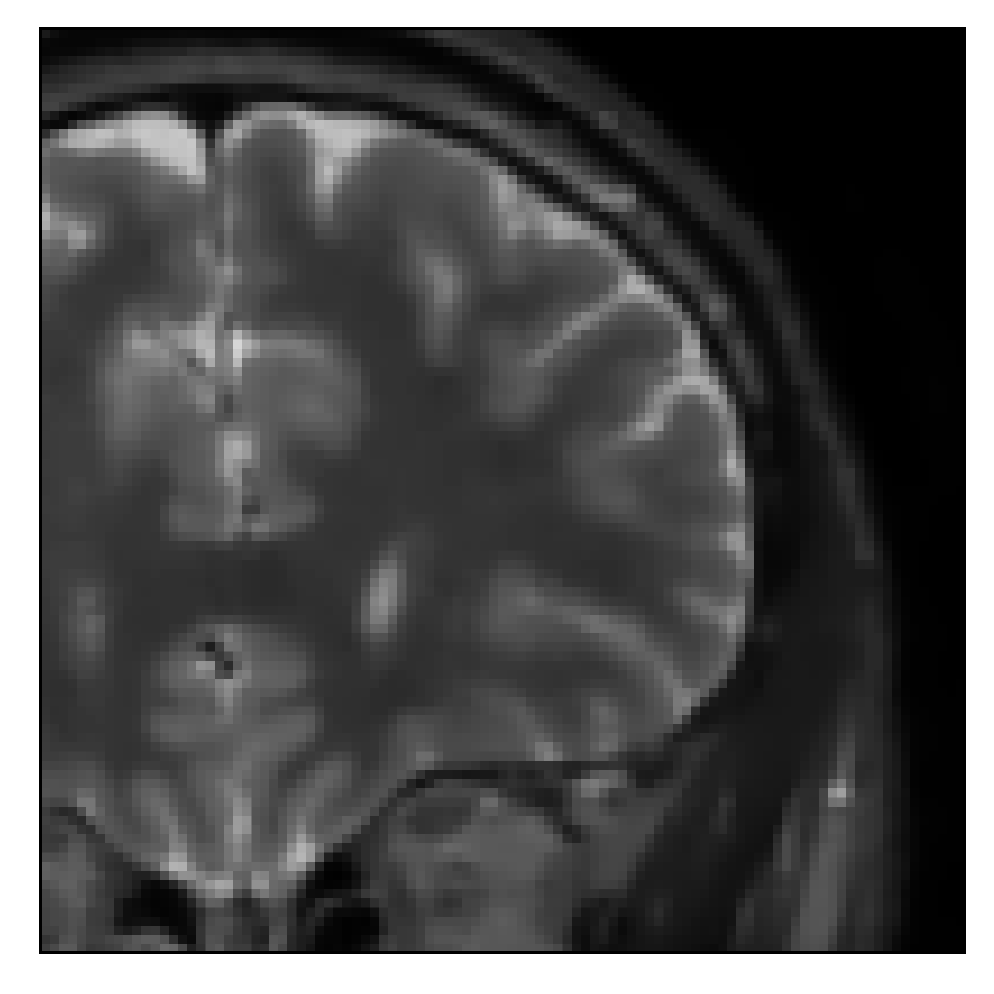}%
        \end{subfigure}%
    \end{subfigure}
    \caption{Reconstruction results for volunteer 2. The volunteer is instructed to move five times during the scan. The corrupted contrast is T1-weighted, while the reference contrast is T2-weighted. The proposed correction scheme is agnostic about the choice of corrupted/reference contrast combinations with similar spectral content.}\label{fig:prior}
\end{figure}
\begin{figure}[!htb]
    \centering
    \begin{subfigure}{\textwidth}
        \centering
        \rotatebox{90}{\hspace{2.5em}Sagittal} %
        \begin{subfigure}[b]{0.24\textwidth}
            \caption*{Corrupted}\vspace{-0.5em}%
            \begin{overpic}[width=\textwidth]{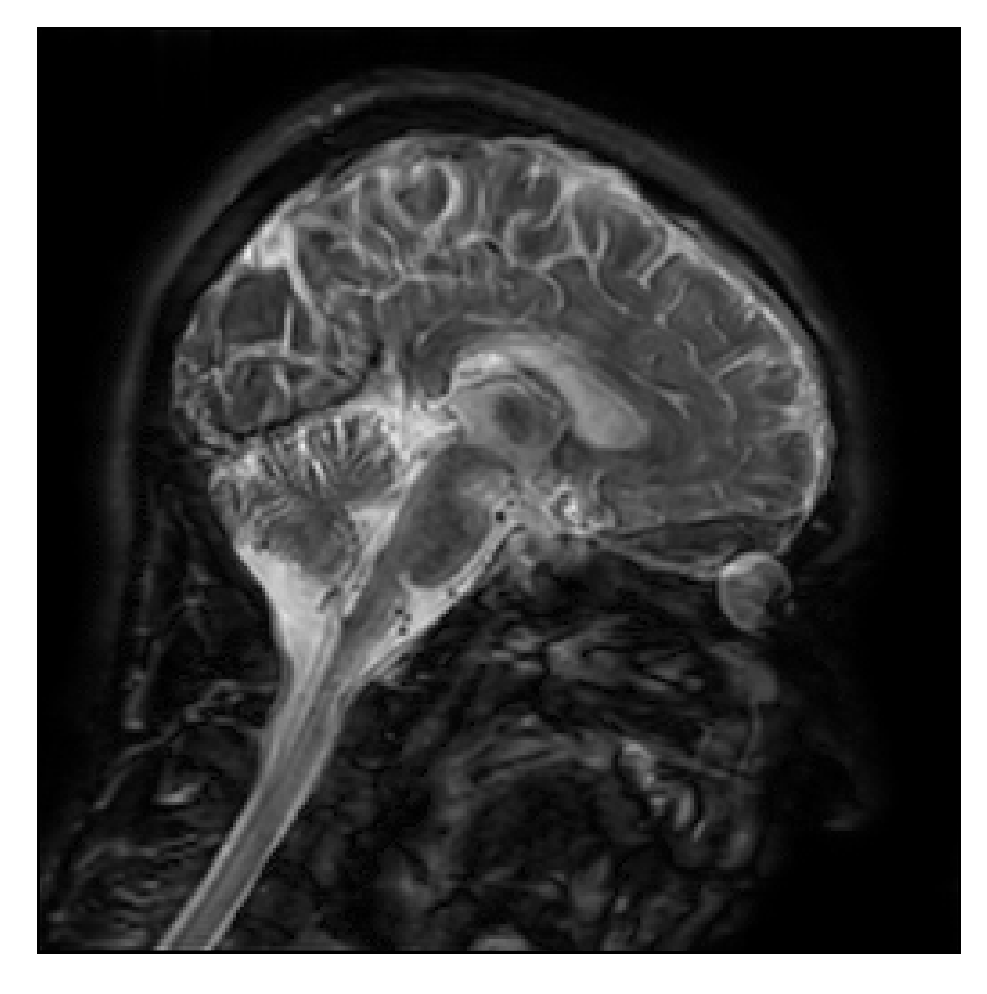}%
                \linethickness{2pt}
                \put(5,13){{\tiny\color{white}PSNR: 22.26}}
                \put(5,6){{\tiny\color{white}SSIM: 0.6963}}
            \end{overpic}%
        \end{subfigure}%
        \begin{subfigure}[b]{0.24\textwidth}
            \caption*{Corrected}\vspace{-0.5em}%
            \begin{overpic}[width=\textwidth]{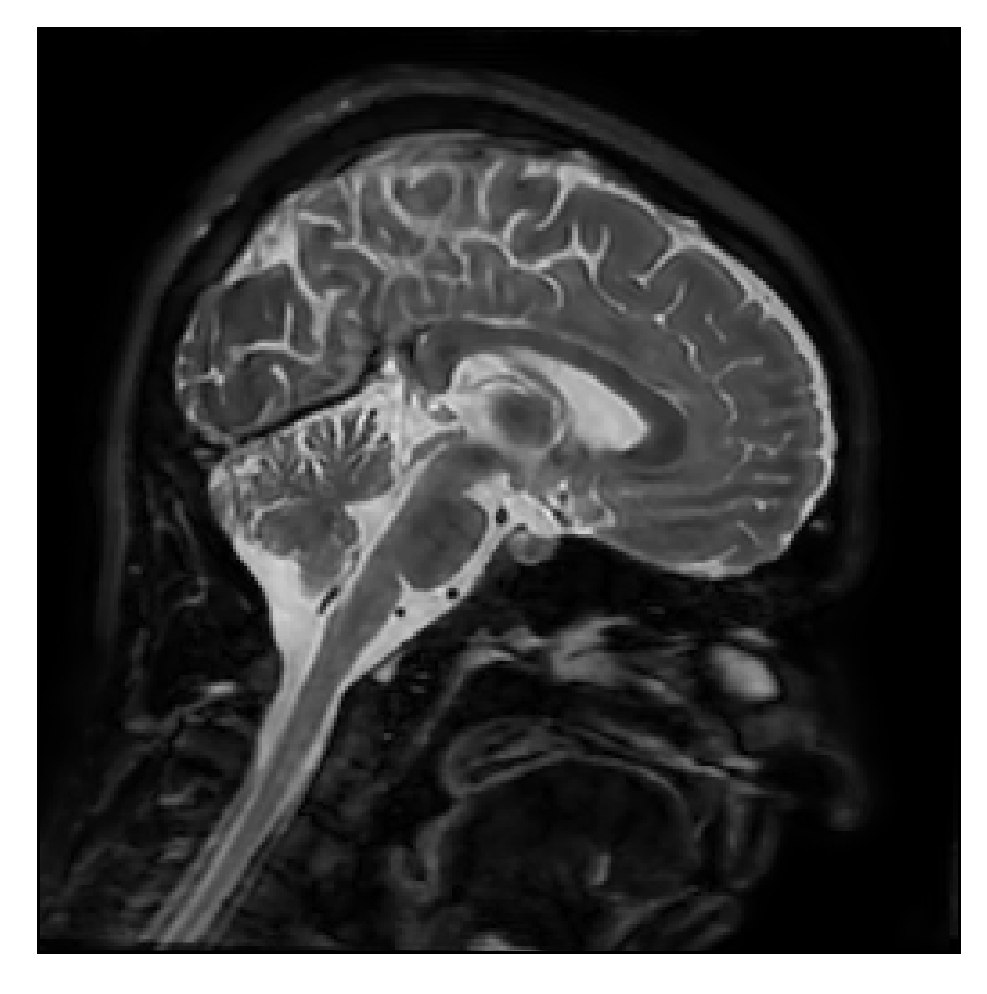}%
                \linethickness{2pt}
                \put(5,13){{\tiny\color{white}PSNR: 27.54}}
                \put(5,6){{\tiny\color{white}SSIM: 0.8409}}
            \end{overpic}%
        \end{subfigure}%
        \begin{subfigure}[b]{0.24\textwidth}
            \caption*{Ground truth}\vspace{-0.5em}%
            \includegraphics[width=\textwidth]{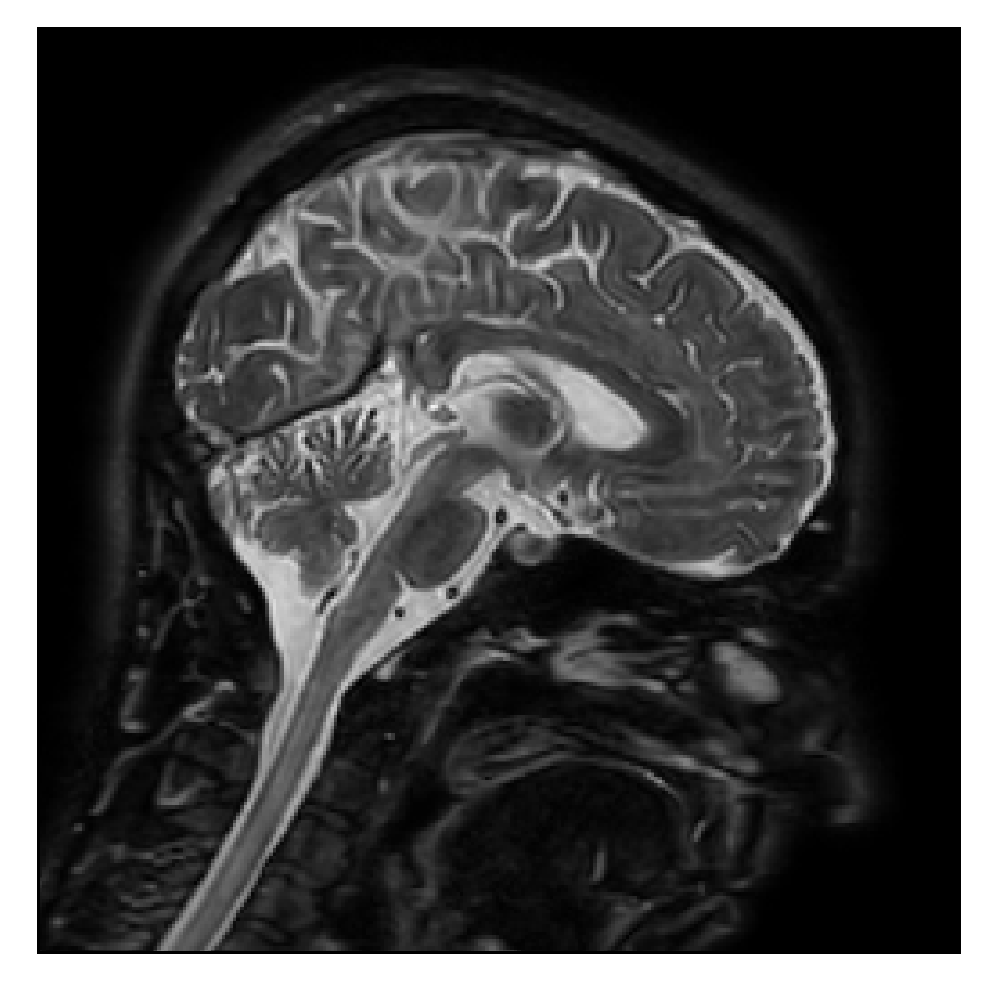}%
        \end{subfigure}%
        \begin{subfigure}[b]{0.24\textwidth}
            \caption*{Reference}\vspace{-0.5em}%
            \includegraphics[width=\textwidth]{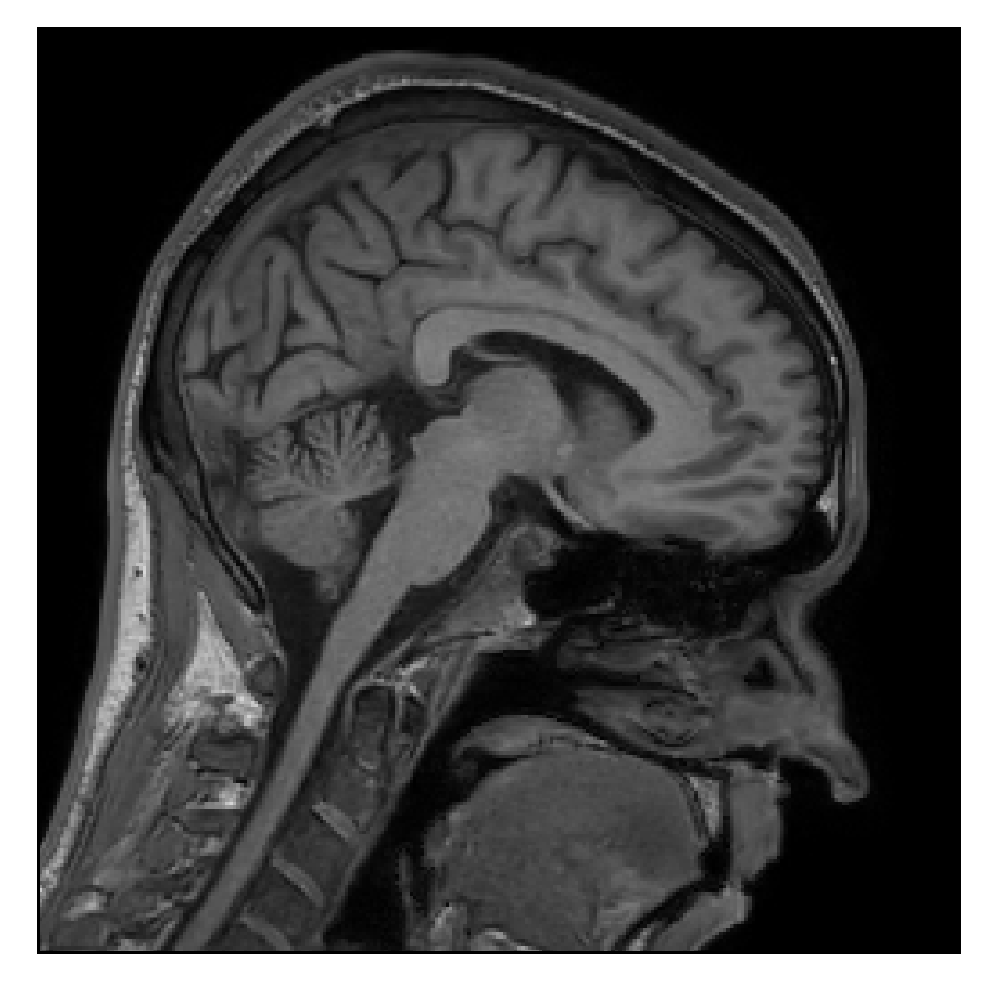}%
        \end{subfigure}%
    \end{subfigure}
    \begin{subfigure}{\textwidth}
        \centering
        \rotatebox{90}{\hspace{2.5em}Coronal} %
        \begin{subfigure}[b]{0.24\textwidth}
            \begin{overpic}[width=\textwidth]{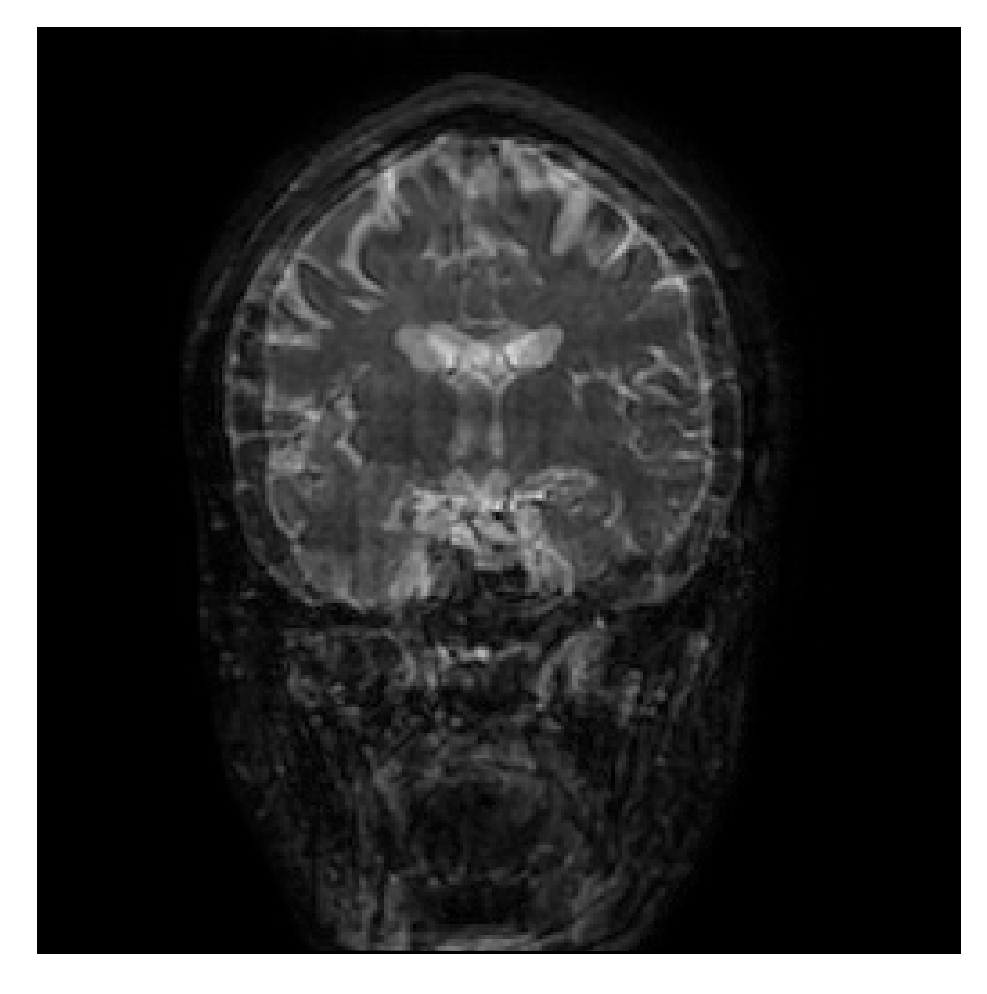}%
                \linethickness{2pt}
                \put(5,13){{\tiny\color{white}PSNR: 23.46}}
                \put(5,6){{\tiny\color{white}SSIM: 0.7321}}
            \end{overpic}%
        \end{subfigure}%
        \begin{subfigure}[b]{0.24\textwidth}
            \begin{overpic}[width=\textwidth]{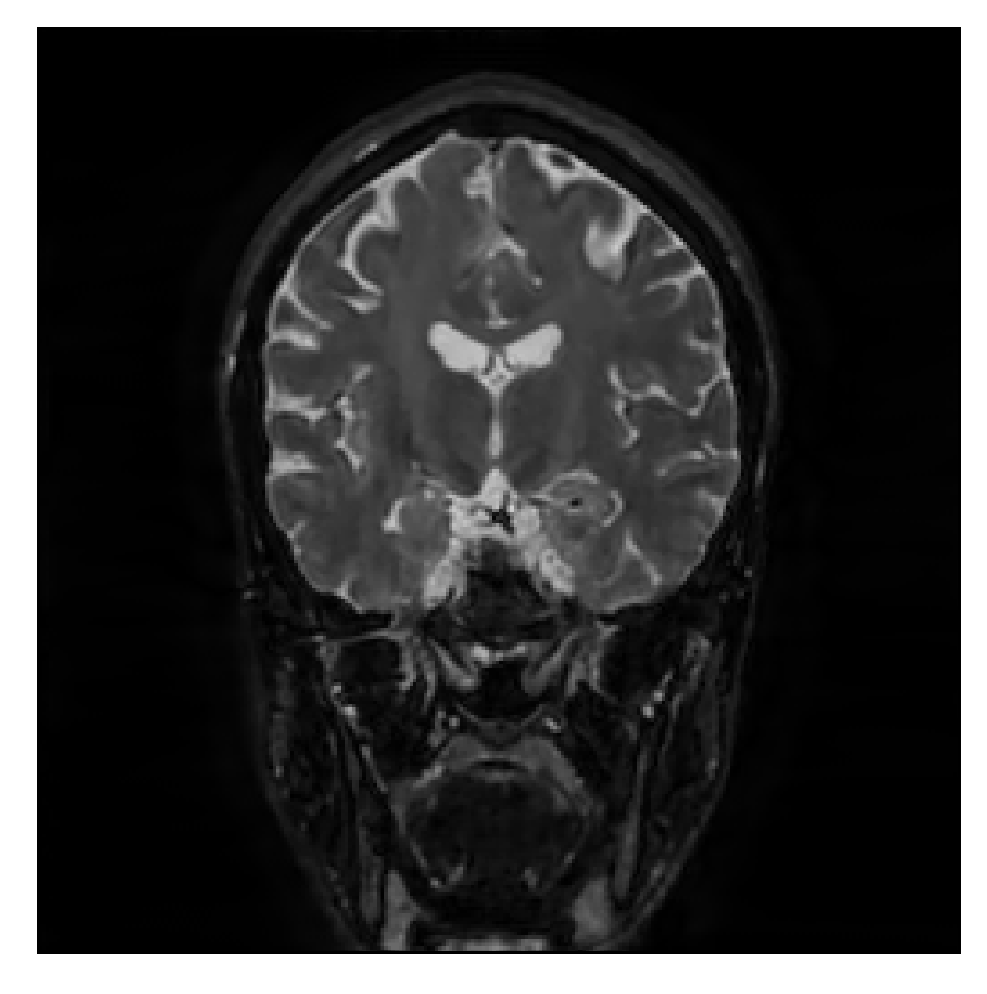}%
                \linethickness{2pt}
                \put(5,13){{\tiny\color{white}PSNR: 31.65}}
                \put(5,6){{\tiny\color{white}SSIM: 0.8370}}
            \end{overpic}%
        \end{subfigure}%
        \begin{subfigure}[b]{0.24\textwidth}
            \includegraphics[width=\textwidth]{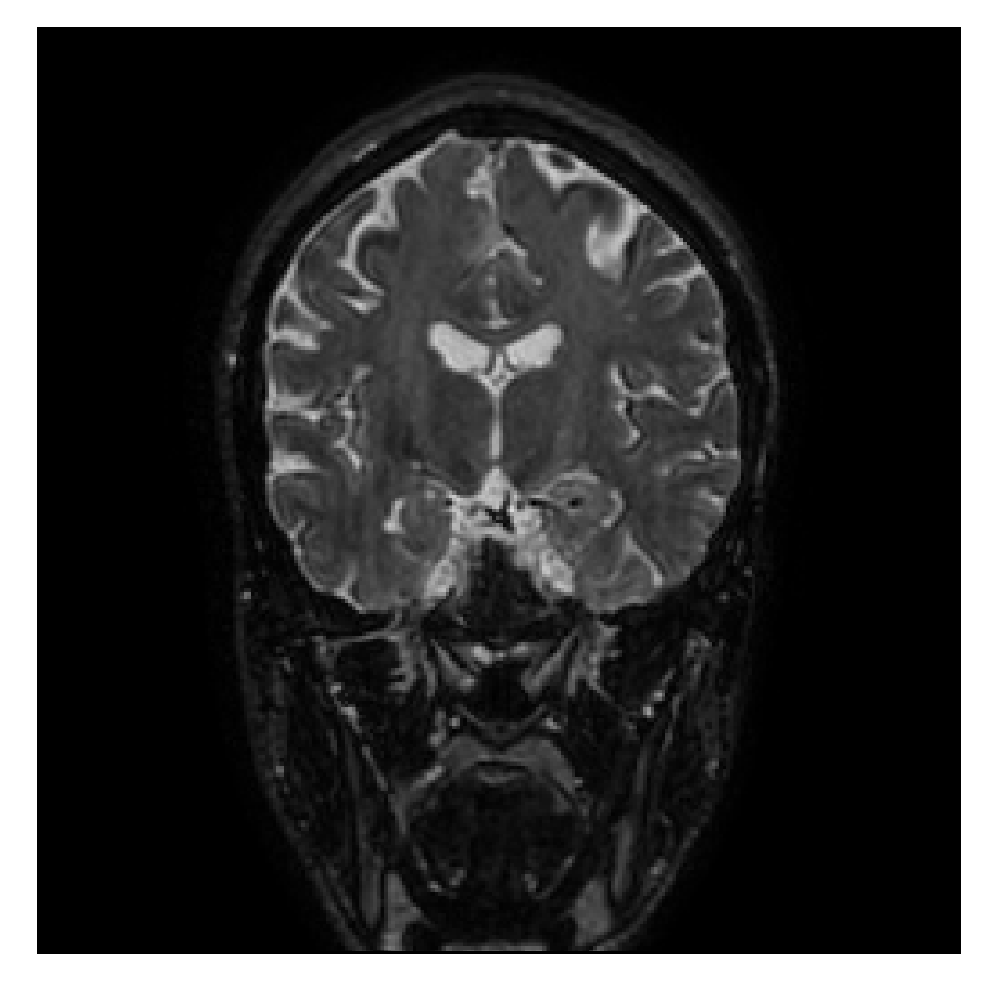}%
        \end{subfigure}%
        \begin{subfigure}[b]{0.24\textwidth}
            \includegraphics[width=\textwidth]{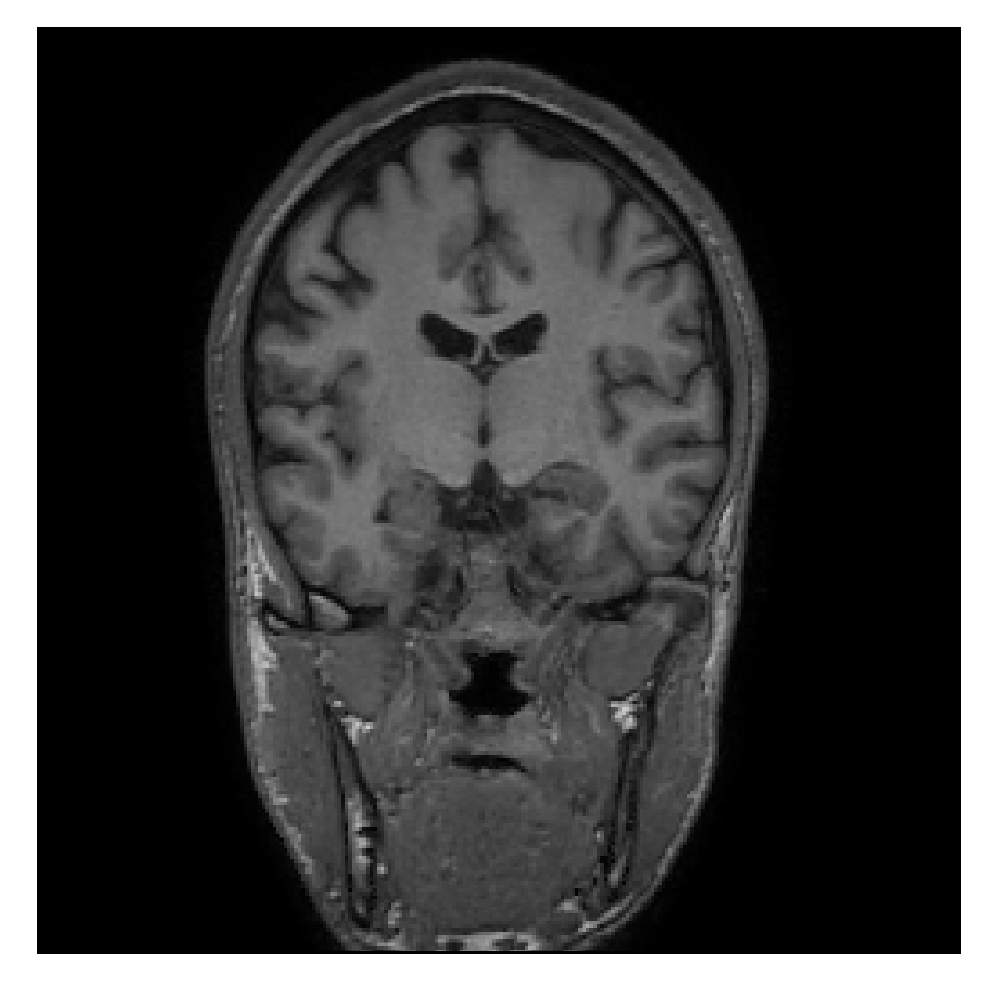}%
        \end{subfigure}%
    \end{subfigure}
    \begin{subfigure}{\textwidth}
        \centering
        \rotatebox{90}{\hspace{3em}Axial} %
        \begin{subfigure}[b]{0.24\textwidth}
            \begin{overpic}[width=\textwidth]{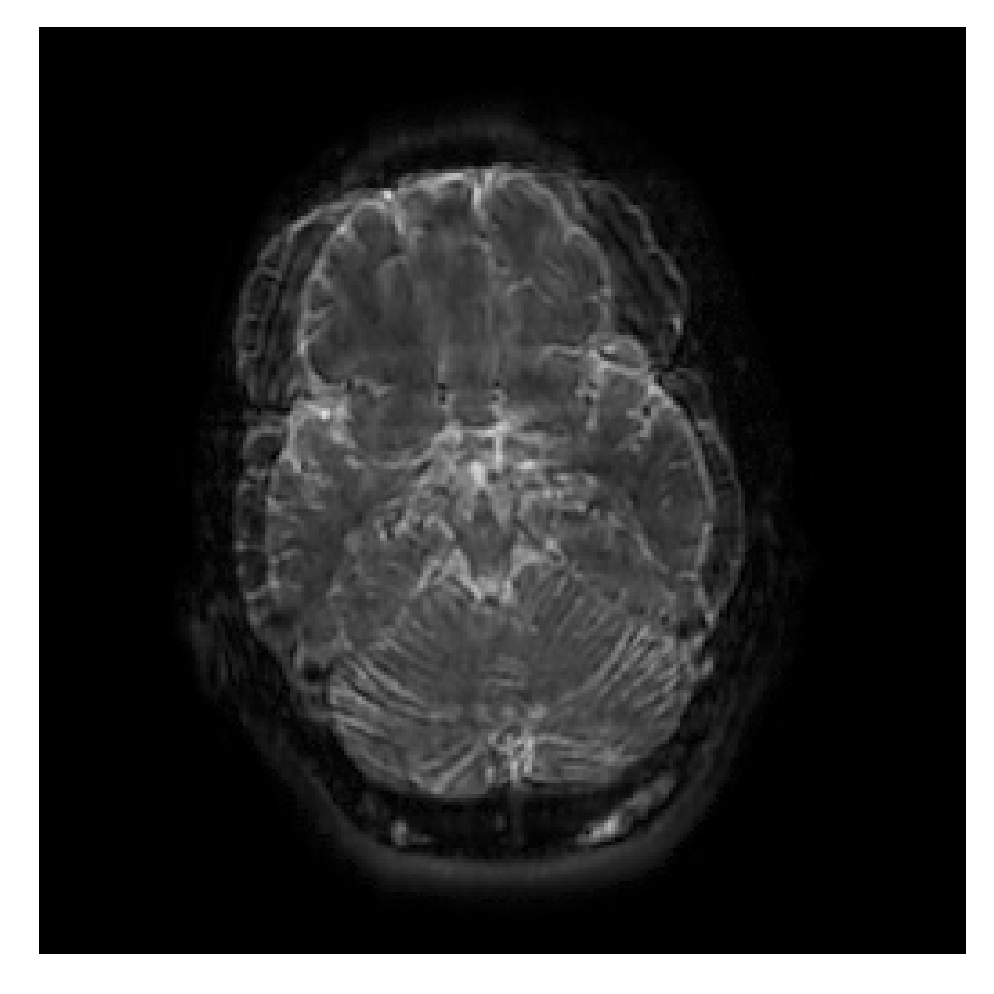}%
                \linethickness{2pt}
                \put(5,13){{\tiny\color{white}PSNR: 24.55}}
                \put(5,6){{\tiny\color{white}SSIM: 0.7895}}
            \end{overpic}%
        \end{subfigure}%
        \begin{subfigure}[b]{0.24\textwidth}
            \begin{overpic}[width=\textwidth]{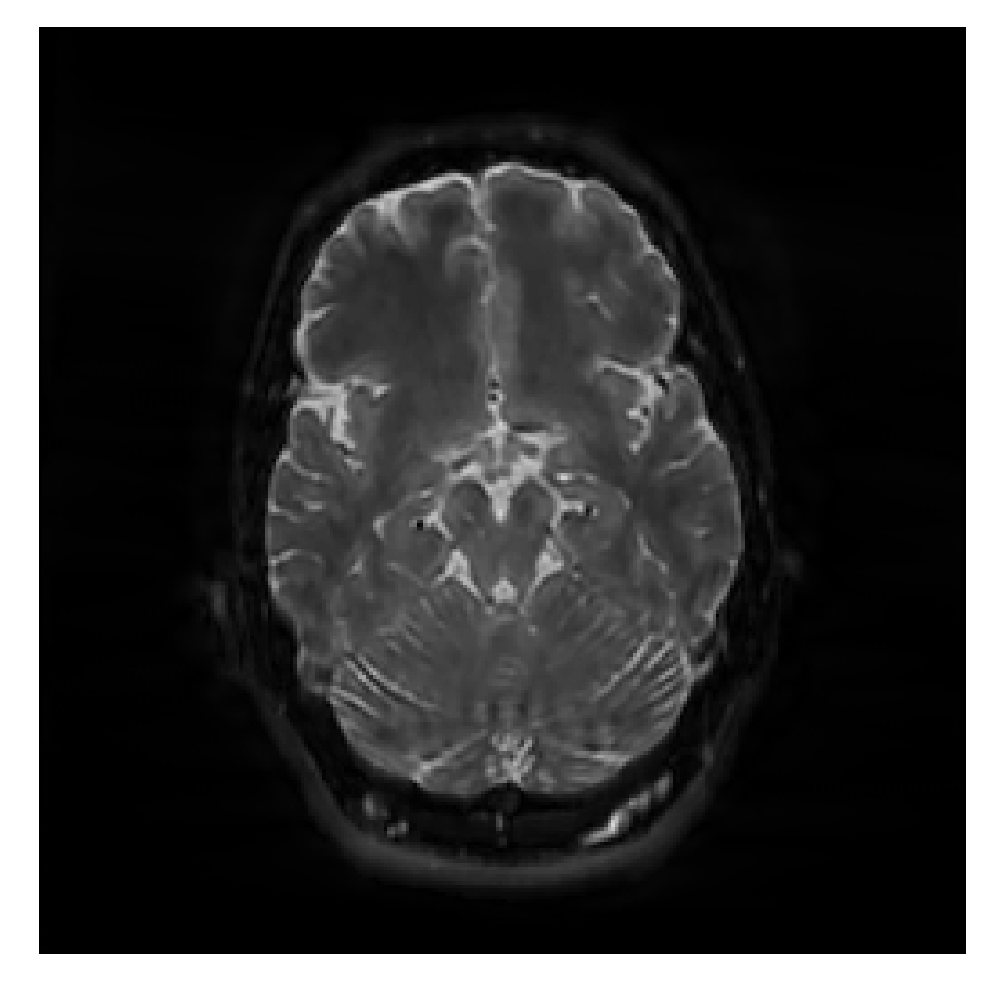}%
                \linethickness{2pt}
                \put(5,13){{\tiny\color{white}PSNR: 32.33}}
                \put(5,6){{\tiny\color{white}SSIM: 0.8144}}
            \end{overpic}
        \end{subfigure}%
        \begin{subfigure}[b]{0.24\textwidth}
            \includegraphics[width=\textwidth]{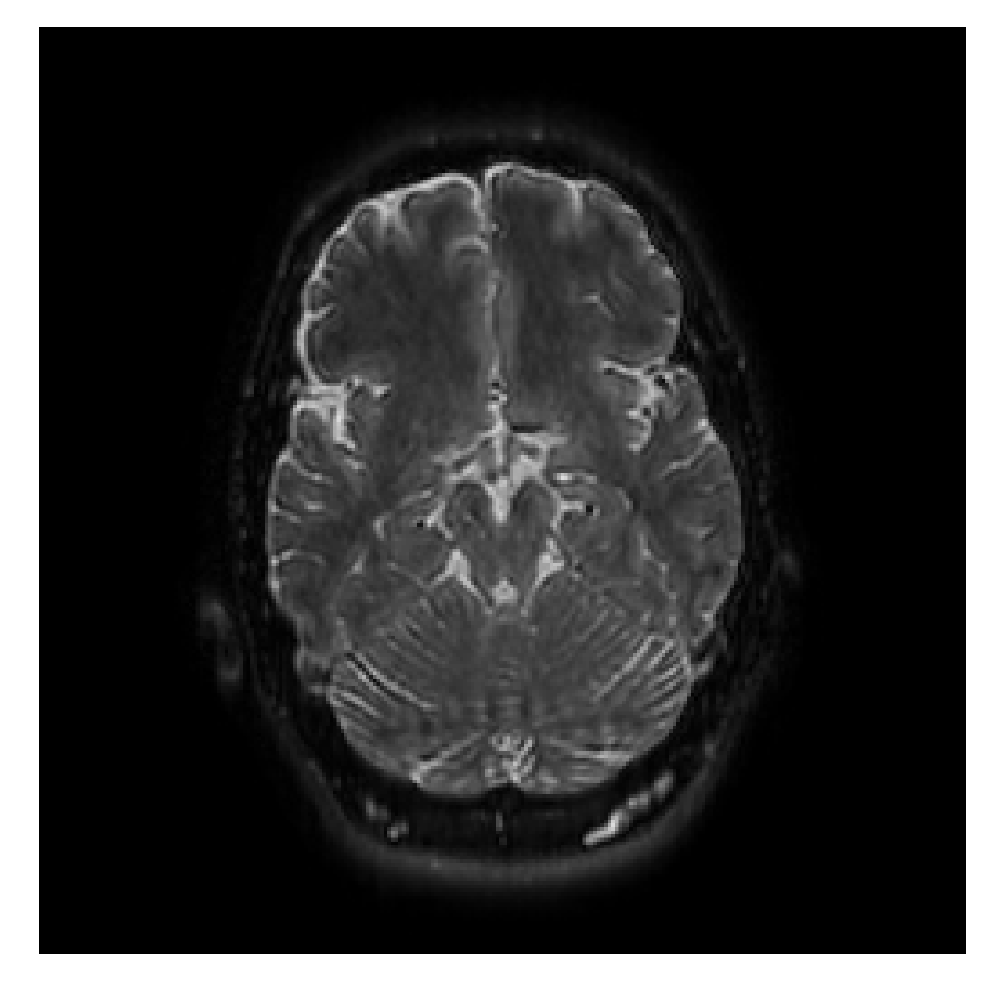}%
        \end{subfigure}%
        \begin{subfigure}[b]{0.24\textwidth}
            \includegraphics[width=\textwidth]{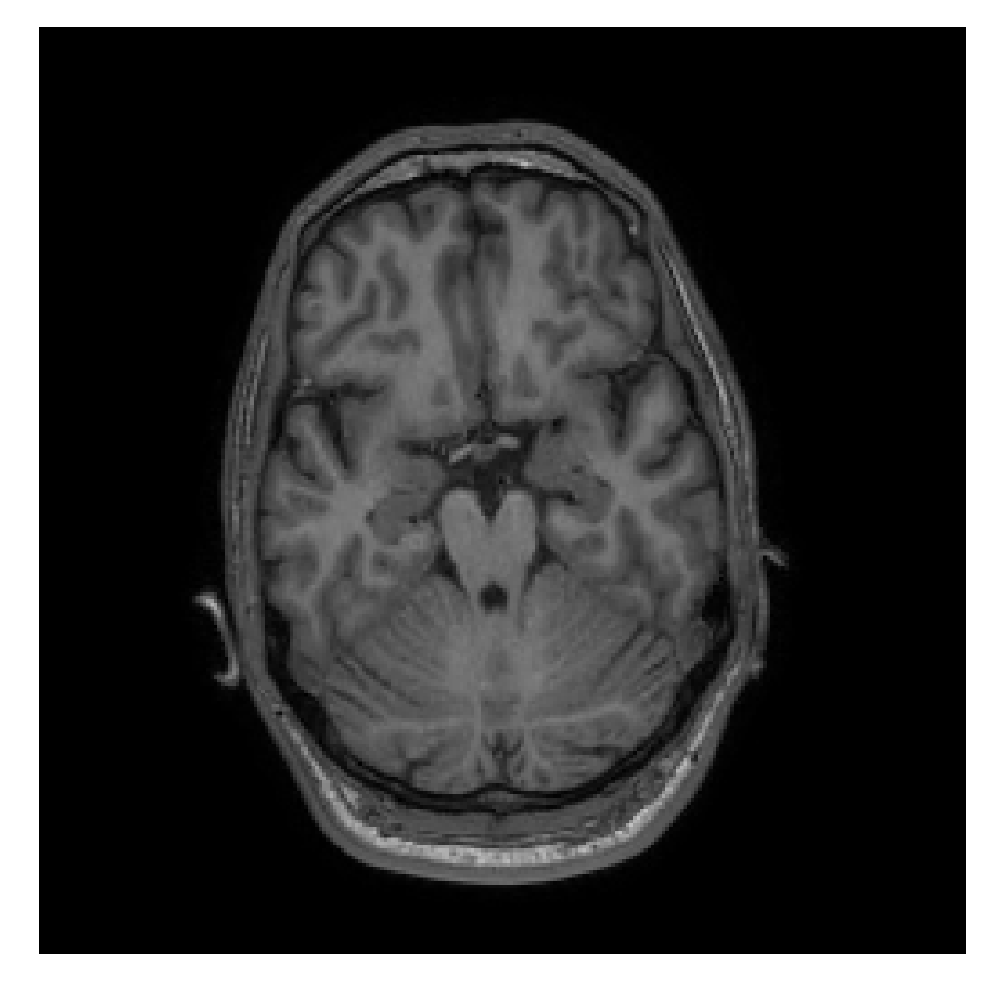}%
        \end{subfigure}%
    \end{subfigure}
    \begin{subfigure}{\textwidth}
        \centering
        \rotatebox{90}{\hspace{1.3em}Axial detail} %
        \begin{subfigure}[b]{0.24\textwidth}
            \includegraphics[width=\textwidth]{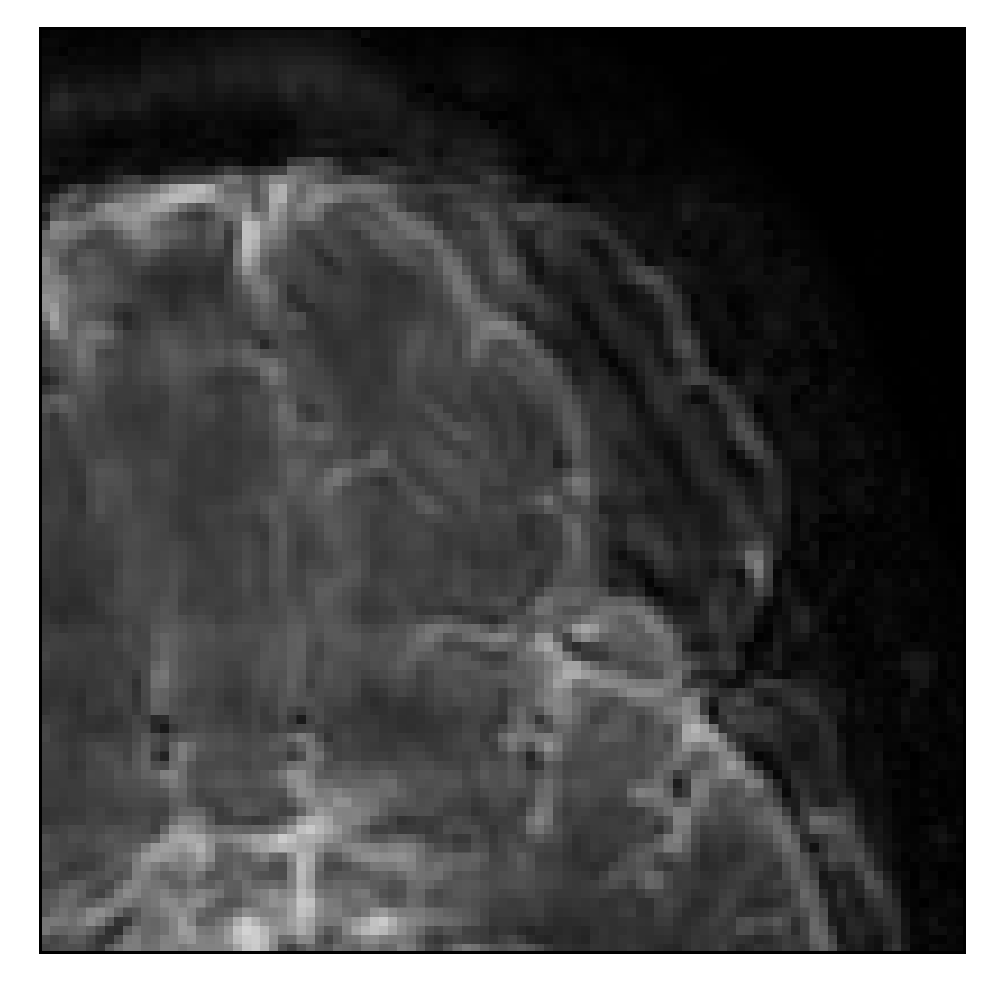}%
        \end{subfigure}%
        \begin{subfigure}[b]{0.24\textwidth}
            \includegraphics[width=\textwidth]{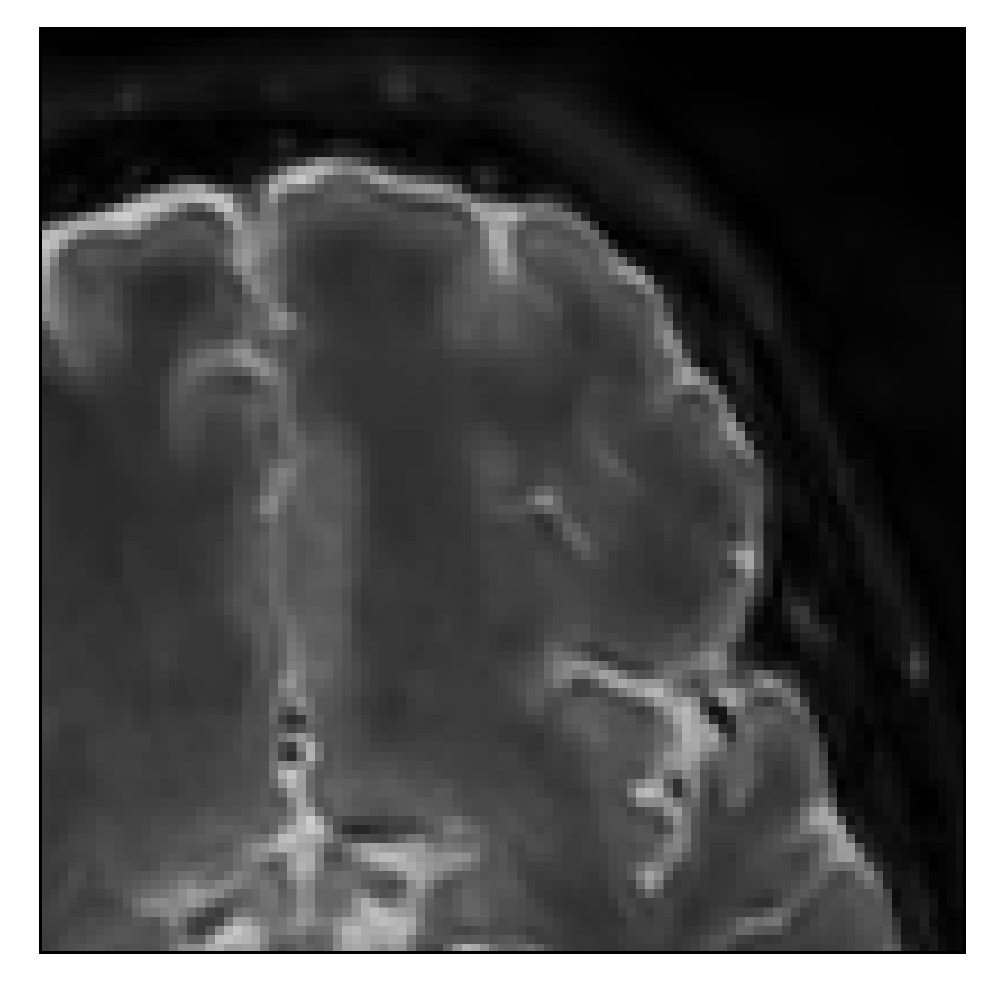}%
        \end{subfigure}%
        \begin{subfigure}[b]{0.24\textwidth}
            \includegraphics[width=\textwidth]{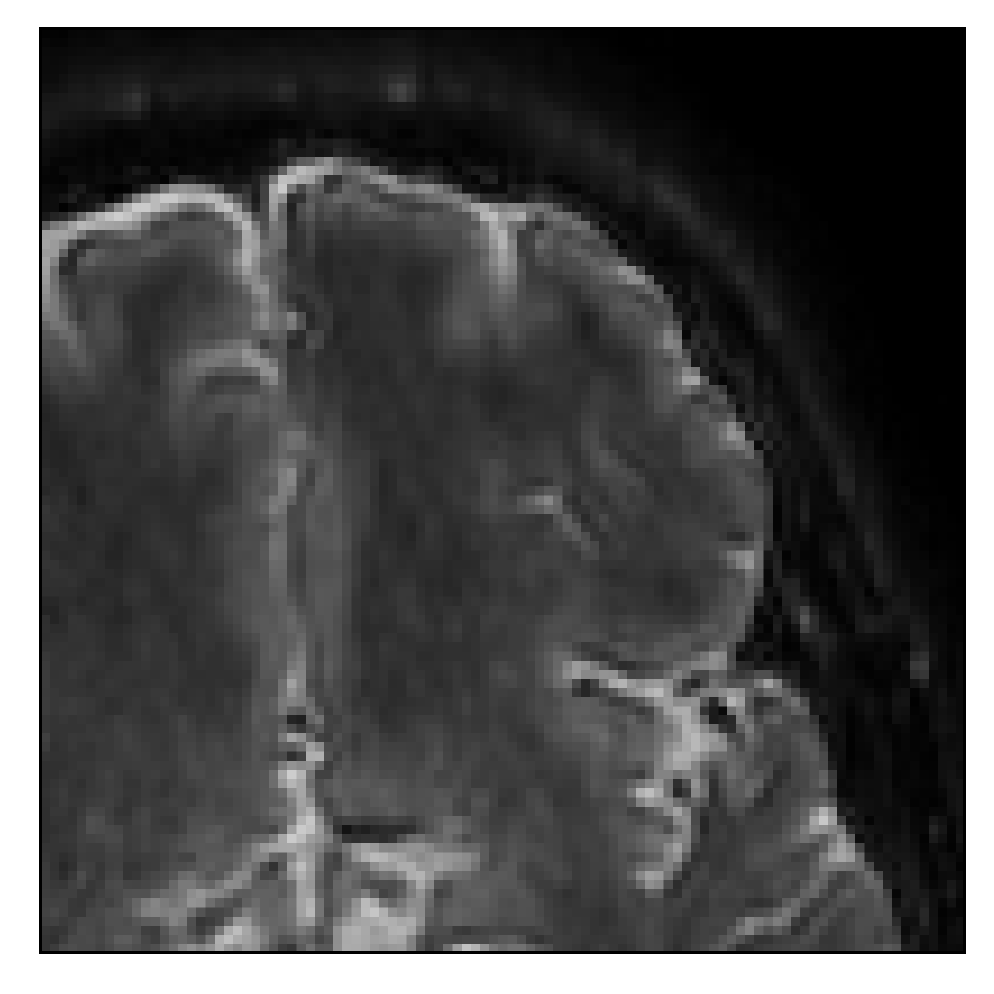}%
        \end{subfigure}%
        \begin{subfigure}[b]{0.24\textwidth}
            \includegraphics[width=\textwidth]{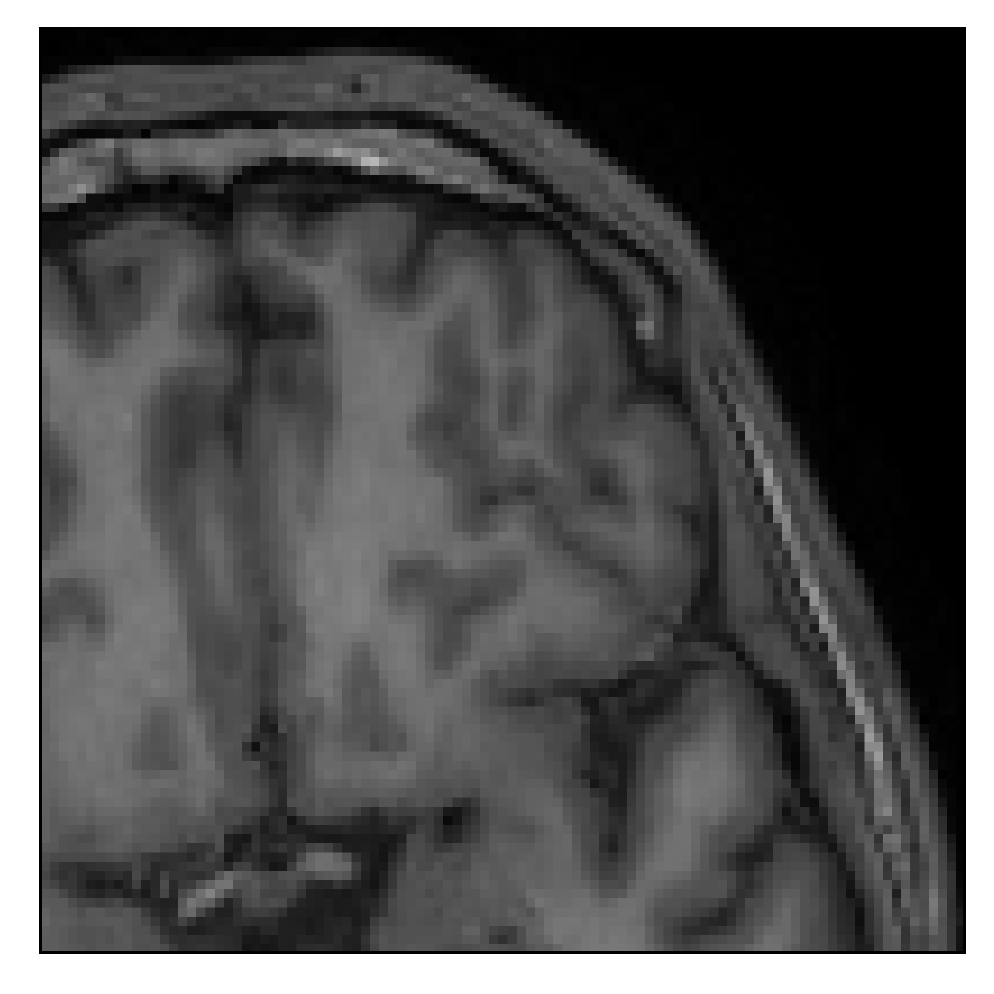}%
        \end{subfigure}%
    \end{subfigure}
    \caption{Reconstruction results for volunteer 3. The volunteer is instructed to move once, halfway through the scan (the two overlapping positions are clearly visible in the corrupted slices). The corrupted contrast is T2-weighted, while the reference contrast is T1-weighted. In this case, the input data for the correction algorithm is directly extracted from the scanner reconstruction in DICOM format (comprising both amplitude and phase). The acquisition scheme for the T2-weighted contrast follows a linear filling pattern in $k$-space. The proposed method successfully removes the motion artifacts because the scanner reconstruction is obtained through a conventional SENSE reconstruction (cf. Figure \ref{fig:scanreconvsrawdata2}).}\label{fig:scanreconvsrawdata1}
\end{figure}
\begin{figure}[!htb]
    \centering
    \begin{subfigure}{\textwidth}
        \centering
        \rotatebox{90}{\hspace{2.5em}Sagittal} %
        \begin{subfigure}[b]{0.24\textwidth}
            \caption*{Corrupted}\vspace{-0.5em}%
            \begin{overpic}[width=\textwidth]{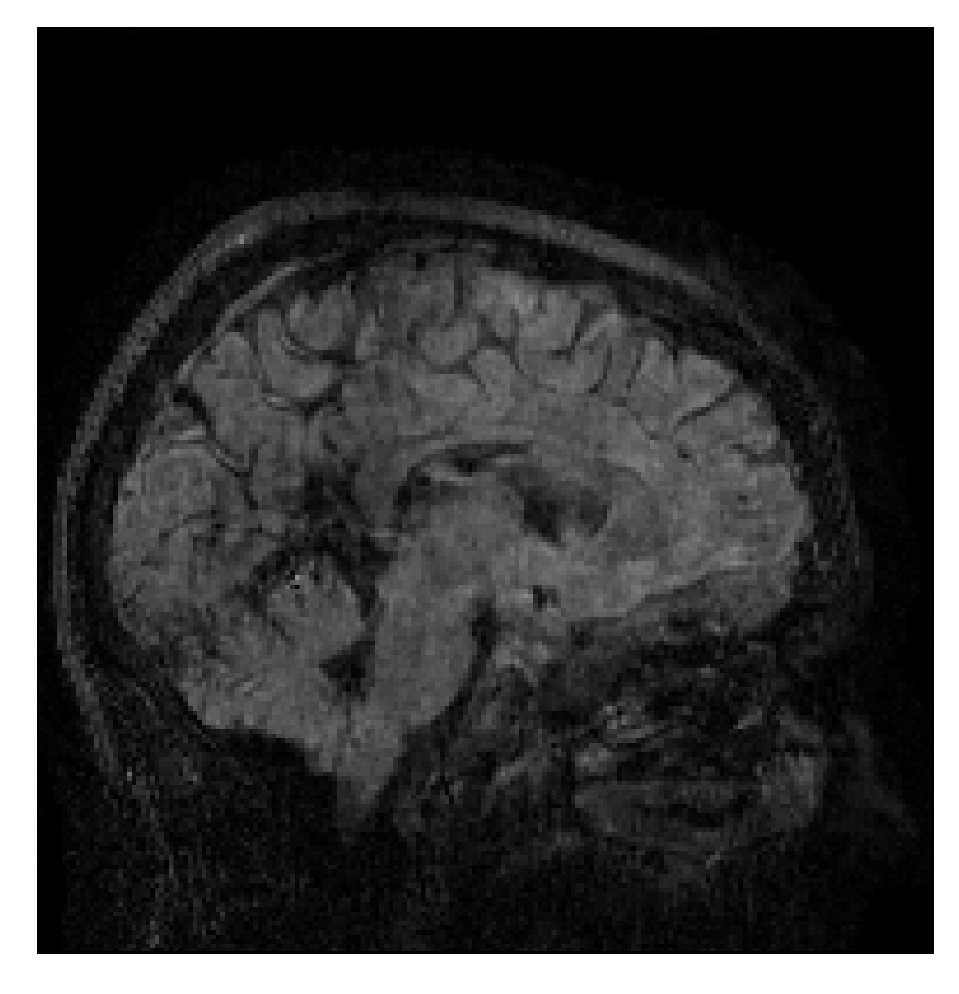}%
                \linethickness{2pt}
                \put(5,13){{\tiny\color{white}PSNR: 24.72}}
                \put(5,6){{\tiny\color{white}SSIM: 0.6762}}
            \end{overpic}%
        \end{subfigure}%
        \begin{subfigure}[b]{0.24\textwidth}
            \caption*{Corrected}\vspace{-0.5em}%
            \begin{overpic}[width=\textwidth]{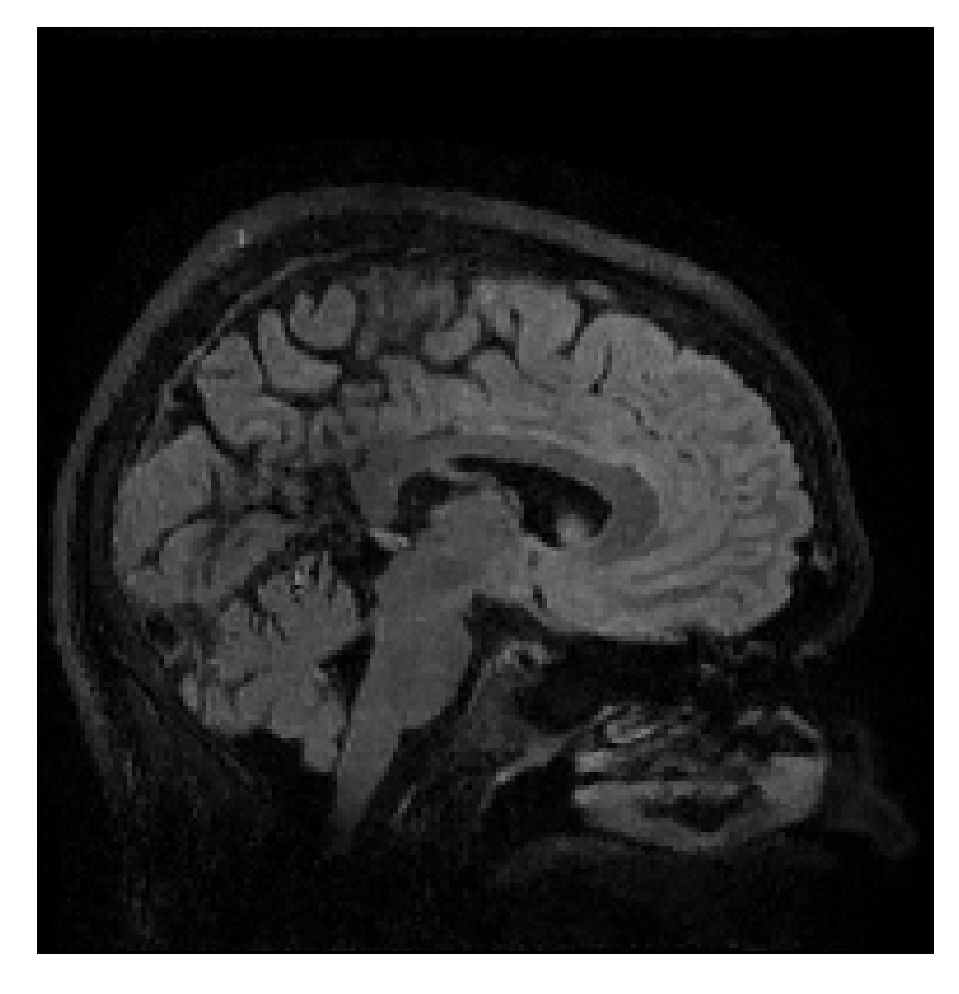}%
                \linethickness{2pt}
                \put(5,13){{\tiny\color{white}PSNR: 28.76}}
                \put(5,6){{\tiny\color{white}SSIM: 0.7818}}
            \end{overpic}%
        \end{subfigure}%
        \begin{subfigure}[b]{0.24\textwidth}
            \caption*{Ground truth}\vspace{-0.5em}%
            \includegraphics[width=\textwidth]{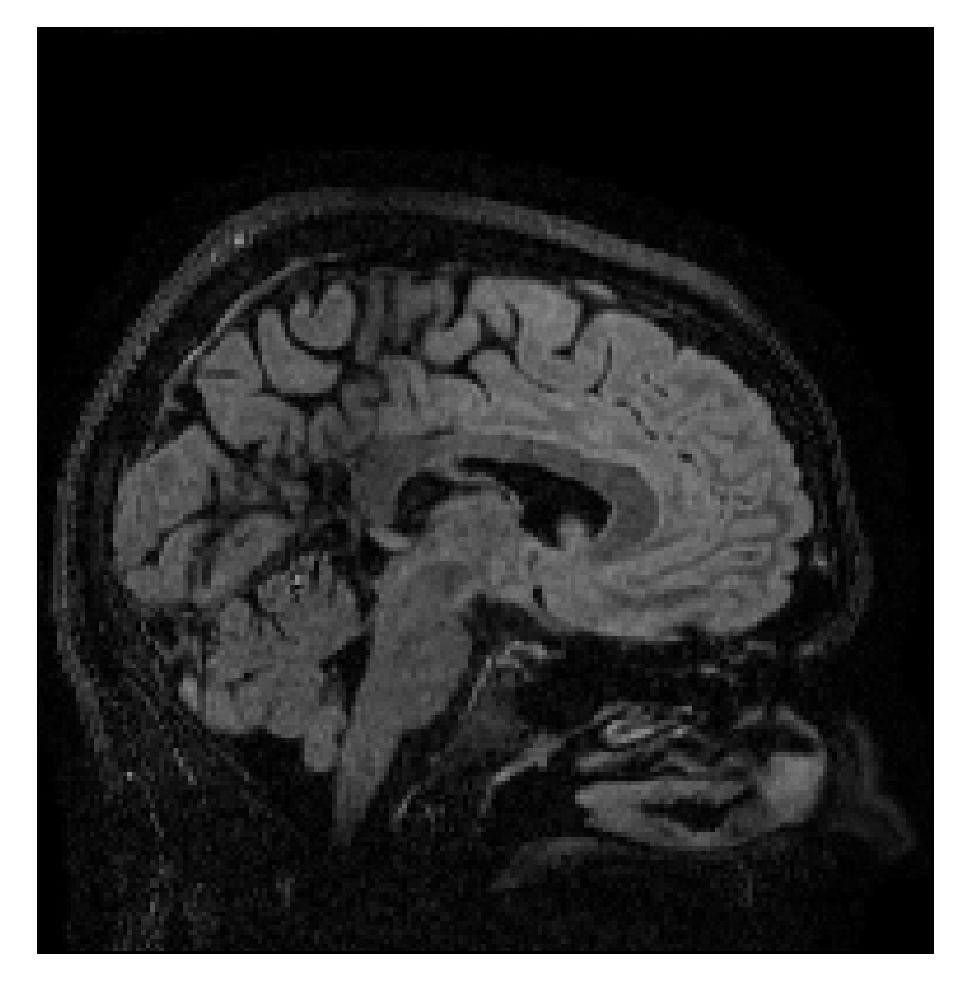}%
        \end{subfigure}%
        \begin{subfigure}[b]{0.24\textwidth}
            \caption*{Reference}\vspace{-0.5em}%
            \includegraphics[width=\textwidth]{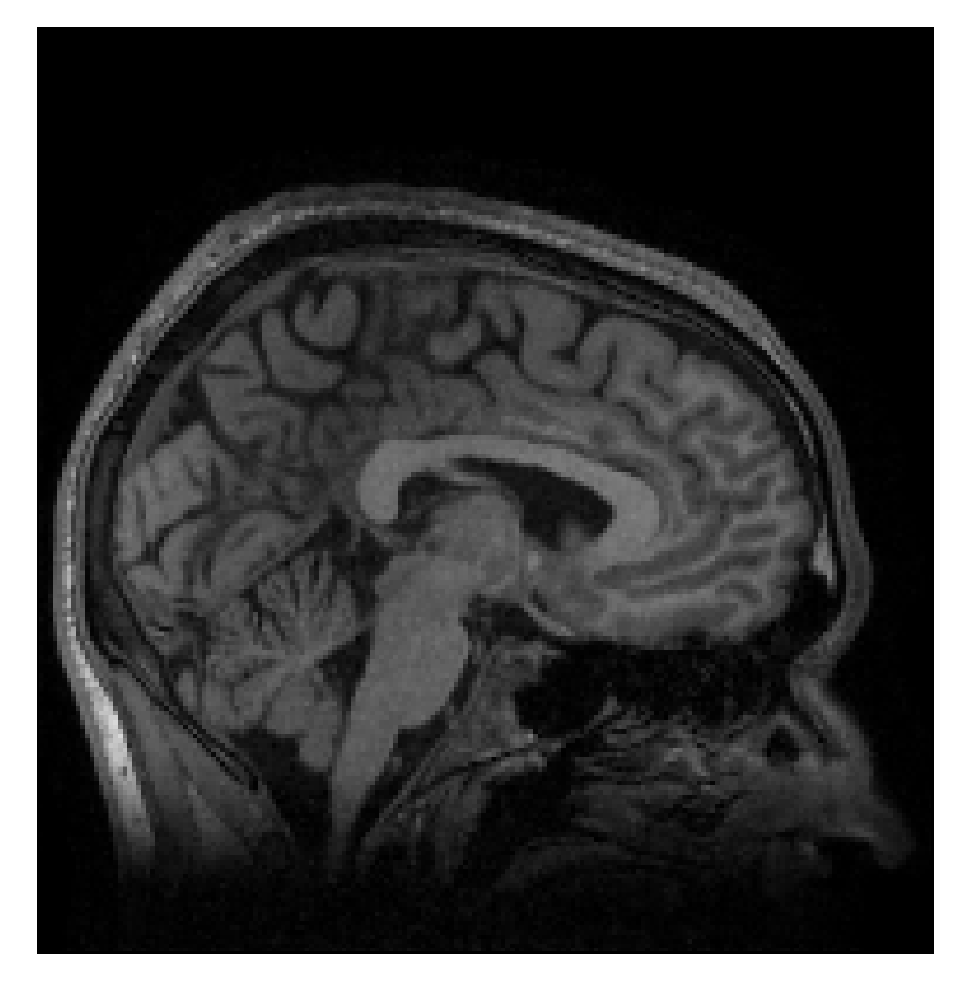}%
        \end{subfigure}%
    \end{subfigure}
    \begin{subfigure}{\textwidth}
        \centering
        \rotatebox{90}{\hspace{2.5em}Coronal} %
        \begin{subfigure}[b]{0.24\textwidth}
            \begin{overpic}[width=\textwidth]{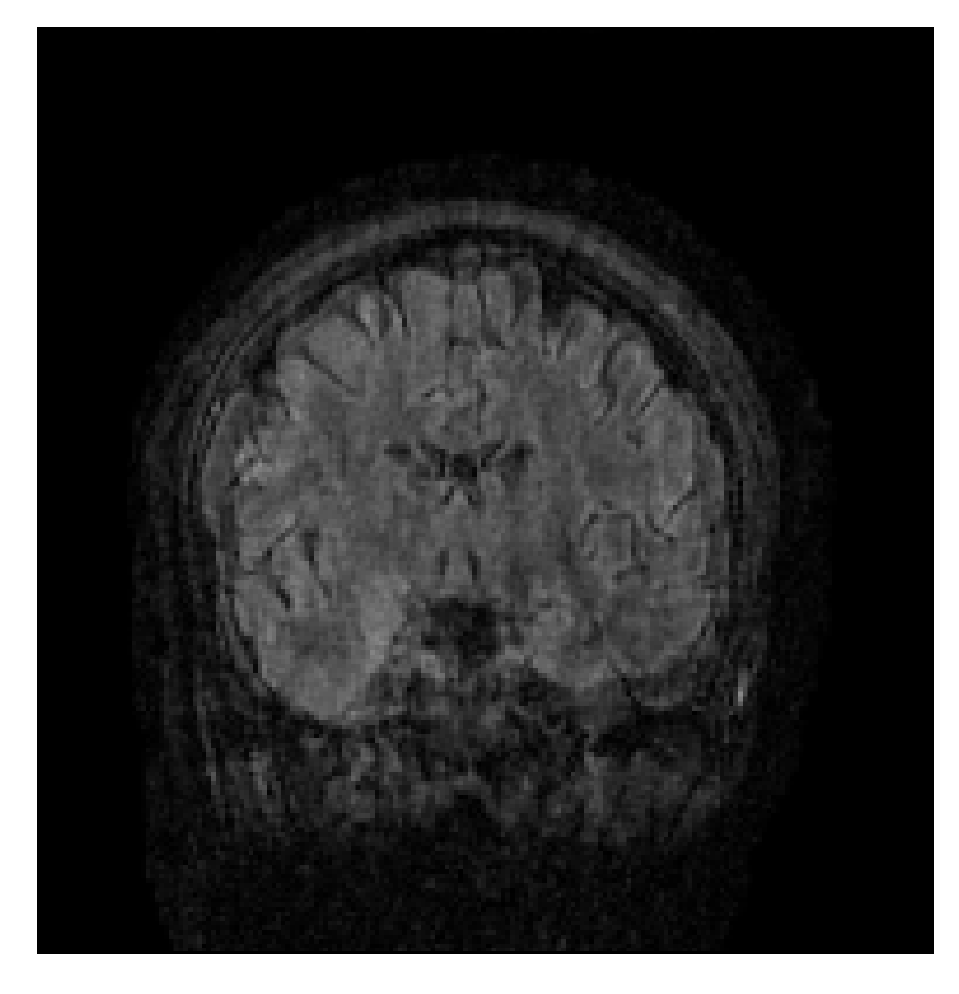}%
                \linethickness{2pt}
                \put(5,13){{\tiny\color{white}PSNR: 25.95}}
                \put(5,6){{\tiny\color{white}SSIM: 0.7238}}
            \end{overpic}%
        \end{subfigure}%
        \begin{subfigure}[b]{0.24\textwidth}
            \begin{overpic}[width=\textwidth]{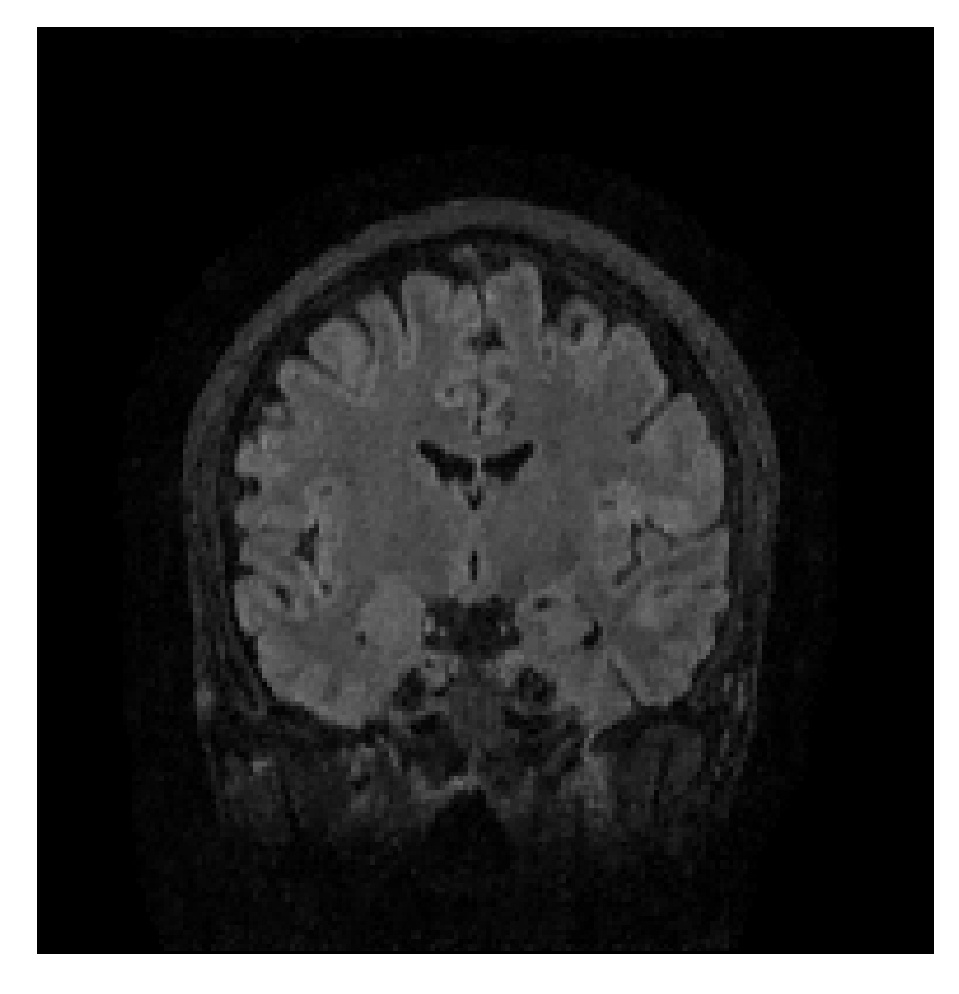}%
                \linethickness{2pt}
                \put(5,13){{\tiny\color{white}PSNR: 29.54}}
                \put(5,6){{\tiny\color{white}SSIM: 0.8107}}
            \end{overpic}%
        \end{subfigure}%
        \begin{subfigure}[b]{0.24\textwidth}
            \includegraphics[width=\textwidth]{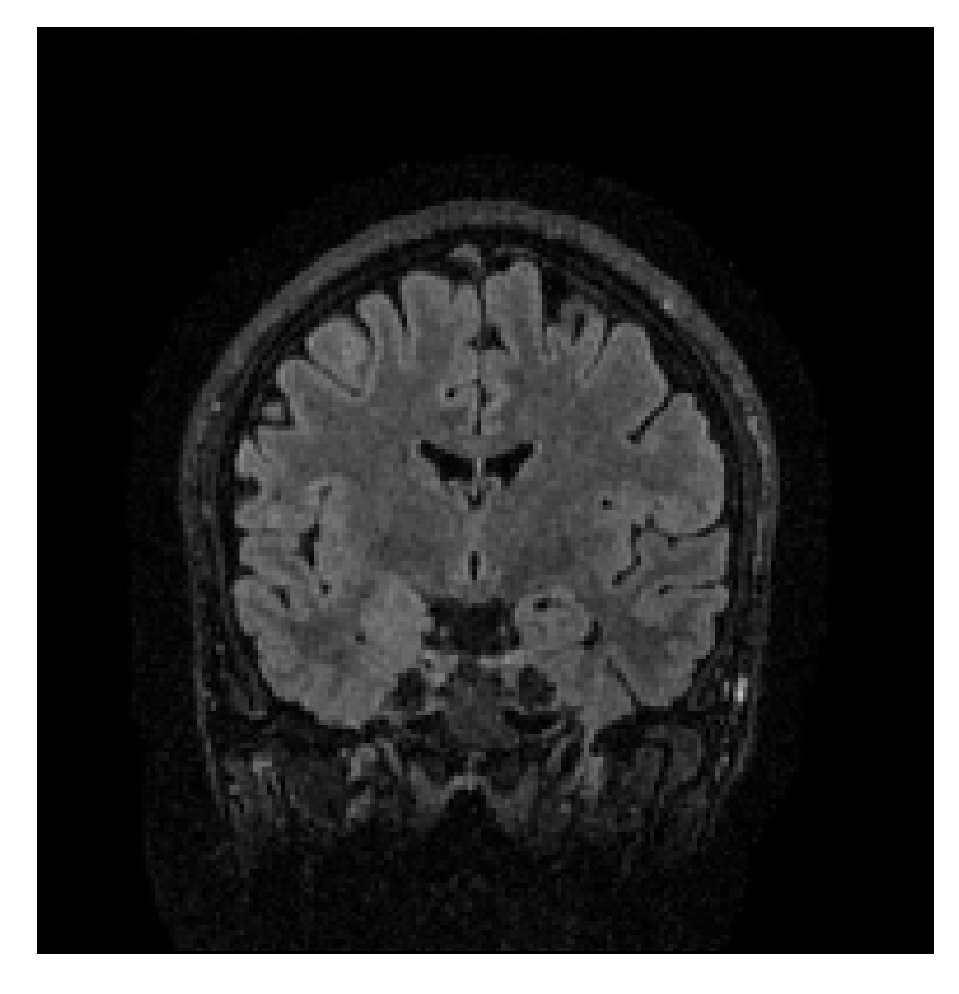}%
        \end{subfigure}%
        \begin{subfigure}[b]{0.24\textwidth}
            \includegraphics[width=\textwidth]{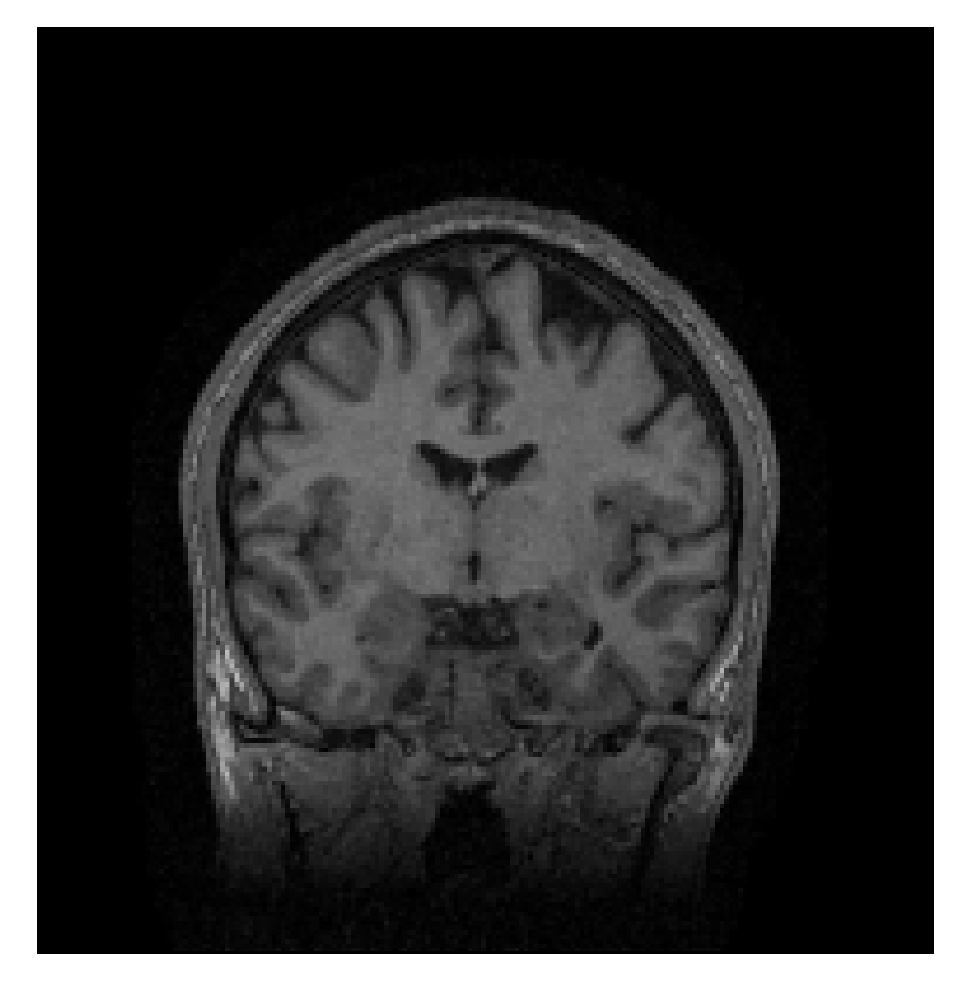}%
        \end{subfigure}%
    \end{subfigure}
    \begin{subfigure}{\textwidth}
        \centering
        \rotatebox{90}{\hspace{3em}Axial} %
        \begin{subfigure}[b]{0.24\textwidth}
            \begin{overpic}[width=\textwidth]{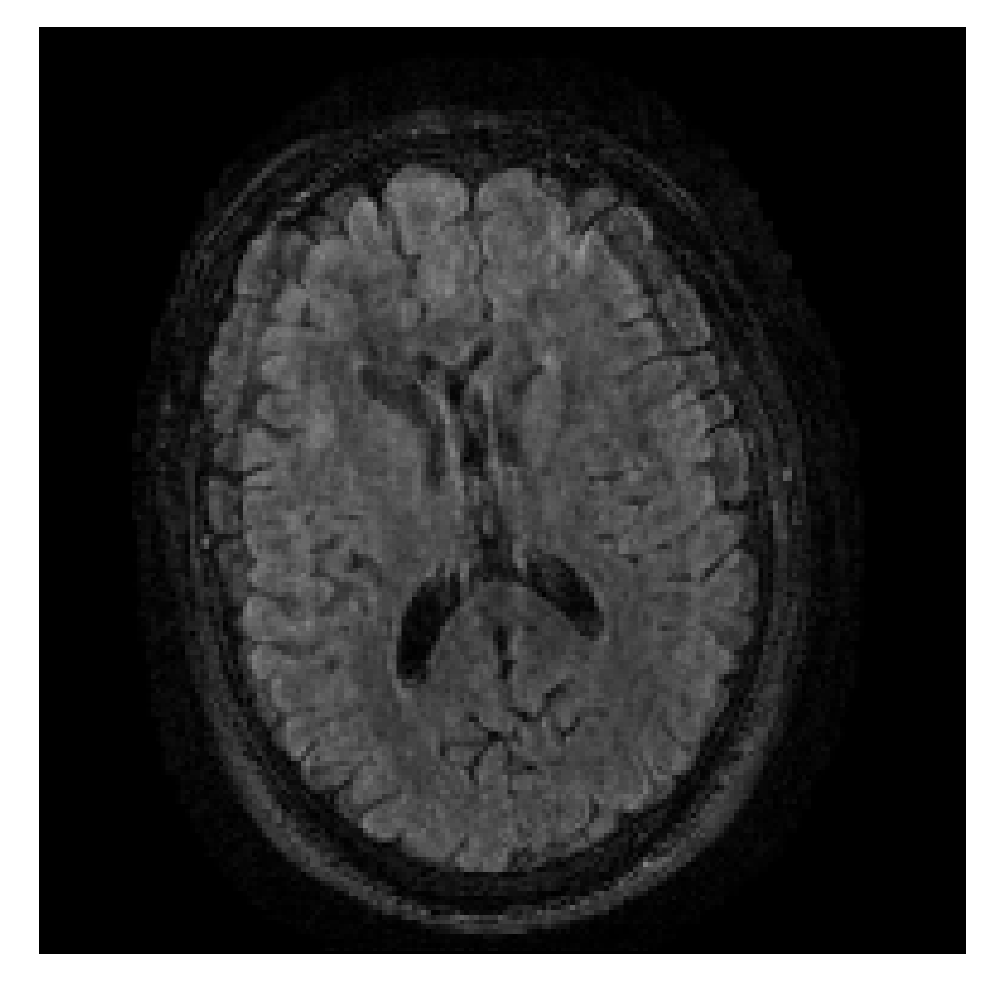}%
                \linethickness{2pt}
                \put(5,13){{\tiny\color{white}PSNR: 25.08}}
                \put(5,6){{\tiny\color{white}SSIM: 0.7263}}
            \end{overpic}%
        \end{subfigure}%
        \begin{subfigure}[b]{0.24\textwidth}
            \begin{overpic}[width=\textwidth]{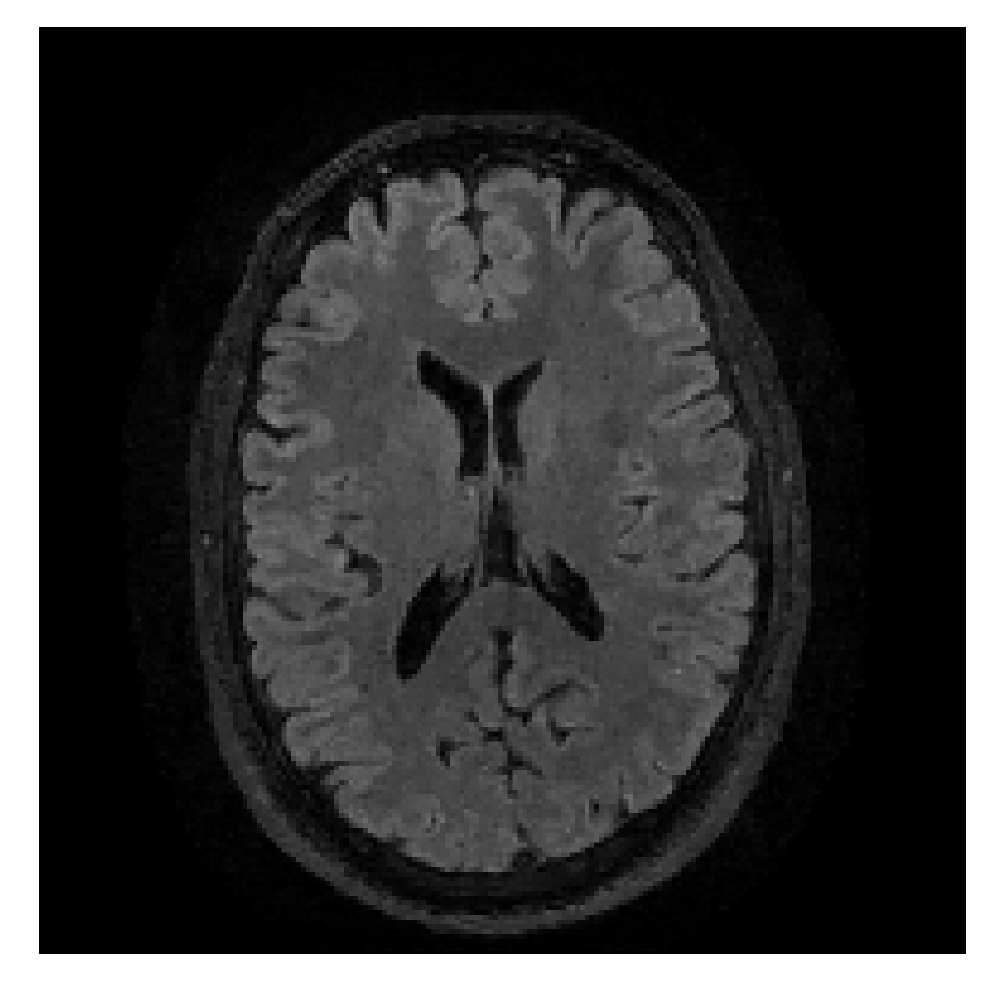}%
                \linethickness{2pt}
                \put(5,13){{\tiny\color{white}PSNR: 29.59}}
                \put(5,6){{\tiny\color{white}SSIM: 0.8407}}
            \end{overpic}
        \end{subfigure}%
        \begin{subfigure}[b]{0.24\textwidth}
            \includegraphics[width=\textwidth]{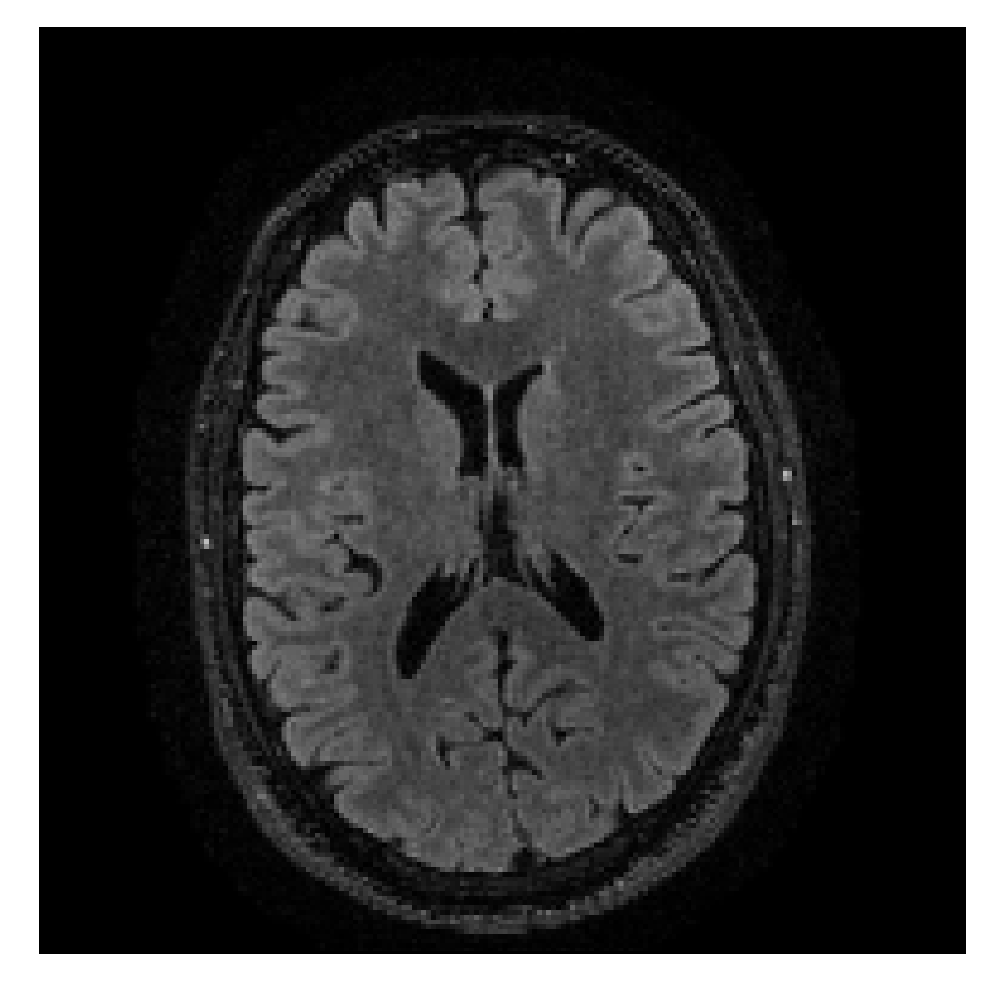}%
        \end{subfigure}%
        \begin{subfigure}[b]{0.24\textwidth}
            \includegraphics[width=\textwidth]{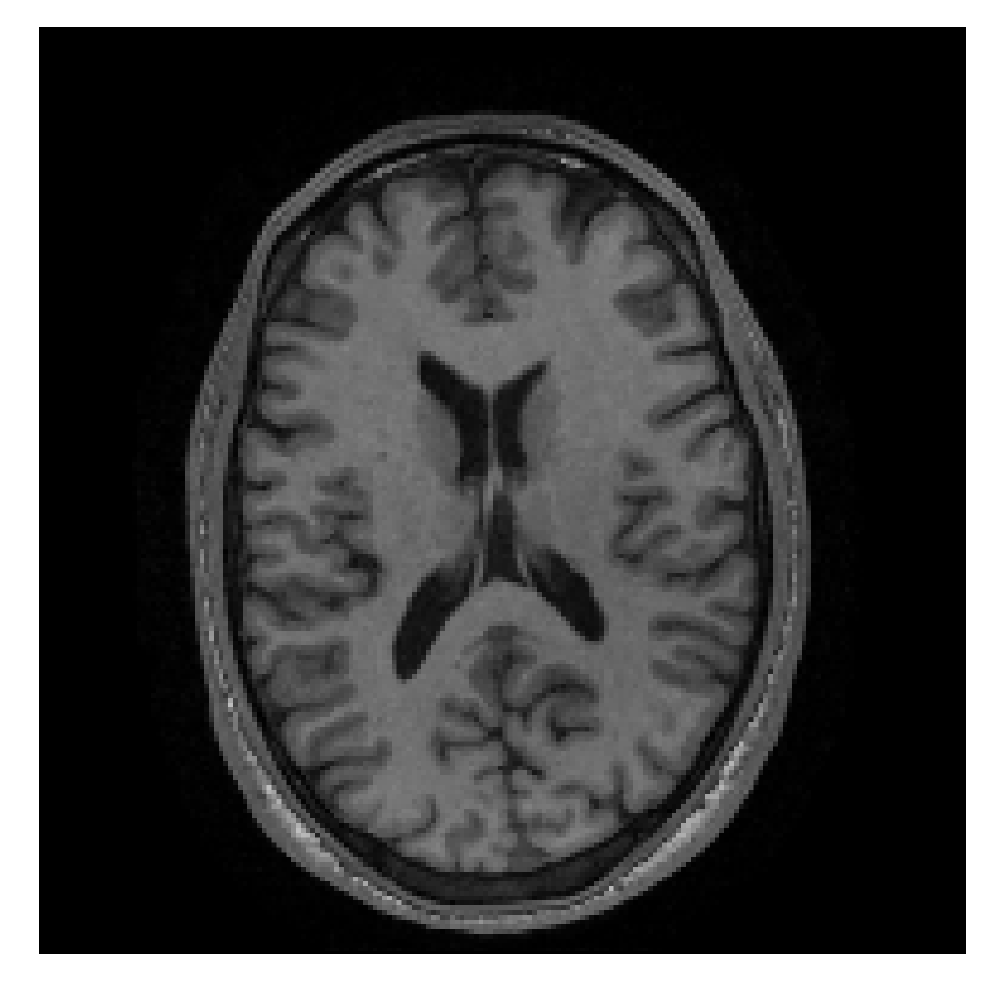}%
        \end{subfigure}%
    \end{subfigure}
    \begin{subfigure}{\textwidth}
        \centering
        \rotatebox{90}{\hspace{1.3em}Axial detail} %
        \begin{subfigure}[b]{0.24\textwidth}
            \includegraphics[width=\textwidth]{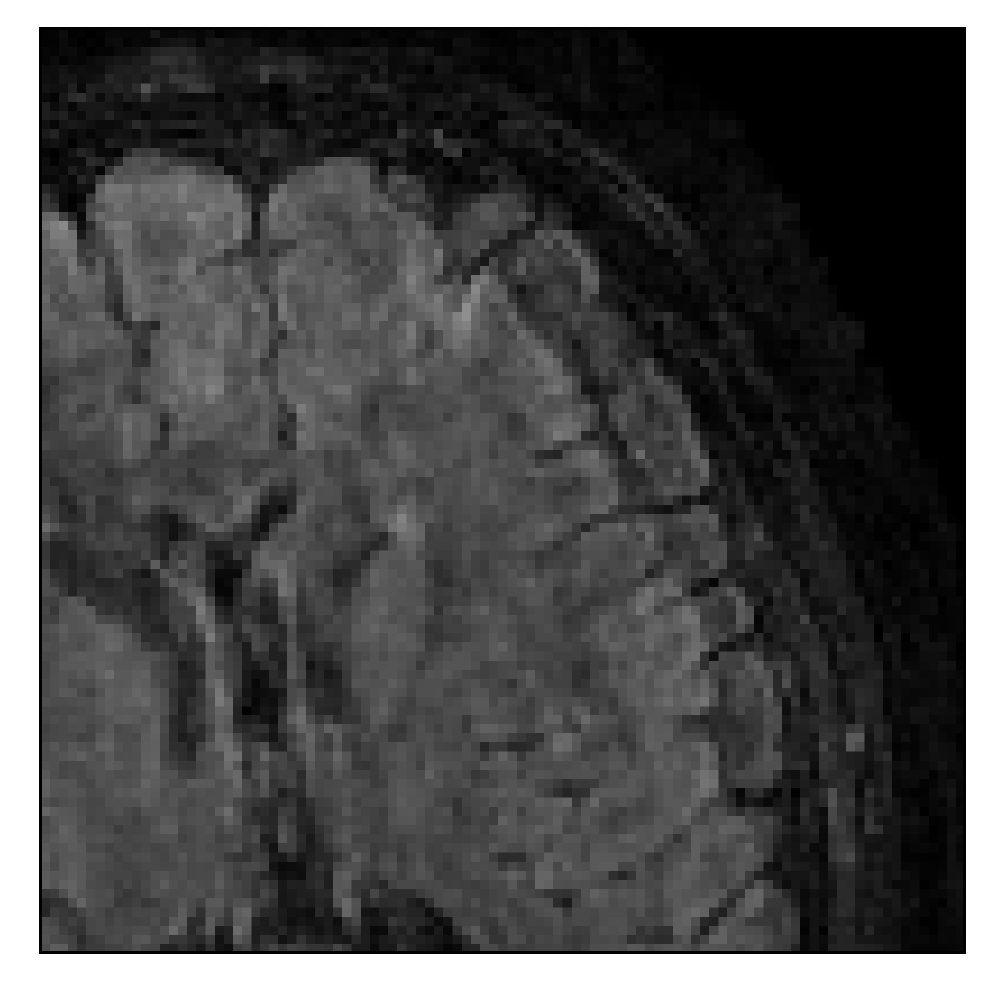}%
        \end{subfigure}%
        \begin{subfigure}[b]{0.24\textwidth}
            \includegraphics[width=\textwidth]{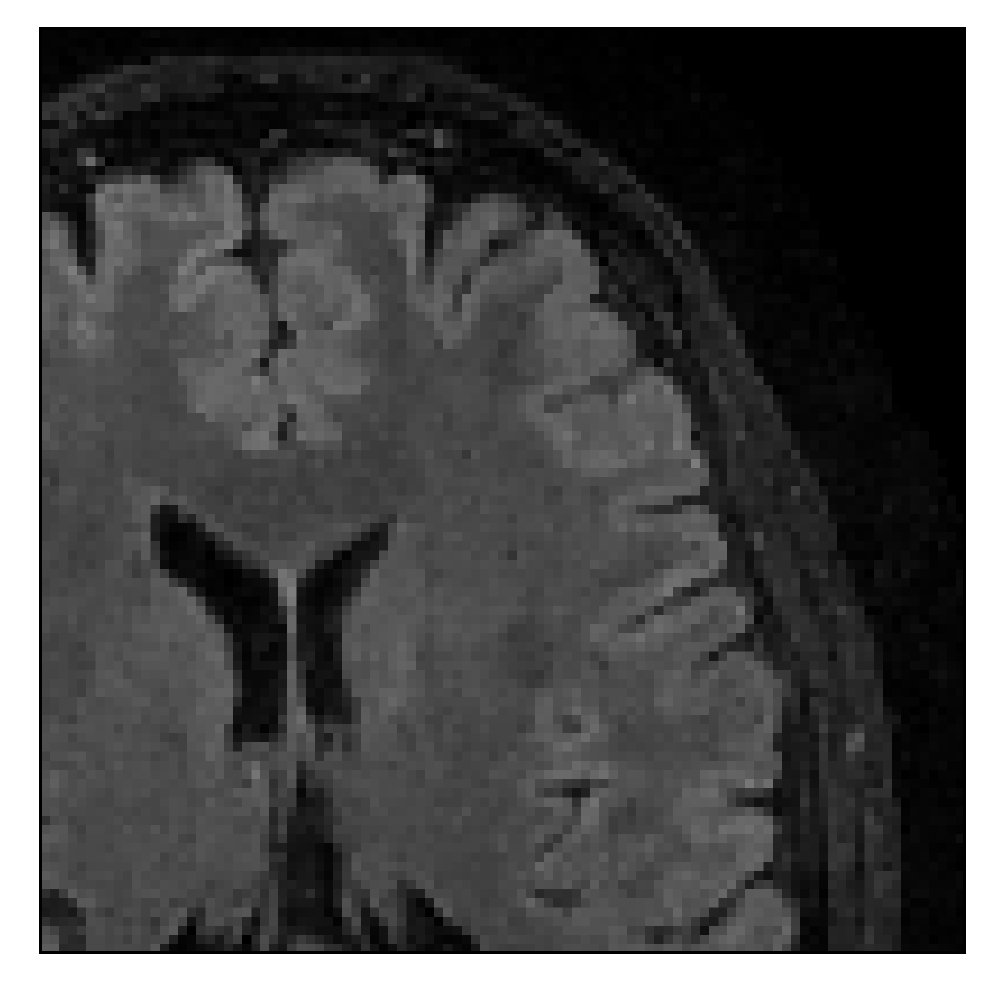}%
        \end{subfigure}%
        \begin{subfigure}[b]{0.24\textwidth}
            \includegraphics[width=\textwidth]{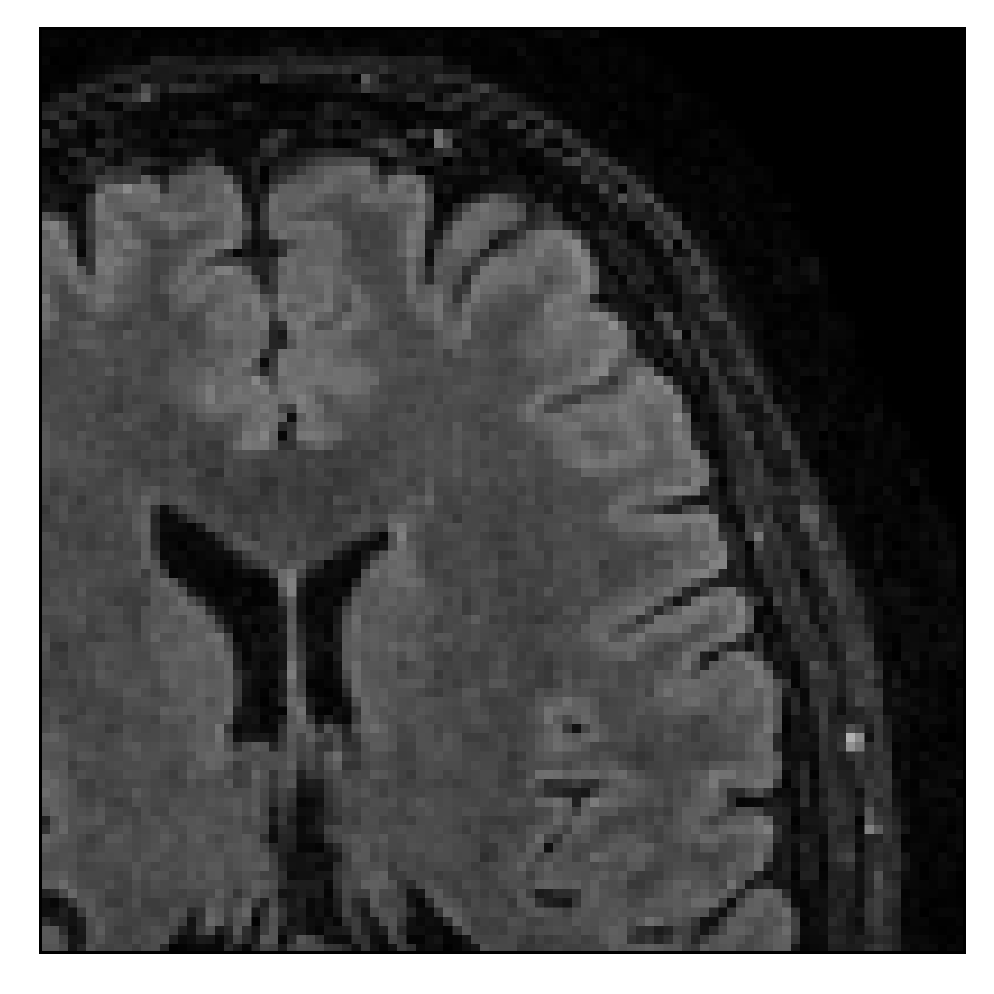}%
        \end{subfigure}%
        \begin{subfigure}[b]{0.24\textwidth}
            \includegraphics[width=\textwidth]{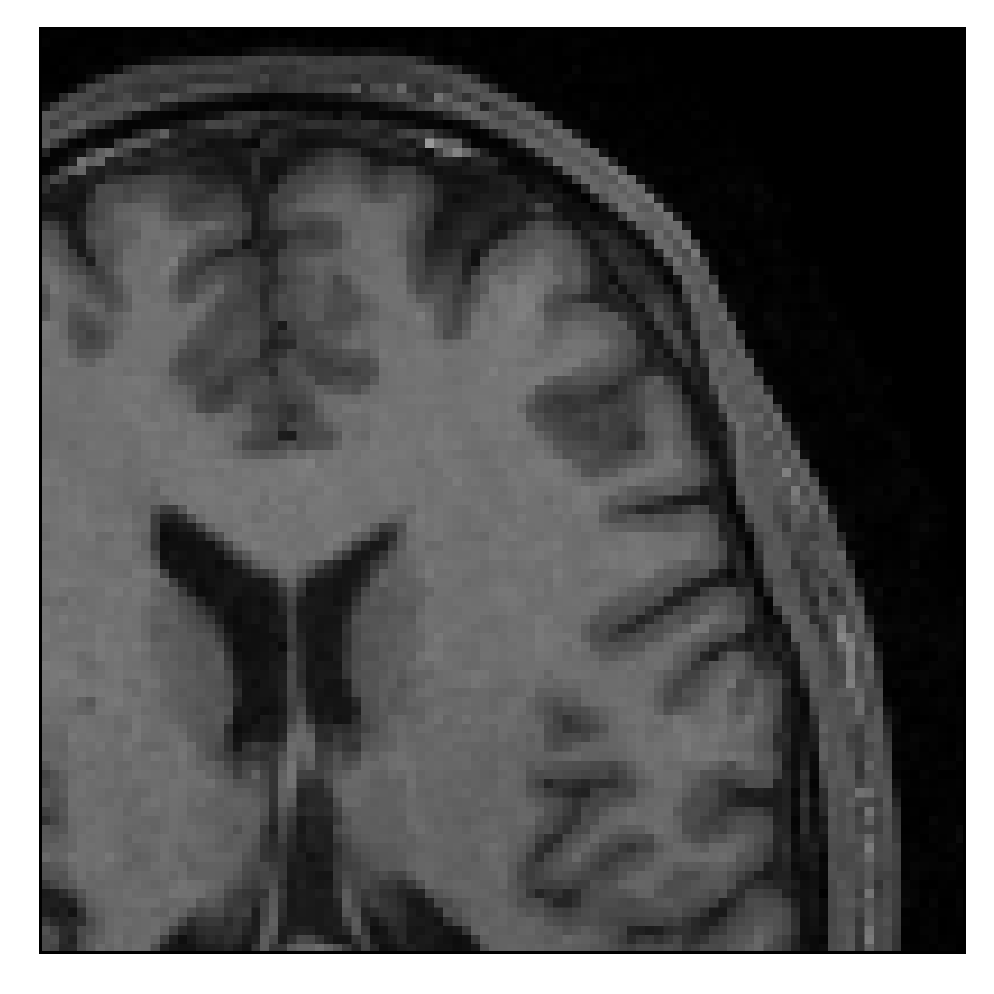}%
        \end{subfigure}%
    \end{subfigure}
    \caption{Reconstruction results for volunteer 3. The volunteer is instructed to move once, halfway through the scan (the two overlapping positions are clearly visible in the corrupted slices). The corrupted contrast is T2-FLAIR-weighted, while the reference contrast is T1-weighted. Unlike Figure \ref{fig:scanreconvsrawdata1}, the input data for the correction algorithm is not extracted from the scanner, but is obtained via SENSE reconstruction of the raw $k$-space data. The acquisition scheme for the T2-FLAIR-weighted contrast is randomized in $k$-space. When the scanner reconstruction is directly used as input data, the motion correction is highly suboptimal (cf. Figure \ref{fig:scanreconvsrawdata3} in Appendix \ref{app:scanreconvsrawdata3}).}\label{fig:scanreconvsrawdata2}
\end{figure}

\newpage
\phantom{.}
\newpage
\phantom{.}
\newpage
\phantom{.}
\newpage
\phantom{.}
\newpage
\phantom{.}
\newpage
\phantom{.}
\newpage
\phantom{.}
\newpage
\bibliography{rizzuti_2022_3dmc}

\newpage
\appendix

\section{Inadequate motion correction with scanner reconstruction as input data}\label{app:scanreconvsrawdata3}

\begin{figure}[!htb]
    \centering
    \begin{subfigure}{\textwidth}
        \centering
        \rotatebox{90}{\hspace{3em}Axial} %
        \begin{subfigure}[b]{0.24\textwidth}
            \caption*{Corrupted}\vspace{-0.5em}%
            \includegraphics[width=\textwidth]{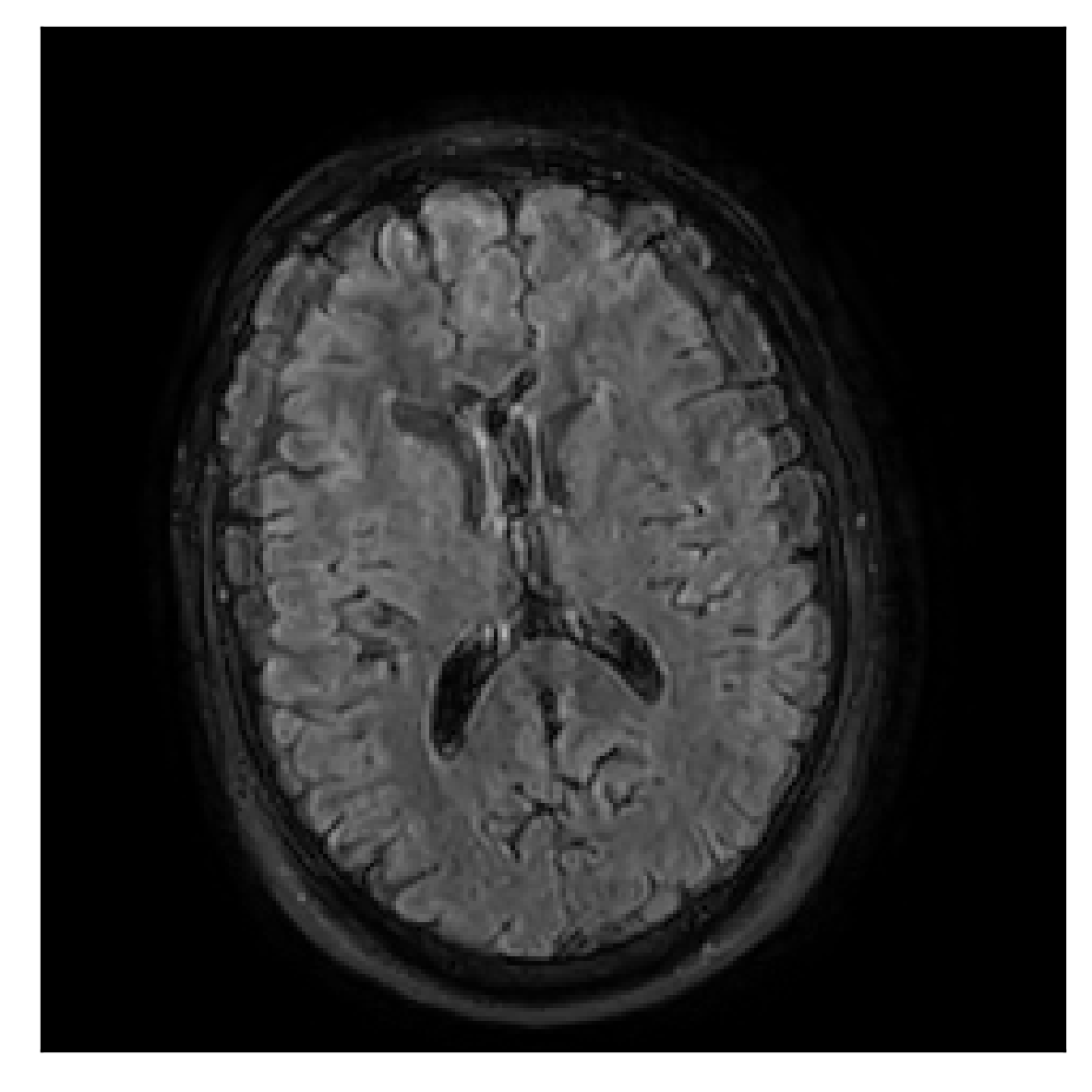}%
        \end{subfigure}%
        \begin{subfigure}[b]{0.24\textwidth}
            \caption*{Corrected}\vspace{-0.5em}%
            \includegraphics[width=\textwidth]{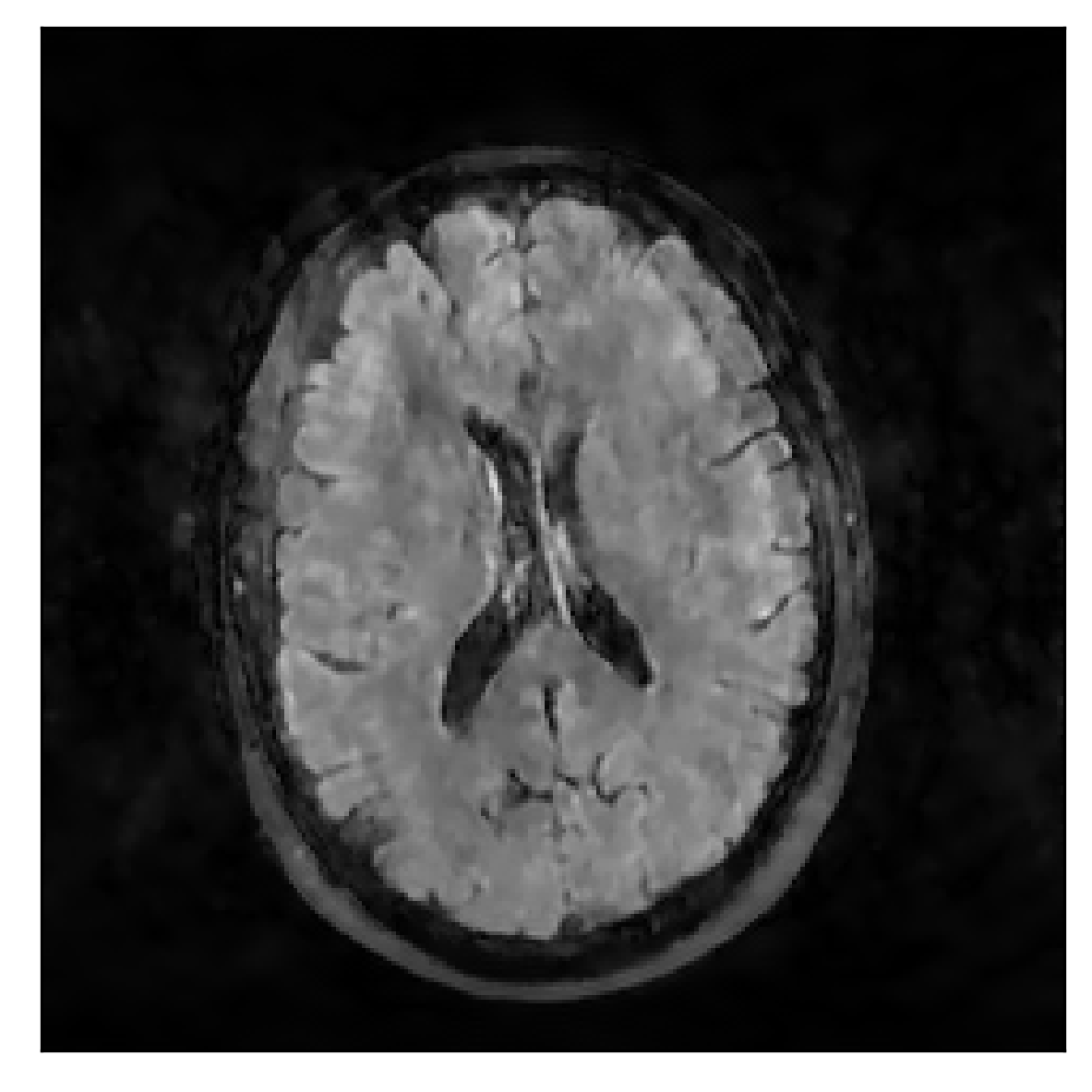}%
        \end{subfigure}%
        \begin{subfigure}[b]{0.24\textwidth}
            \caption*{Ground truth}\vspace{-0.5em}%
            \includegraphics[width=\textwidth]{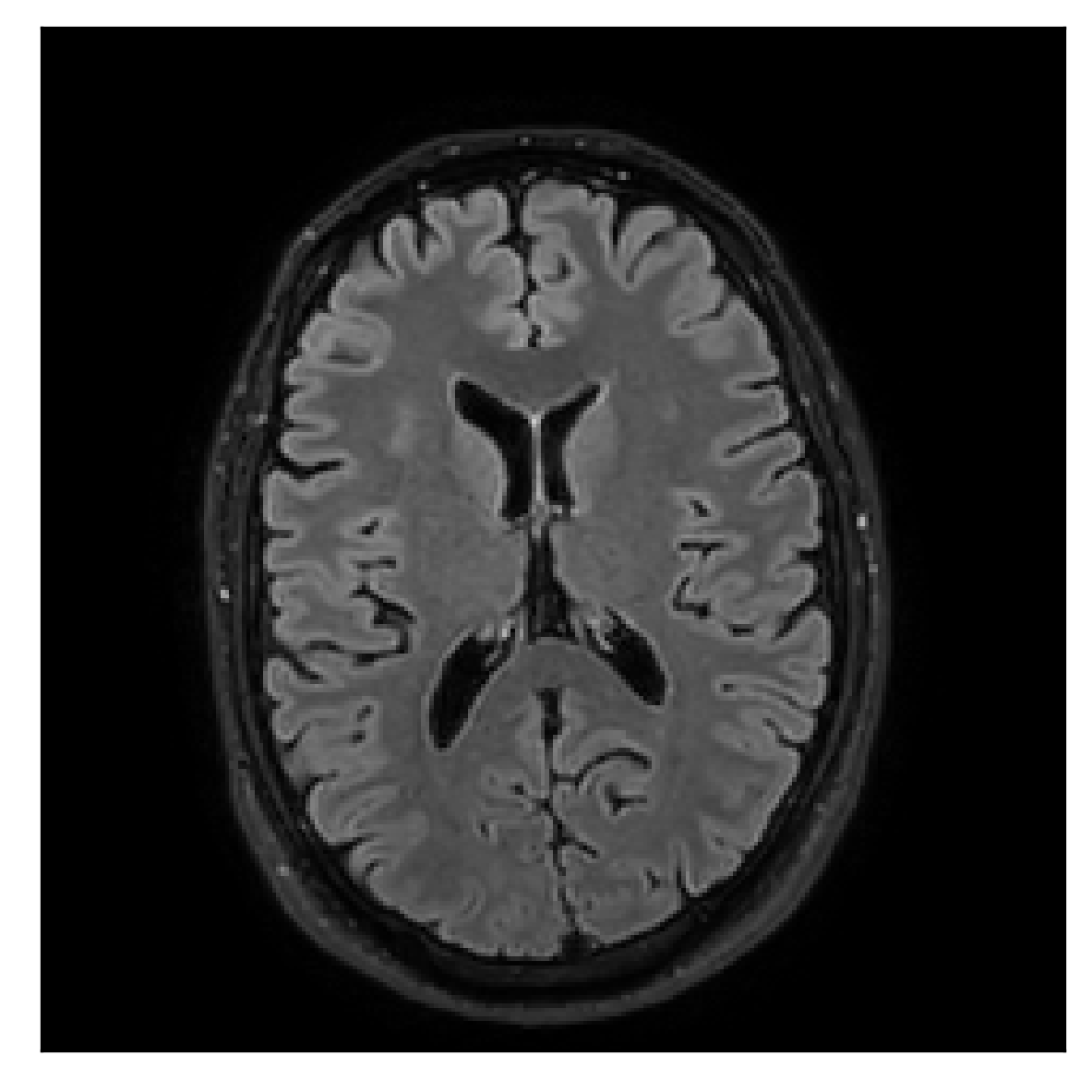}%
        \end{subfigure}%
        \begin{subfigure}[b]{0.24\textwidth}
            \caption*{Reference}\vspace{-0.5em}%
            \includegraphics[width=\textwidth]{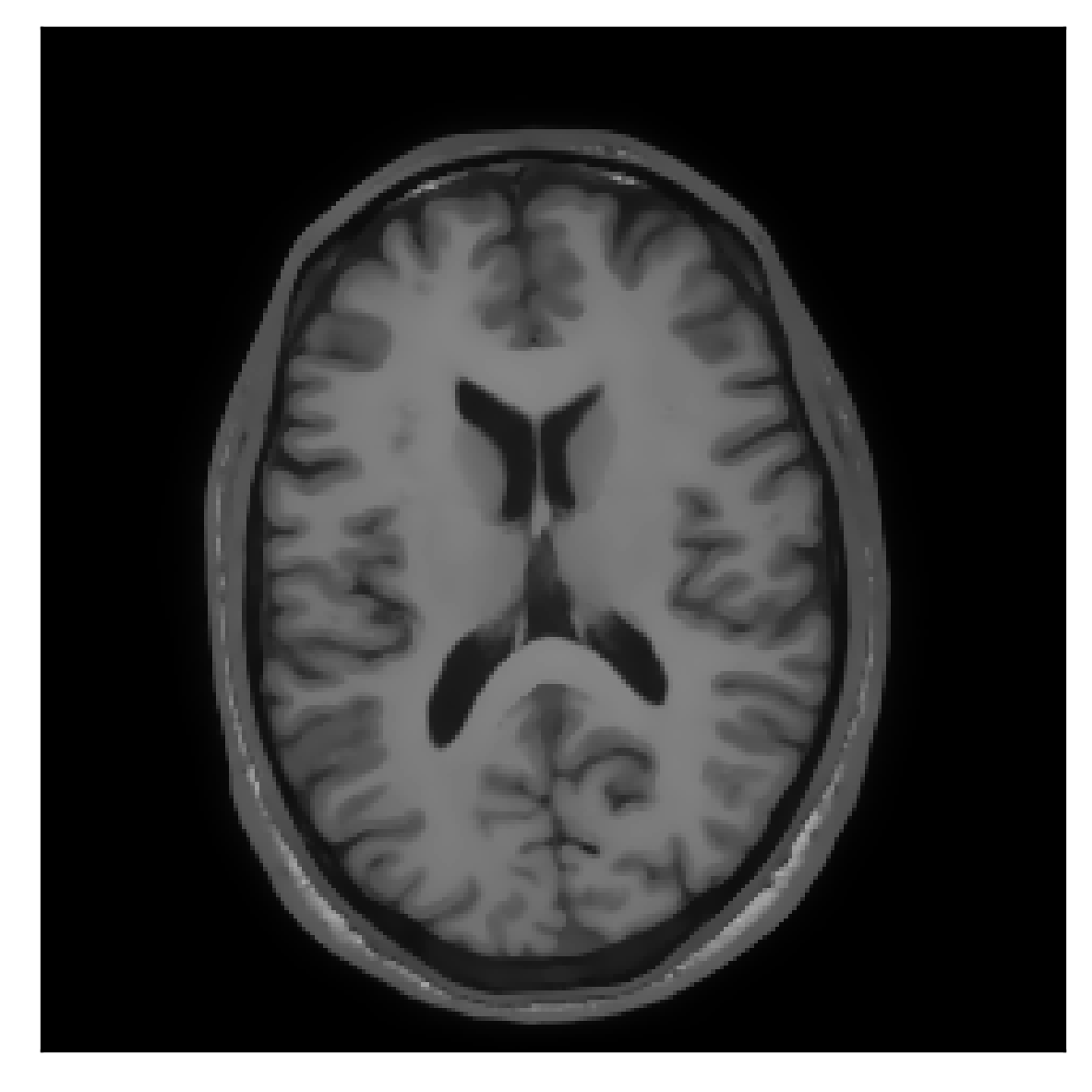}%
        \end{subfigure}%
    \end{subfigure}
    \caption{Reconstruction results for volunteer 3. The volunteer is instructed to move once, halfway through the scan. The corrupted contrast is T2-FLAIR-weighted, while the reference contrast is T1-weighted. In this experiment, the proposed motion correction scheme processes the scanner reconstruction directly. Since the reconstruction algorithm implemented in the scanner destroys the coherence of the rigid motion artifact, the proposed method cannot properly recover the correct reconstruction by simply estimating the motion parameters. With contrasts obtained by randomized acquisitions, we advise to use raw $k$-space data instead (cf. Figure \ref{fig:scanreconvsrawdata2}).}\label{fig:scanreconvsrawdata3}
\end{figure}

As anticipated in Section \ref{exp:data}, directly using the scanner reconstruction (extracted as DICOM files of both the amplitude and phase of the reconstruction) as input data for the proposed motion correction scheme may degrade the performance when compressed-sensing reconstruction tools have been employed in the reconstruction process. To motivate this conclusion, we setup an experiment with the same setting as described in the second experiment in Section \ref{exp:data}, the only difference being in how the input data is generated. In this case, the input data consist of the Fourier transform of the extracted scanner reconstruction. The related suboptimal correction is quite evident when comparing Figure \ref{fig:scanreconvsrawdata3} with Figure \ref{fig:scanreconvsrawdata2}.

\section{Comparison of motion correction with and without a reference guide}\label{app:baseline}

The reference-guided motion-correction algorithm described in Section \ref{sec:theory} is compared with a standard retrospective motion-correction algorithm based on the TV regularization.

We note that most retrospective motion-correction methods follows the basic mathematical framework detailed in Section \ref{sec:theory} (see, for example, \citet{Loktyushin2013} or \citet{Cordero-Grande2020}), where the main mathematical difference consists in the choice of the regularization term $g_u$, in equation \eqref{eq:opt}. Hence, in order to assess the effect of the reference contrast, we adopt the same formulation described in Section \ref{sec:theory} with a simple TV regularization term $g_u(\bu)=\sum_{\bx}\norm{\nabla\bu\rvert_{\bx}}$ (cf. equation \ref{eq:wtv} for the reference-guided version of TV).

\begin{figure}[!htb]
    \centering
    \begin{subfigure}{\textwidth}
        \centering
        \rotatebox{90}{\hspace{2em}Move once} %
        \begin{subfigure}[b]{0.24\textwidth}
            \caption*{Corrupted}\vspace{-0.5em}%
            \begin{overpic}[width=\textwidth]{./figs/robustness/vol1/corrupted1_vol1_robustness_slice3-eps-converted-to.pdf}%
                \linethickness{2pt}
                \put(5,13){{\tiny\color{white}PSNR: 25.40}}
                \put(5,6){{\tiny\color{white}SSIM: 0.7616}}
            \end{overpic}%
        \end{subfigure}%
        \begin{subfigure}[b]{0.24\textwidth}
            \caption*{Corrected (ours)}\vspace{-0.5em}%
            \begin{overpic}[width=\textwidth]{./figs/robustness/vol1/corrected1_vol1_robustness_slice3-eps-converted-to.pdf}%
                \linethickness{2pt}
                \put(5,13){{\tiny\color{white}PSNR: 30.16}}
                \put(5,6){{\tiny\color{white}SSIM: 0.8490}}
            \end{overpic}
        \end{subfigure}%
        \begin{subfigure}[b]{0.24\textwidth}
            \caption*{Corrected (baseline)}\vspace{-0.5em}%
            \begin{overpic}[width=\textwidth]{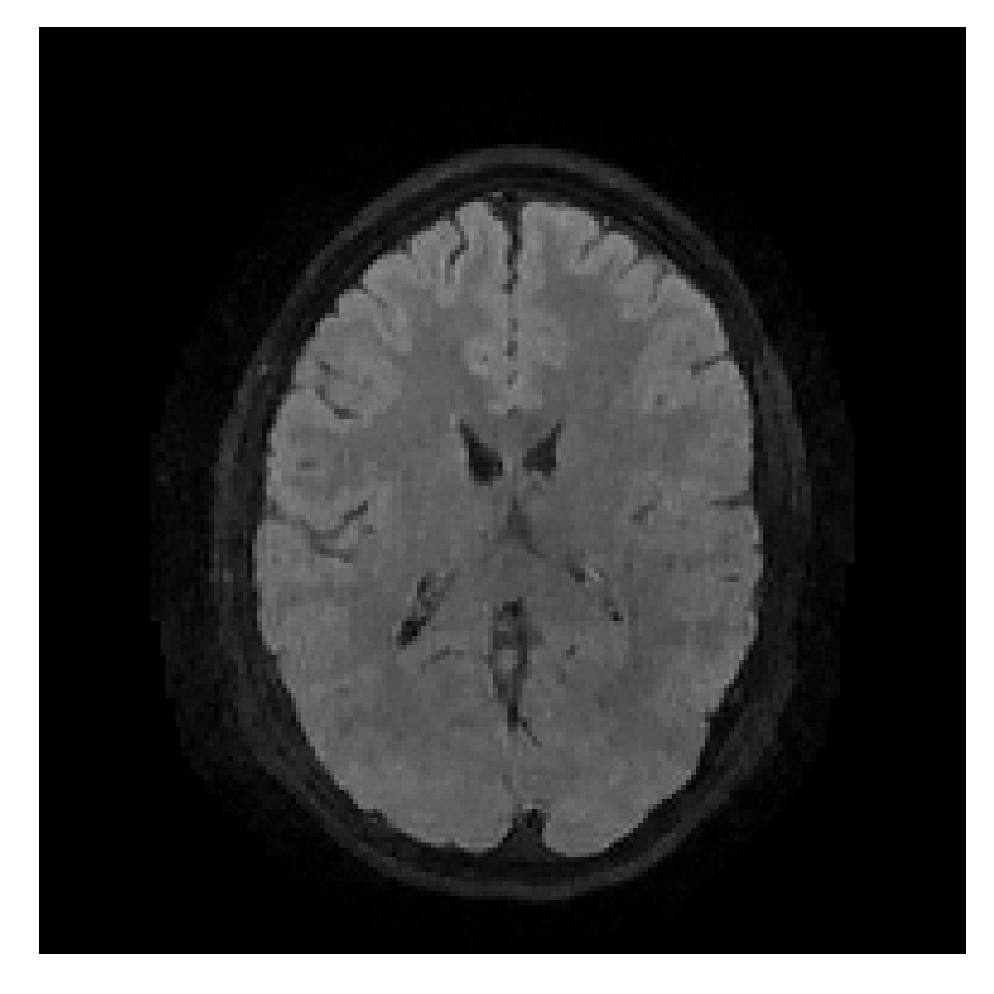}%
                \linethickness{2pt}
                \put(5,13){{\tiny\color{white}PSNR: 27.54}}
                \put(5,6){{\tiny\color{white}SSIM: 0.8007}}
            \end{overpic}
        \end{subfigure}%
        \begin{subfigure}[b]{0.24\textwidth}
            \caption*{Ground truth}\vspace{-0.5em}%
            \includegraphics[width=\textwidth]{./figs/robustness/vol1/ground_truth_vol1_robustness_slice3-eps-converted-to.pdf}%
        \end{subfigure}%
    \end{subfigure}
    \begin{subfigure}{\textwidth}
        \centering
        \rotatebox{90}{\hspace{2em}Move twice} %
        \begin{subfigure}[b]{0.24\textwidth}
            \begin{overpic}[width=\textwidth]{./figs/robustness/vol1/corrupted2_vol1_robustness_slice3-eps-converted-to.pdf}%
                \linethickness{2pt}
                \put(5,13){{\tiny\color{white}PSNR: 27.79}}
                \put(5,6){{\tiny\color{white}SSIM: 0.8104}}
            \end{overpic}%
        \end{subfigure}%
        \begin{subfigure}[b]{0.24\textwidth}
            \begin{overpic}[width=\textwidth]{./figs/robustness/vol1/corrected2_vol1_robustness_slice3-eps-converted-to.pdf}%
                \linethickness{2pt}
                \put(5,13){{\tiny\color{white}PSNR: 29.70}}
                \put(5,6){{\tiny\color{white}SSIM: 0.8362}}
            \end{overpic}
        \end{subfigure}%
        \begin{subfigure}[b]{0.24\textwidth}
            \begin{overpic}[width=\textwidth]{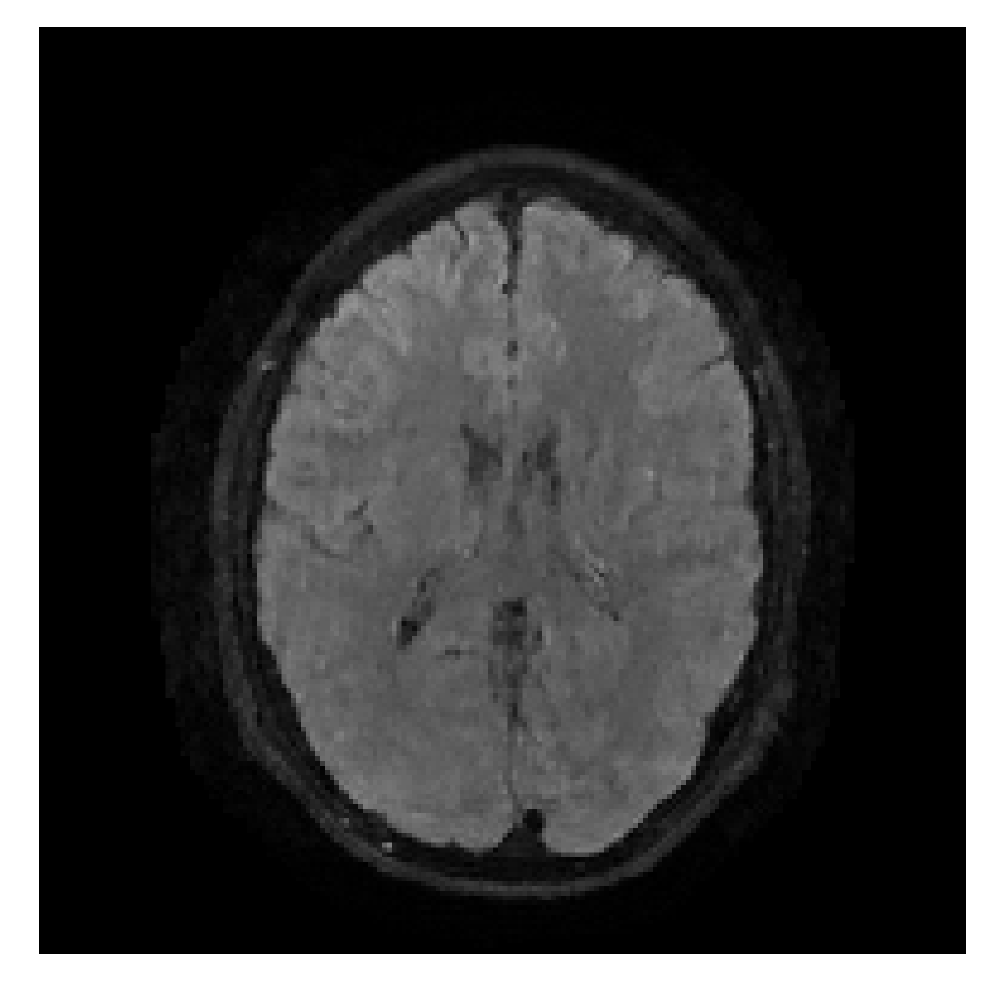}%
                \linethickness{2pt}
                \put(5,13){{\tiny\color{white}PSNR: 26.57}}
                \put(5,6){{\tiny\color{white}SSIM: 0.7708}}
            \end{overpic}
        \end{subfigure}%
        \begin{subfigure}[b]{0.24\textwidth}
            \includegraphics[width=\textwidth]{./figs/robustness/vol1/ground_truth_vol1_robustness_slice3-eps-converted-to.pdf}%
        \end{subfigure}%
    \end{subfigure}
    \begin{subfigure}{\textwidth}
        \centering
        \rotatebox{90}{\hspace{1.5em}Move five times} %
        \begin{subfigure}[b]{0.24\textwidth}
            \begin{overpic}[width=\textwidth]{./figs/robustness/vol1/corrupted3_vol1_robustness_slice3-eps-converted-to.pdf}%
                \linethickness{2pt}
                \put(5,13){{\tiny\color{white}PSNR: 24.15}}
                \put(5,6){{\tiny\color{white}SSIM: 0.7086}}
            \end{overpic}%
        \end{subfigure}%
        \begin{subfigure}[b]{0.24\textwidth}
            \begin{overpic}[width=\textwidth]{./figs/robustness/vol1/corrected3_vol1_robustness_slice3-eps-converted-to.pdf}%
                \linethickness{2pt}
                \put(5,13){{\tiny\color{white}PSNR: 27.66}}
                \put(5,6){{\tiny\color{white}SSIM: 0.8298}}
            \end{overpic}
        \end{subfigure}%
        \begin{subfigure}[b]{0.24\textwidth}
            \begin{overpic}[width=\textwidth]{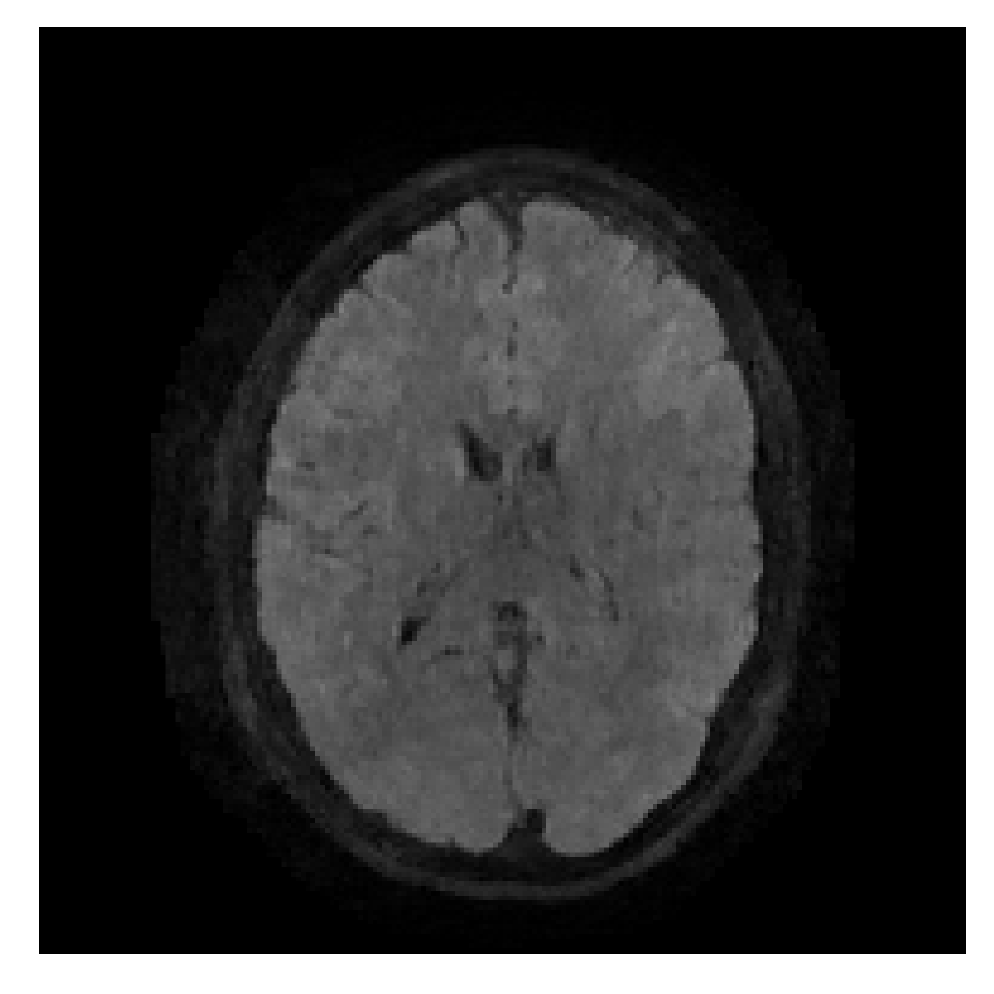}%
                \linethickness{2pt}
                \put(5,13){{\tiny\color{white}PSNR: 24.96}}
                \put(5,6){{\tiny\color{white}SSIM: 0.7562}}
            \end{overpic}
        \end{subfigure}%
        \begin{subfigure}[b]{0.24\textwidth}
            \includegraphics[width=\textwidth]{./figs/robustness/vol1/ground_truth_vol1_robustness_slice3-eps-converted-to.pdf}%
        \end{subfigure}%
    \end{subfigure}
    \caption{Comparison of the reconstruction results for volunteer 1 with a reference-guided (ours) and a baseline motion-correction method (e.g., not guided by a reference contrast). The volunteer is instructed to move a variable number of times during the scan in order to test the robustness of the proposed correction schemes with respect the motion complexity. The corrupted images are increasingly affected by motion artifacts. The decrease in reconstruction quality for the baseline method is substantially more pronounced than the results obtained with our reference-guided correction (see also Figures \ref{fig:robustness1_vol2}--\ref{fig:robustness3_vol2}).}\label{fig:baseline}
\end{figure}

For the comparison with the baseline method, we use the same experimental settings in Section \ref{exp:robust}. Once again, the motion artifacts are prospectively induced by prompting the volunteer to move during the scan. The results are summarized in Figure \ref{fig:baseline} and Table \ref{tab:baseline}.%
\begin{table}
    \centering
    \resizebox{\textwidth}{!}{%
    \begin{tabular}{|l|l|ccc|ccc|}
        \hline
        Experiment & Slice orientation & \multicolumn{3}{c|}{PSNR ($\uparrow$)} & \multicolumn{3}{c|}{SSIM ($\uparrow$)}\\
        & & Corrupted & \multicolumn{2}{c|}{Corrected} & Corrupted & \multicolumn{2}{c|}{Corrected}\\
        & & & Ours & Baseline & & Ours & Baseline\\
        \hline
        Move once & Sagittal & 23.94 & \textbf{27.95} & 25.27 & 0.7068 & \textbf{0.7936} & 0.7529\\
                  & Coronal  & 26.66 & \textbf{29.82} & 27.61 & 0.7653 & \textbf{0.8332} & 0.7818\\
                  & Axial    & 25.40 & \textbf{30.16} & 27.54 & 0.7616 & \textbf{0.8490} & 0.8007\\
        \hline
        Move twice & Sagittal & 25.78 & \textbf{27.76} & 24.13 & 0.7263 & \textbf{0.7816} & 0.6925\\
                   & Coronal  & 28.19 & \textbf{29.73} & 26.68 & 0.7847 & \textbf{0.8244} & 0.7448\\
                   & Axial    & 27.79 & \textbf{29.70} & 26.57 & 0.8104 & \textbf{0.8362} & 0.7708\\
        \hline
        Move five times & Sagittal & 22.45 & \textbf{25.28} & 22.10 & 0.6116 & \textbf{0.7661} & 0.6719\\
                        & Coronal  & 24.54 & \textbf{27.40} & 24.72 & 0.6734 & \textbf{0.8060} & 0.7327\\
                        & Axial    & 24.15 & \textbf{27.66} & 24.96 & 0.7086 & \textbf{0.8298} & 0.7562\\     
        \hline
    \end{tabular}%
    }%
    \caption{Comparison of the motion-correction results for the baseline and reference-guided methods in terms of PSNR and SSIM. The experiment setup is described in Section \ref{exp:robust}.}\label{tab:baseline}
\end{table}

We note that the difference in performance between the reference-guided and blind motion correction is even more pronounced in this example than what was previously shown in \citet{rizzutiMOCO} (which was limited to 2D synthetic data). This might depend on the fact that the problem is substantially more ill-posed in 3D with randomized sampling than the 2D full-acquisition setup considered in \citet{rizzutiMOCO}. It is also worth noting that, in our experience, the results for blind motion correction depend more sensibly on the choice of the hyper-parameters in equation \eqref{eq:opt} than the proposed reference-based version.

\section{Motion parameter estimation}\label{app:motionparameters}

The proposed motion correction algorithm described in Section \ref{sec:theory} estimates the rigid motion that the object of interest undergoes during the scan, in order to undo its effect on the reconstructed 3D image. In 3D, the rigid motion is performed by: a plane rotation $\theta_{xy}$ in the corresponding plane $xy$, a plane rotation $\theta_{xz}$ in the $xz$ plane, a plane rotation $\theta_{yz}$ in the $yz$ plane, a translation $\tau_x$ in the $x$ direction, a translation $\tau_y$ in the $y$ direction, and a translation $\tau_z$ in the $z$ direction (in this order). We adopt the following convention: the $x$ direction corresponds to the left-right direction, $y$ to the posterior-anterior direction, and $z$ to the inferior-superior direction, the $xy$ plane corresponds to the axial plane, $xz$ to the coronal plane, and $yz$ to the sagittal plane. Left/right, anterior/posterior, and inferior/superior are meant from the patient perspective. The orientation of the rotation planes is determined by the right-hand rule.

\begin{figure}[!htb]
    \centering
    \includegraphics[width=0.7\textwidth]{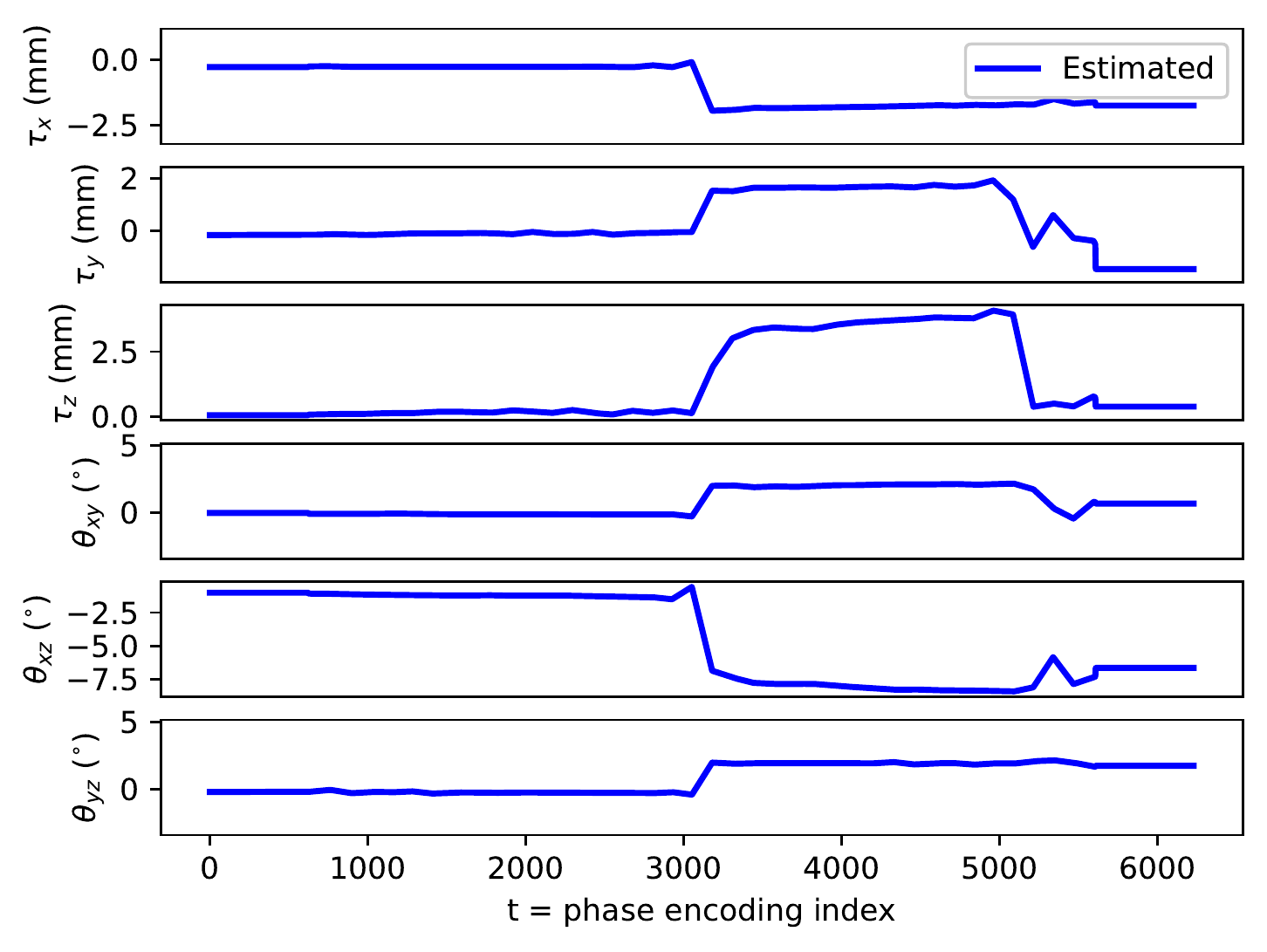}%
    \caption{Estimated rigid motion parameters for the experiment described in Sections \ref{exp:robust}, \ref{res:robust}, with motion-correction results in Figure \ref{fig:robustness1_vol2}. The volunteer was asked to move once during the scan.}\label{fig:mpars_robust1}
\end{figure}

\begin{figure}[!htb]
    \centering
    \includegraphics[width=0.7\textwidth]{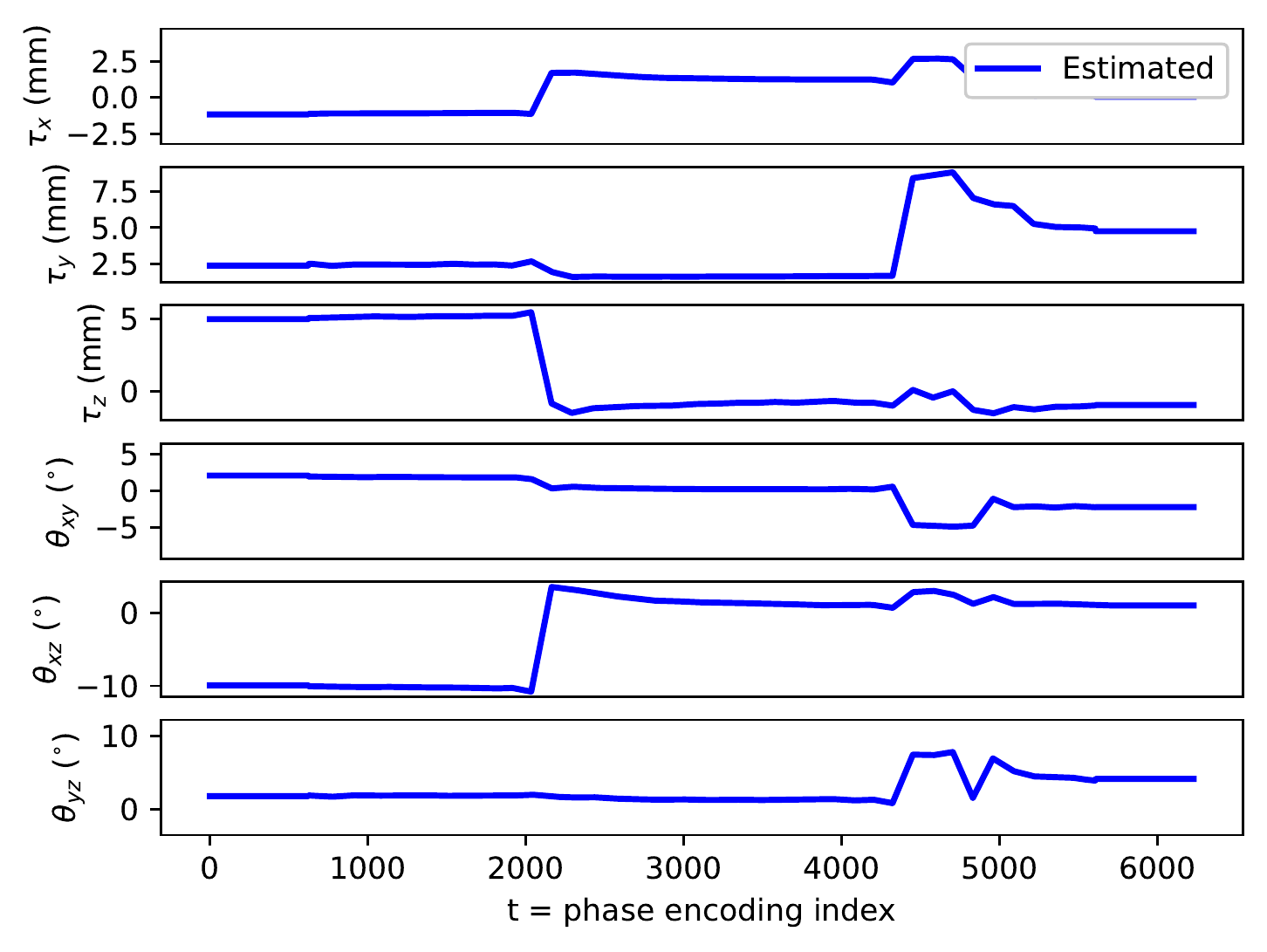}%
    \caption{Estimated rigid motion parameters for the experiment described in Sections \ref{exp:robust}, \ref{res:robust}, with motion-correction results in Figure \ref{fig:robustness2_vol2}. The volunteer was asked to move twice during the scan.}\label{fig:mpars_robust2}
\end{figure}

\begin{figure}[!htb]
    \centering
    \includegraphics[width=0.7\textwidth]{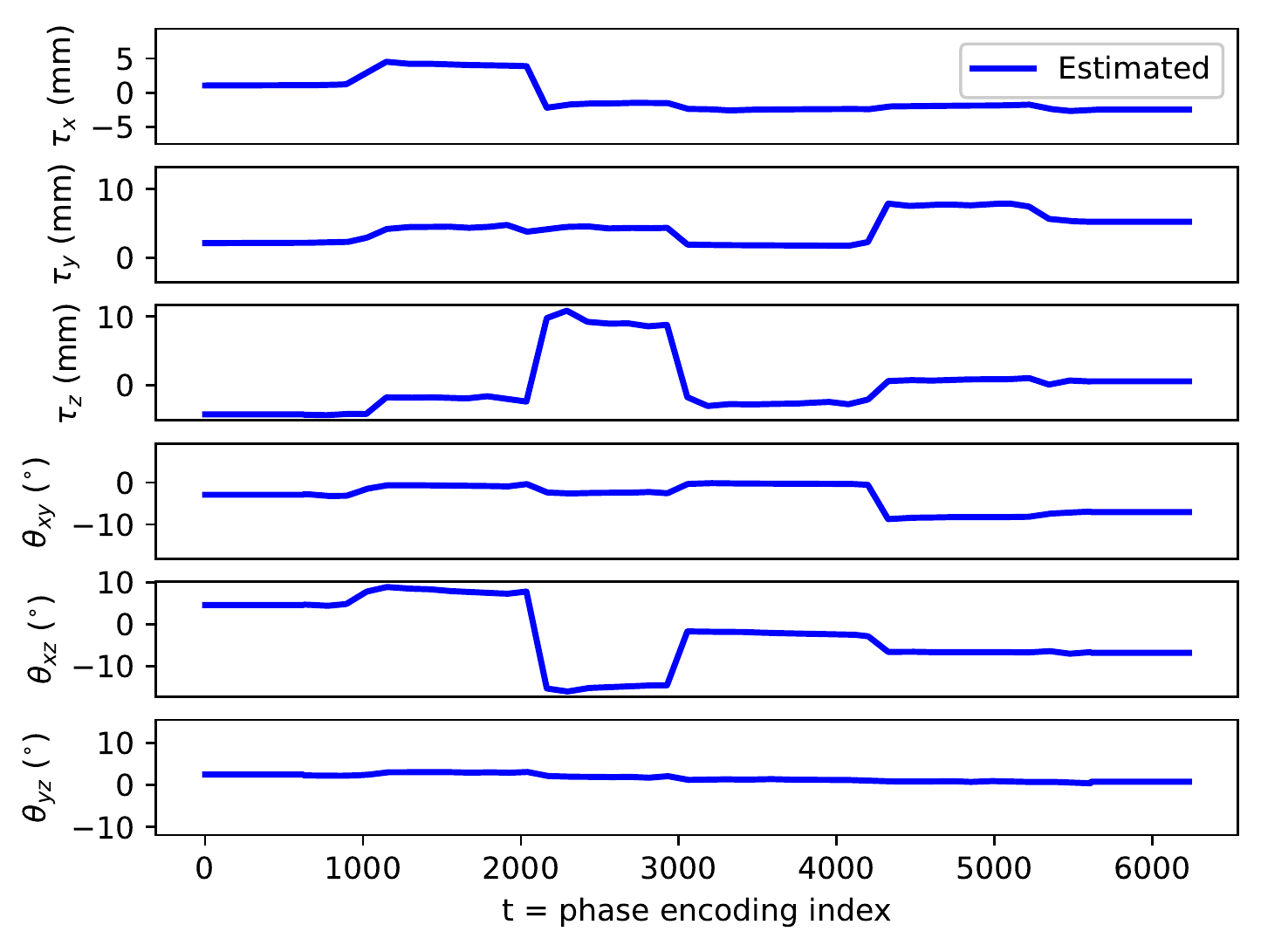}%
    \caption{Estimated rigid motion parameters for the experiment described in Sections \ref{exp:robust}, \ref{res:robust}, with motion-correction results in Figure \ref{fig:robustness3_vol2}. The volunteer was asked to move five times during the scan.}\label{fig:mpars_robust3}
\end{figure}

\begin{figure}[!htb]
    \centering
    \includegraphics[width=0.7\textwidth]{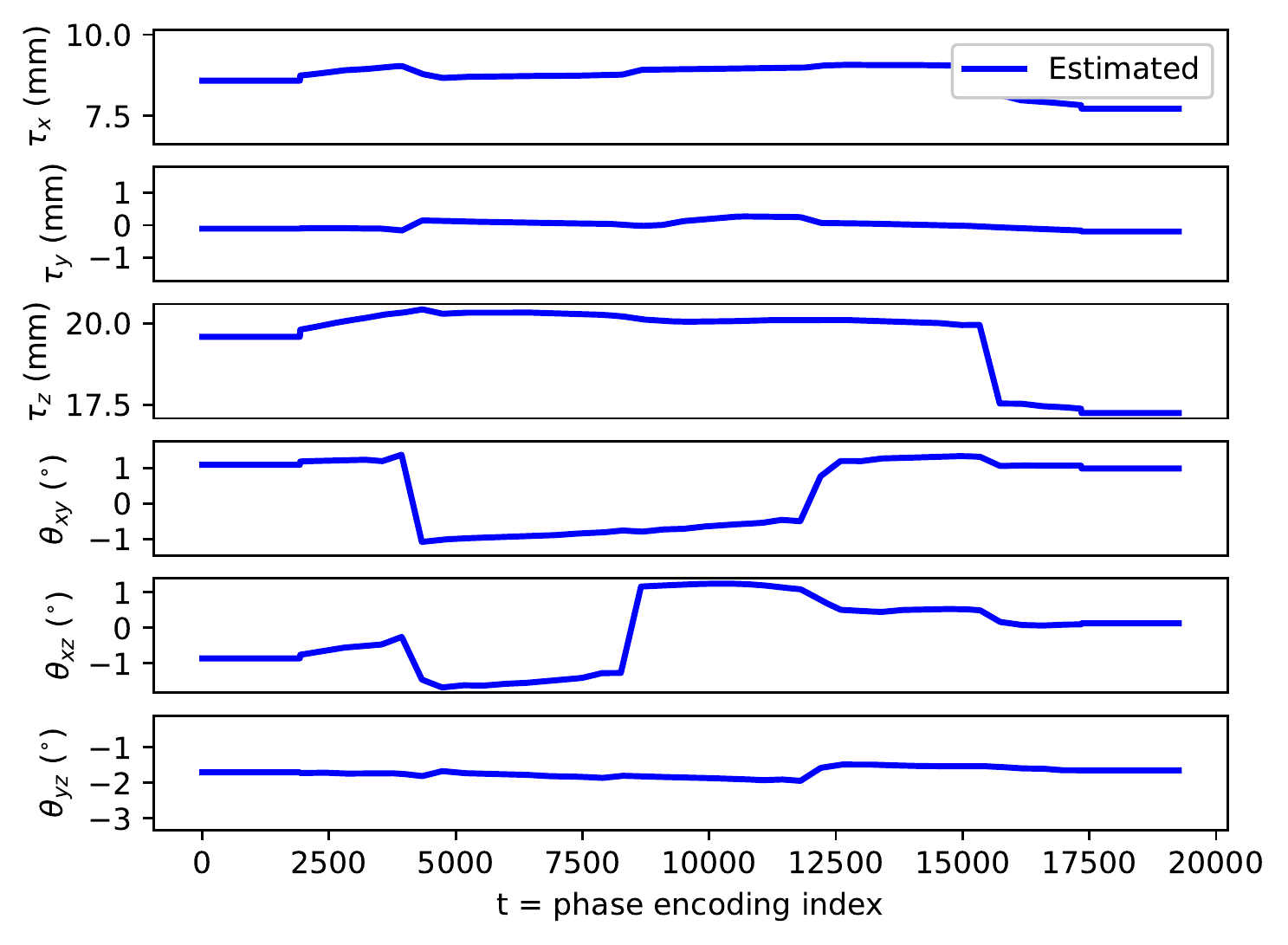}%
    \caption{Estimated rigid motion parameters for the experiment described in Sections \ref{exp:prior}, \ref{res:prior}, with motion-correction results in Figure \ref{fig:prior}}\label{fig:mpars_prior}
\end{figure}

\begin{figure}[!htb]
    \centering
    \includegraphics[width=0.7\textwidth]{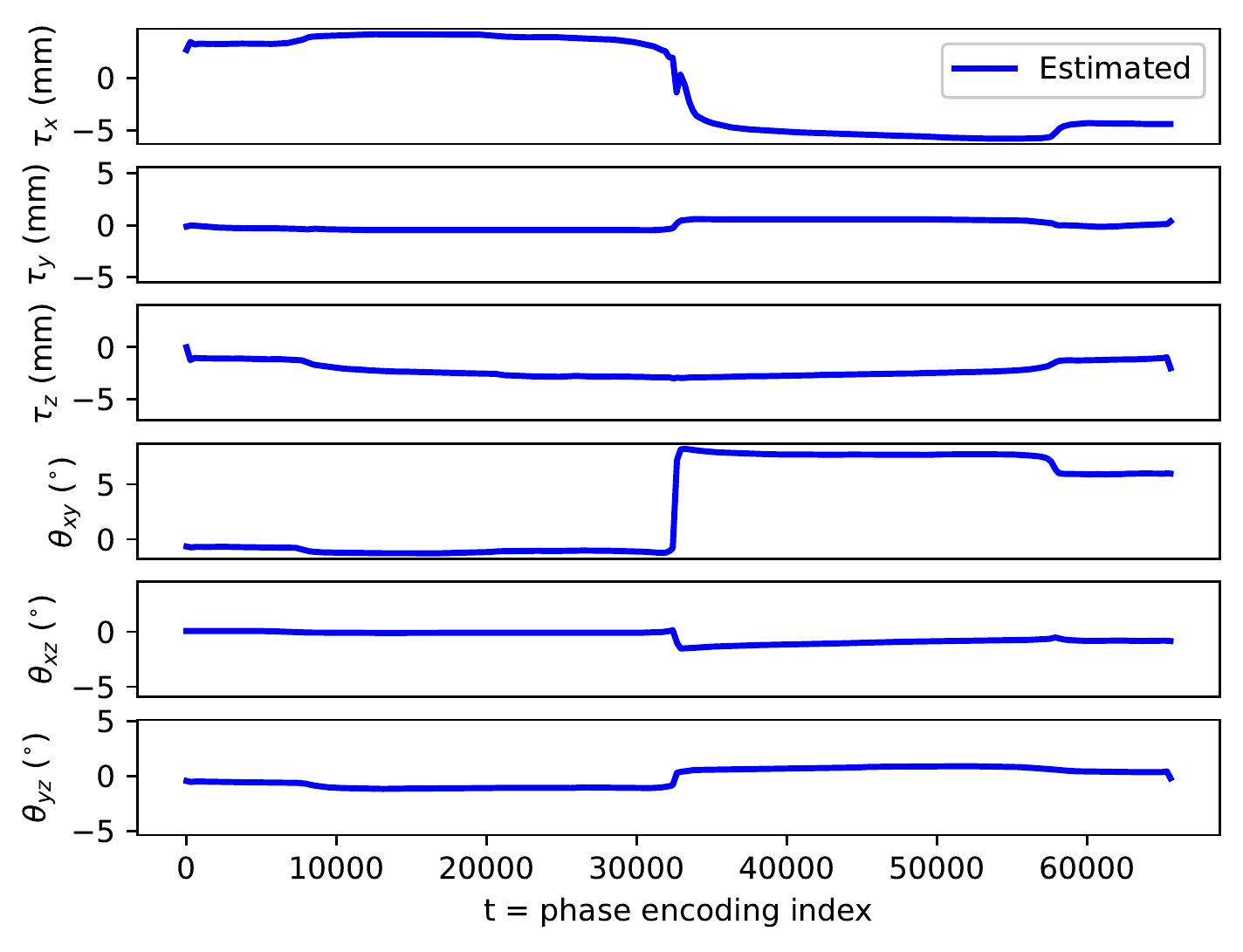}%
    \caption{Estimated rigid motion parameters for the experiment described in Sections \ref{exp:data}, \ref{res:data}, with motion-correction results in Figure \ref{fig:scanreconvsrawdata1}}\label{fig:mpars_data1}
\end{figure}

\begin{figure}[!htb]
    \centering
    \includegraphics[width=0.7\textwidth]{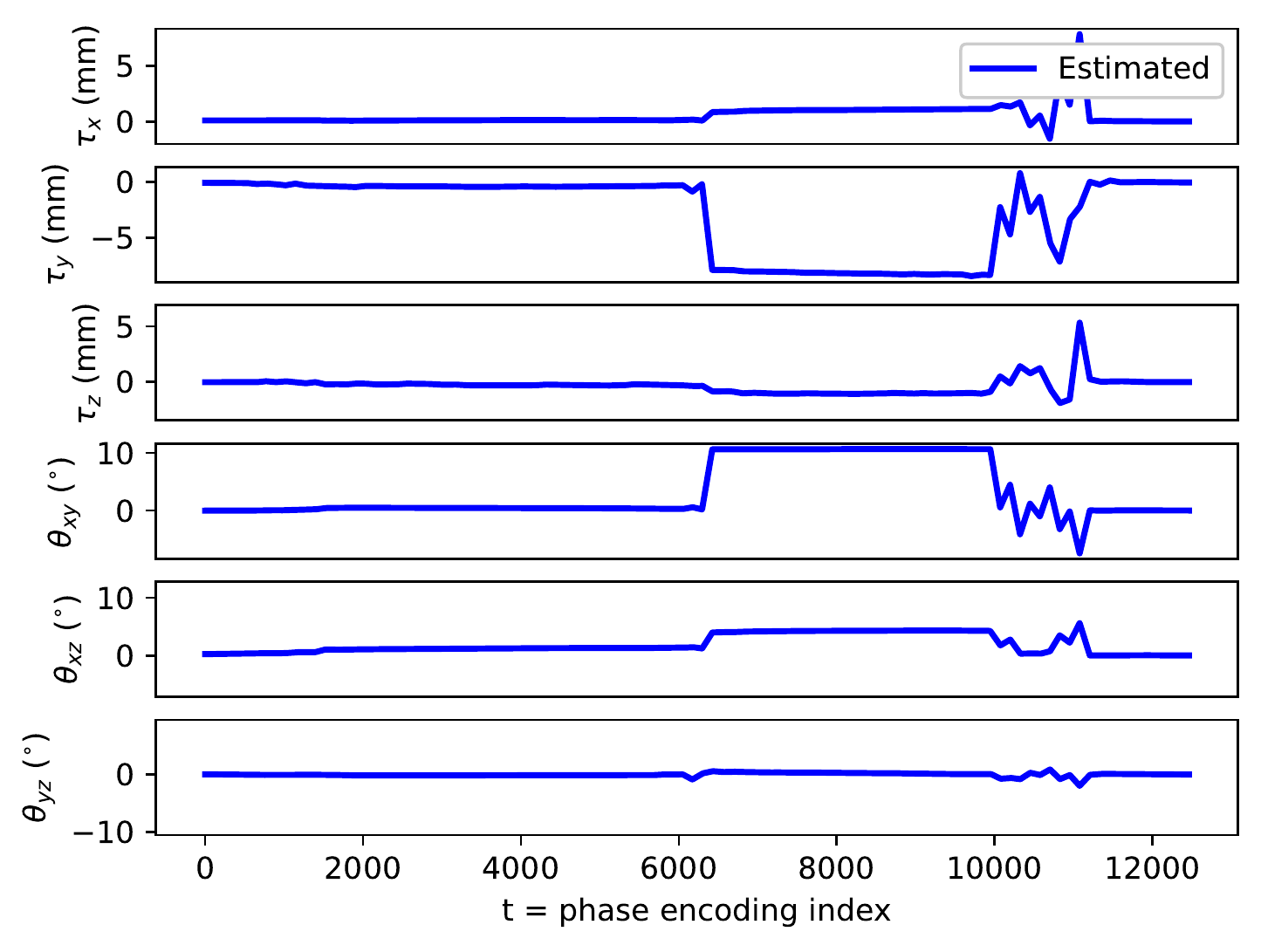}%
    \caption{Estimated rigid motion parameters for the experiment described in Sections \ref{exp:data}, \ref{res:data}, with motion-correction results in Figure \ref{fig:scanreconvsrawdata2}}\label{fig:mpars_data2}
\end{figure}

By design, the prospectively-induced motion for all the experiments detailed in Section \ref{sec:exp} follows a step-wise behavior (each step corresponding to a change of pose). In this appendix, we gather the estimated rigid motion parameters for the results shown in Section \ref{sec:results}, as a function of time. As noted in the main body of the paper, time is equated to the phase-encoding plane coordinate index, ordered by the corresponding acquisition ordering. We display the estimated motion parameters in Figure \ref{fig:mpars_robust1} (see Sections \ref{exp:robust}, \ref{res:robust}, Figure \ref{fig:robustness1_vol2}), Figure \ref{fig:mpars_robust2} (see Sections \ref{exp:robust}, \ref{res:robust}, Figure \ref{fig:robustness2_vol2}), Figure \ref{fig:mpars_robust3} (see Sections \ref{exp:robust}, \ref{res:robust}, Figure \ref{fig:robustness3_vol2}), Figure \ref{fig:mpars_prior} (see Sections \ref{exp:prior}, \ref{res:prior}, Figure \ref{fig:prior}), Figure \ref{fig:mpars_data1} (see Sections \ref{exp:data}, \ref{res:data}, Figure \ref{fig:scanreconvsrawdata1}), and Figure \ref{fig:mpars_data2} (see Sections \ref{exp:data}, \ref{res:data}, Figure \ref{fig:scanreconvsrawdata2}).

\end{document}